# Visible Imaging of Incoherent 1200-nm Light *via* Triplet–Triplet Annihilation Upconversion

*Pournima Narayanan[1,2], Rabeeya Hamid[3], Linda Pucurimay[1,4], Ona Segura Lecina[1], Ben P. Carwithen[5], Jacob Schopp[3], Justin S. Edwards[3,6], Oluwaseun Noah Adeyeye[1], Demeng Feng[3], Diptarka Hait[2,7,8,9], Todd J. Martínez[2,7], Timothy W. Schmidt[5], Michael P. Nielsen[10], Murad J. Y. Tayebjee[10], Mikhail A. Kats[3,6], Daniel N. Congreve[1*]*

[1] Department of Electrical Engineering, Stanford University, Stanford, CA 94305, USA

[2] Department of Chemistry, Stanford University, Stanford, CA 94305, USA

[3] Department of Electrical and Computer Engineering, University of Wisconsin–Madison, Madison, WI 53706, USA

[4] Department of Material Science Engineering, Stanford University, Stanford, CA 94305, USA

[5] School of Chemistry, University of New South Wales, Sydney, NSW 2052, Australia

[6] Department of Physics, University of Wisconsin–Madison, Madison, 53706, USA, USA

[7] The PULSE Institute, SLAC National Accelerator Laboratory, Menlo Park, CA 94025, USA

[8] Department of Chemistry, Columbia University, New York, NY 10027, USA

[9] Initiative for Computational Catalysis, Flatiron Institute, New York, NY 10010, USA

[10] School of Photovoltaic and Renewable Energy Engineering, University of New South Wales, Sydney, NSW 2052, Australia

*Corresponding author email: congreve@stanford.edu

ABSTRACT. Upconversion of low-energy photons to higher-energy photons provides an opportunity to surpass traditional limitations in fields such as 3D printing, photovoltaics, and photocatalysis. Triplet-triplet annihilation upconversion (TTA-UC) is particularly appealing for such applications as it can efficiently upconvert low-intensity, incoherent light. However, previously demonstrated thin-film TTA systems are simultaneously constrained by modest efficiencies and limited reach into the near infrared (NIR). Here, we design a single-layer thin-film bulk heterojunction that integrates PbS quantum dots (QDs) as tunable NIR absorbers within an organic semiconductor matrix of TES-ADT, achieving large anti-Stokes shifts up to ~500 nm and high internal quantum efficiencies across the NIR-I and NIR-II windows (800-1200 nm). Through the incorporation of 5-tetracene carboxylic acid ligands on the PbS QD surface, the yield of sensitized triplets was boosted, as confirmed by transient absorption and time-resolved photoluminescence measurements. The resulting films demonstrated a 15-fold improvement in UC efficiency. Further, we demonstrate visible imaging of incoherent 1200 nm light via thin-film TTA-UC at incident intensities at the imaging mask as low as ~20 $mWcm^{-2}$, marking a significant advance toward practical implementation of solid-state NIR-to-visible upconversion.

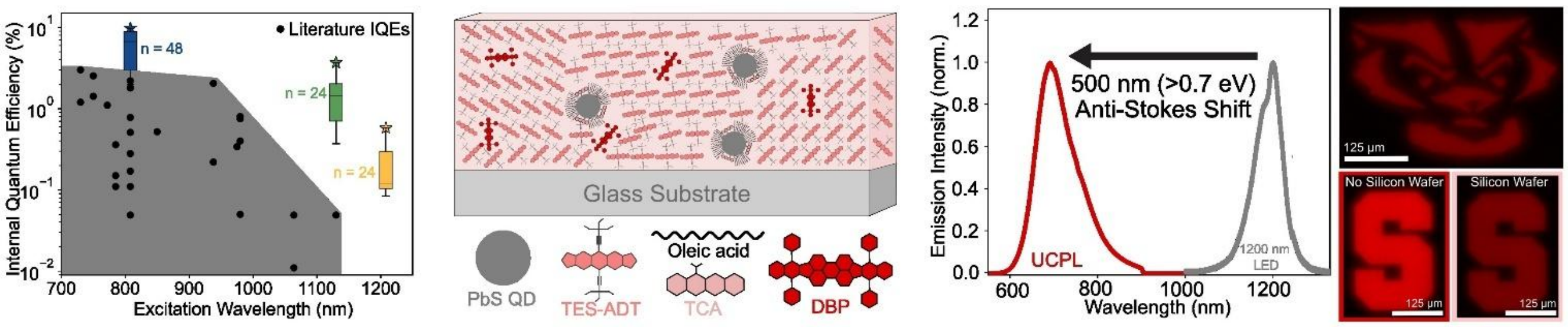

ToC TEXT. Here, thin-film bulk heterojunctions which integrate PbS quantum dots within an organic semiconductor matrix of TES-ADT are reported to achieve high upconversion quantum efficiencies across the near infrared I and II windows (800-1200 nm). Upconversion performance was improved 15-fold through the incorporation of 5-tetracene carboxylic acid ligands, thereby enabling visible imaging of low-intensity, incoherent 1200 nm light.

## MAIN TEXT

By combining the energy of two low-energy photons, upconversion (UC) can generate higher-energy photons *in-situ*, opening pathways in diverse applications including photovoltaics,[1,2] additive manufacturing,[3,4] night vision,[5] and photocatalysis.[6–8] UC of near infrared (NIR) photons into the visible is particularly appealing for night vision, below-bandgap harvesting for solar cells, and NIR imaging. Several mechanisms, including photon avalanching,[9] excited-state absorption,[10,11] and triplet-triplet annihilation (TTA),[12] have been employed for the NIR. Among these, TTA-UC is particularly promising due to its relatively high UC efficiencies under low-intensity illumination, eliminating the need for pulsed-laser excitation and allowing UC of incoherent light.[12,13] However, the implementation of TTA-based NIR UC in these contexts remains limited by modest efficiencies and relatively limited reach into the NIR.

TTA-UC relies on photosensitizers to absorb low-energy light, followed by Dexter energy transfer to generate triplet states in organic semiconductor molecules known as annihilators (Figure 1a).[14–16] Two triplet-excited annihilators can interact in a concerted, spin-allowed process through which one annihilator is promoted to a higher-energy singlet state while the other relaxes to the ground state.[12,17,18] The resulting singlet-excited annihilator can then relax by emitting a higher-energy photon. To enhance UC efficiency, highly emissive dopants are often incorporated into the UC matrix, serving as singlet sinks that extract singlet excitons from the annihilator to mitigate triplet-pair separation.[15,19,20] In thin films, rubrene has been widely employed as an annihilator for NIR-to-visible TTA-UC, with DBP (tetraphenyldibenzoperiflanthene) as a dopant emitter. The rubrene/DBP system has shown UC internal quantum efficiencies (IQEs, defined as the ratio of emitted photons to absorbed photons) of up to a few percent when paired with a range of sensitizers including organic semiconductors,[21–23] lead sulfide quantum dots (PbS QDs),[20,24–30] transition metal dichalcogenides,[31,32] porphyrin complexes,[33] osmium complexes,[1,34] and perovskites.[30,35] Cyano-derivates of rubrene with PdPc as sensitizer showed IQE up to 3% at 730 nm excitation.[33] ITIC-Cl/rubrene and Y6/rubrene have shown UC efficiency as high as 2.53% at 750 nm and 0.51% at 808 nm, respectively.[22,23] Osmium complexes with direct S-T excitation have also been deployed with rubrene to show up to 2.05% IQE at 730 nm and have been reported to upconvert as deep as 938 nm.[1] However, the lack of tunability of these organic and organometallic sensitizers limits their use for deeper NIR UC. Among sensitizer options, PbS QDs are particularly attractive as sensitizers for NIR-to-visible UC due to their high tunability, enabling absorption from 700 nm to 2000 nm, depending on the QD size.[36,37]

While PbS QD/rubrene-based systems have demonstrated IQEs as high as 1.6% at 808 nm (~1.53 eV),[24,28] their efficiency rolls off sharply at longer wavelengths due to rubrene's triplet energy (~1.14 eV).[20] This has left applications like night vision, photovoltaics, and bio-imaging out of reach due to poor conversion of photons in the NIR II window (1000-1700 nm).[38,39] While violanthrone 79 is promising annihilator which has shown UC from 1140 nm in solutions, it is has not been deployed in thin films.[40] TES-ADT (5,11-bis(triethylsilylethynyl)anthradithiophene) has been identified as a promising thin-film alternative to rubrene, offering a lower triplet energy (1.08 eV, ~1150 nm) to achieve UC from as deep as 980 nm when paired with DBP as an emitter.[7,41–44] Mixtures of PbS QDs, TES-ADT, and DBP were drop-cast to form films,[41,42] achieving IQEs as high as 0.75%.[41] More recently, TES-ADT has been studied to show spin statistics which favor TTA-UC in aggregates and thin films.[45] In 2025, TES-ADT and PbS QDs were deployed in a pseudo solid-state polymer matrix, reaching efficiencies up to 0.011% at 1064 nm.[46] However, UC demonstrations with TES-ADT have not been reported beyond 1064 nm excitation.

In this work, we demonstrate that the PbS/TES-ADT/DBP UC system has been fundamentally limited by triplet energy transfer between PbS and TES-ADT. By introducing 5-tetracene carboxylic acid (TCA) as a mediator ligand to facilitate exciton extraction from PbS QDs, we address this crucial bottleneck and improve the quantum efficiency of the system by 15×, demonstrating a champion IQE of 1.9% at 808 nm excitation of 3.5 W/cm$^2$. Excitingly, the efficiency enhancement is maintained throughout the NIR and allows UC from wavelengths as deep as 1208 nm. Surface functionalization with TCA is confirmed via nuclear magnetic resonance (NMR) spectroscopy and correlated with strongly improved UC efficiencies using kinetics measured using transient absorption and time-resolved photoluminescence. Finally, we demonstrate visible imaging of low-intensity, incoherent 1200-nm light (FWHM = 65 nm) via solid-state, thin-film TTA-UC, highlighting the applicability of this system in imaging, night vision, and photovoltaics.

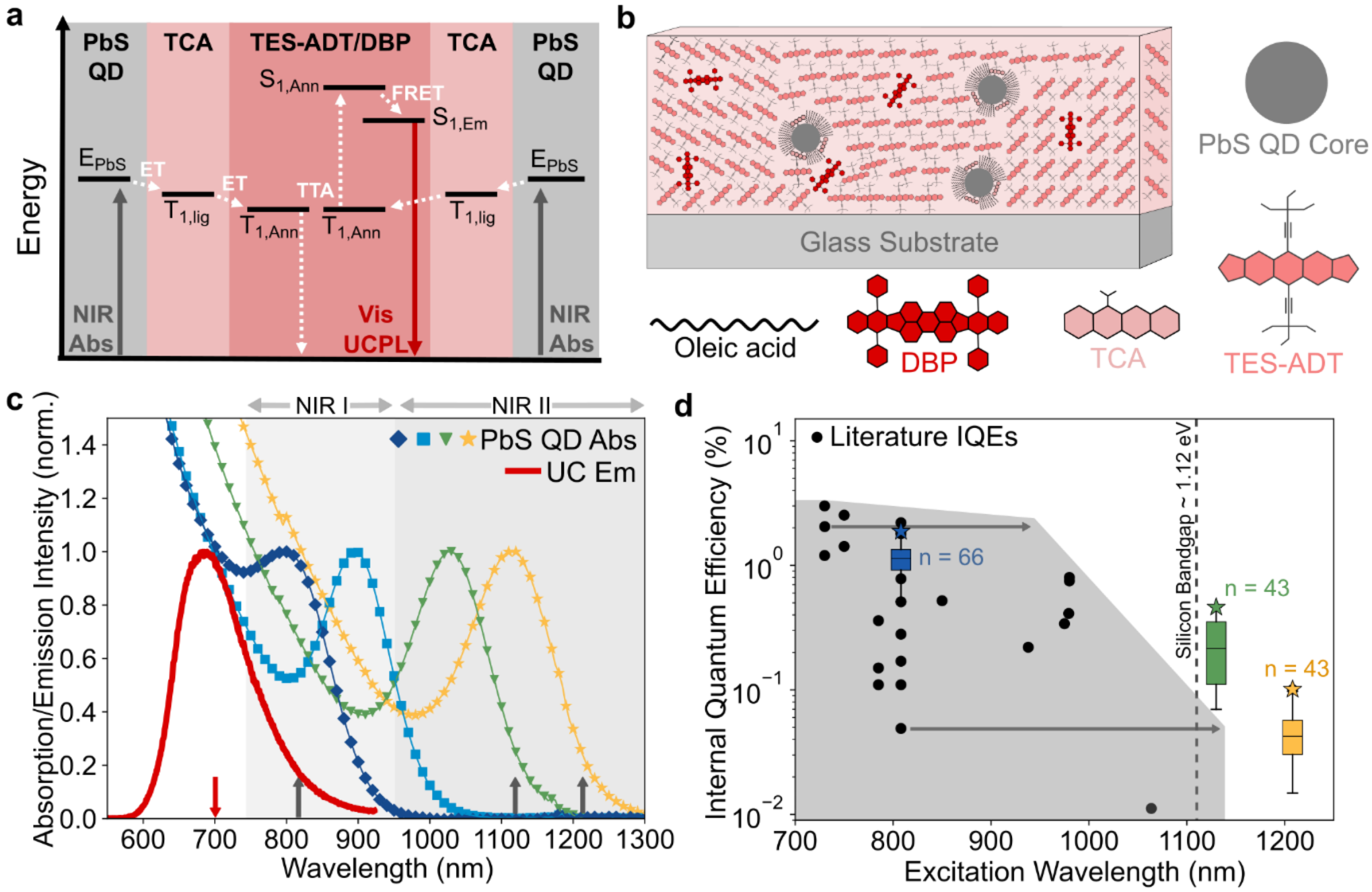

**Figure 1:** Schematic representation and spectral characterization of the TTA-UC system. **(a)** UC energy diagram showing the conversion of absorbed NIR photons into visible photons via absorption, energy transfer (ET), TTA, and Förster resonant energy transfer (FRET) steps. **(b)** Illustrative schematic of the BHJ UC film consisting of TCA-modified PbS QDs, TES-ADT, and DBP. **(c)** Normalized absorption spectra of size-tuned PbS QDs (800 nm – dark blue diamonds,

900 nm – light blue squares, 1020 nm – green triangles, 1120 nm – yellow stars) and normalized UC emission from TES-ADT/DBP (red trace). Gray arrows are used to annotate various laser excitation wavelengths (808, 1130, 1208 nm) used to observe UC. The red arrow highlights the peak wavelength of UC emission (700 nm). **(d)** Representation of IQEs reported in NIR-to-visible TTA-UC systems in literature[1,20–24,26–29,32–35,41,42,46–48] (black circles, see SI Note 1, Table SI 1) and from this work (colored box-and-whisker plots), with champion efficiencies represented by colored stars. The number of measurements used for each box-and-whisker plot is highlighted (3-5 measurements were taken on each of 16, 11, and 10 films for the 808, 1130, 1208 nm excitations). The top gray arrow signifies that although the IQE was measured at 730 nm for the films reported by Kinoshita et al., UC was demonstrated as deep as 938 nm.[1] Similarly, Wu et al. measured IQE at 808 nm, but reported UC till ~1130 nm through an excitation spectrum, as represented by the lower gray arrow.[20]

## RESULTS

### Film Fabrication and TCA Ligand Exchange:

To fabricate the BHJ film architecture shown in Figure 1b, we used a one-step spin-coating process with blended solutions containing the UC components: NIR-absorbing PbS QDs (sensitizers), TES-ADT (annihilator), and DBP (emitter dopant). The PbS QDs with native oleic acid ligands were synthesized using a modified Hines hot-injection protocol, where increasing the injection temperature resulted in larger QDs with decreasing bandgaps.[36] The resulting QDs displayed first excitonic absorption peaks shifting from 760 to 1150 nm (1.63-1.08 eV), enabling UC sensitization for the 808, 1130, and 1208 nm lasers used in this work (Figure 1c). These QDs were employed to make PbS/TES-ADT/DBP films which demonstrated UC photoluminescence (UCPL) at 700 nm (Figure 1c) but initially suffered from non-uniformity (Figure SI 3) and poor performance due to low PbS QD solubility.

To address these limitations, we proposed the introduction of ligands to the surface of the PbS QDs as a strategy to improve QD dispersion in the TES-ADT/DBP film and boost exciton extraction from the QDs. Previously, Gray, et al. explored the influence of similar ligand modifications on organic–inorganic self-assembly for UC and singlet fission and found that the introduction of annihilator-like mediator ligands on PbS QDs enables uniform dispersion of these QDs within the acene-based organic matrix.[25] Here, we introduced TCA to the surface of the PbS QDs in a post-synthetic exchange, which has been previously demonstrated to improve energy transfer in the PbS/rubrene system.[27,49–51] TCA is an appealing ligand due to its carboxylic acid moiety which allows strong binding with PbS QDs upon one-to-one ligand exchange with the native oleic acid ligands. This strong binding can increase the rate of Dexter energy transfer from the PbS QDs. Additionally, its triplet energy level of ~1.3 eV makes TCA appealing for the TES-ADT UC system. Since this triplet energy level is higher than the triplet of TES-ADT (~1.08 eV), it can serve as a mediator for energy transfer from the PbS QDs to the TES-ADT. Our quantum chemical calculations (Methods, SI Table 2) show that upon coordination to $Pb^{2+}$ ions, TCA's triplet energy is lowered further due to charge transfer, with the unpaired electrons occupying the highest occupied molecular orbital of TCA and the 6p orbital of $Pb^{2+}$. Therefore, TCA holds potential to serve as an efficient mediating ligand between PbS QDs and TES-ADT.

We used a solution-state ligand exchange protocol where the PbS QDs in toluene and TCA in acetonitrile were stirred together to facilitate replacement of oleic acid with TCA ligands. These TCA-functionalized QDs were then purified using centrifugation and the ligand exchange was confirmed using proton nuclear magnetic resonance spectroscopy ($^1$H NMR) and absorption spectroscopy. Oleic acid possesses vinylic protons (5.3 ppm, highlighted in gray, Figure 2a) and TCA has aromatic protons (7-9.5 ppm, highlighted in pink, Figure 2a), which serve as distinct labels in $^1$H NMR spectroscopy to validate the ligand exchange and quantify the ligand densities (see SI Note 3). As seen in Figure SI 7 and Table SI 3, ligand exchanges with increasing equivalents of TCA resulted in decreased oleic acid signal and increased TCA signal. Absorption spectra (Figure SI 8, SI 9) of the PbS-TCA QDs also displayed an increase in the TCA absorption bands (400-550 nm) when more TCA was used in the ligand exchange.

Incorporating PbS-TCA QDs improved film morphology (Figure SI 3) and UC efficiency (Figure 2b), consistent with homogeneous films and improved triplet extraction. UC emission was optimized when 4 vol/vol equivalents of TCA were used in the ligand exchange process, which produced PbS QDs with a TCA ligand density of 1.0 ligands/nm$^2$ (Table SI 3). This resulted in a ~15-fold increase in UCPL signal under 1130 nm laser excitation (Figure 2c). The right inset in Figure 2c shows the upconverted image of 1200 nm LED illumination in the pattern of a Stanford logo at <0.4 W/cm$^2$ incident intensity on a PbS-TCA/TES-ADT/DBP UC film (Table SI 8). In contrast, UC films without TCA modification demonstrated very low signal under the same illumination (left inset in Figure 2c). This demonstrates the need for mediator ligands to unlock UC imaging at lower incident intensities.

We note that the reported triplet energy of TCA (~1.3 eV)[27] exceeds the bandgap of 1115 nm QDs (~1.1 eV), suggesting unfeasible exciton extraction by TCA in these systems. However, the introduction of TCA clearly leads to boosted UC performance (Figure 2b). To investigate this behavior, quantum chemical calculations (Methods, SI Table 2) were performed to assess the change in TCA's triplet energy upon coordination to $Pb^{2+}$ ions. These calculations revealed that charge-transfer between TCA anions and $Pb^{2+}$ can lead to lower energy triplet states with the unpaired electrons occupying the highest occupied molecular orbital of TCA and the 6p orbital of Pb (Figure SI 4, SI 5). This indicates that hole transfer from PbS quantum dots to TCA can lead to the formation of lower energy triplet states involving both species, which can permit exciton extraction at longer wavelengths.

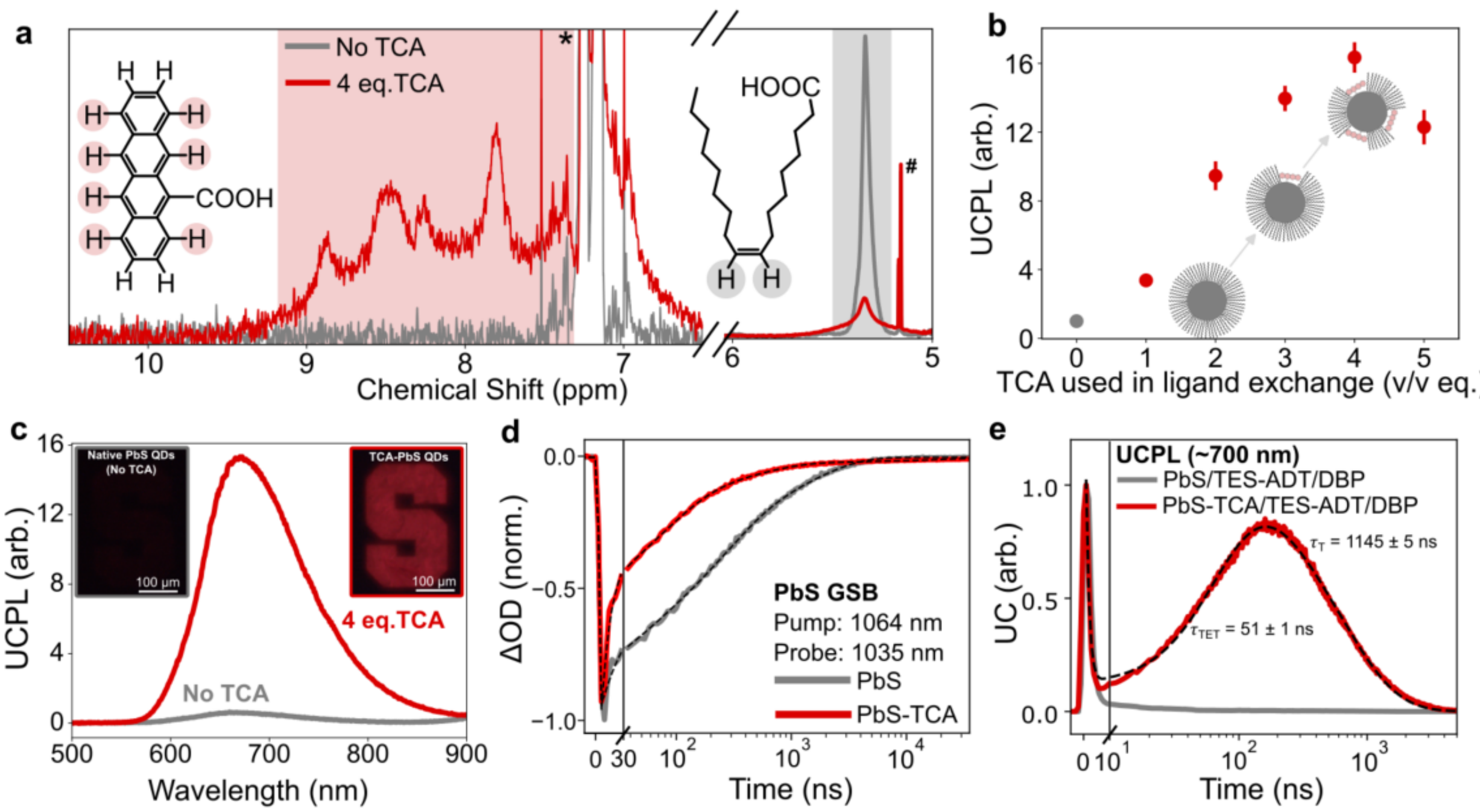

**Figure 2.** Introduction of TCA mediator ligands results in drastic improvements in UC performance. **(a)** NMR spectra of PbS QDs in deuterated chloroform with (red) and without TCA (gray) ligand exchange show a decrease in bound oleic acid (gray region), and an increase in bound TCA (pink region). The inset chemical structures of TCA (left) and oleic acid (right) are highlighted with relevant protons. **(b)** UCPL intensity of films increase with incorporation of TCA ligand on PbS QDs (here, 1115 nm QDs) with an optimum at 4 volume/volume equivalents. The inset shows a schematic representation of increasing TCA ligands on PbS QDs. **(c)** UC spectra of films with optimal TCA (red) and without TCA (black) under same intensity 1130 nm excitation show a 15-fold improvement. Comparison of TCA-based (right inset) and TCA-free (left inset) UC films used for imaging 1200 nm light (full width at half maximum (FWHM) = 65 nm) at identical incident intensity and camera exposure. We note that the non-uniformity in the upconverted image arises from crystallization of the film (see Methods). **(d)** Transient absorption kinetics of the PbS QD GSB at ~1035 nm under 1064 nm excitation for thin films of PbS QDs (gray trace) and PbS-TCA QDs (red trace) shows shorter PbS exciton lifetime when TCA is introduced, indicative of energy transfer. Dashed black traces are Gaussian-convolved multiexponential fits to the data (see Table 1). **(e)** Time-resolved photoluminescence (TRPL) kinetics of emission at ~700 nm from UC films under 1064 nm excitation. Fittings in (d) and (e) are shown in black dashed traces. Dashed black line is squared difference of two exponentials.[40] Further details on fitting and obtained parameters for (d) and (e) can be found in Table SI 5 and SI Note 4.

Transient Absorption and Time-Resolved Photoluminescence (TRPL) Spectroscopy:

To investigate the effect of the addition of TCA on the UC mechanism, we turn to transient spectroscopic studies of films employing PbS QDs with the first excitonic peak at 1030 nm. Figure 2d shows the transient absorption kinetics of the PbS exciton population in QD-only films, as tracked by the PbS ground-state bleach (GSB) at ~1035 nm following excitation at 1064 nm. The addition of TCA results in a shorter exciton lifetime in the PbS QDs, consistent with extraction into the TCA triplet state. Table 1 shows the multiexponential decay fit parameters for these films, highlighting that all lifetime components are shortened for the films with TCA. While we cannot rule out a simultaneous increase in surface trap-mediated recombination rate due to poorer passivation in PbS-TCA, this clearly is not the dominant effect given the much-improved UC efficiency observed.

The TRPL kinetics at ~700 nm following 1064 nm excitation of UC films are shown in Figure 2e. The 51 ns growth in UCPL of the PbS-TCA/TES-ADT/DBP film is commensurate with the decay of the PbS-TCA GSB, confirming its origin in TCA-mediated triplet transfer, and which is further evidenced by the ingrowth of TES-ADT GSB in PbS-TCA UC films (Figure SI 11a). We note that the pulsed excitation source used here results in a high instantaneous light intensity, increasing the probability of direct TES-ADT/DBP excitation via multiphoton absorption. This gives rise to the prompt PL feature observed in the kinetics, but which contributes only a small proportion (<1%) of the time-integrated emission and therefore a vanishingly small contribution to the low instantaneous intensity steady-state measurements presented elsewhere in this work. In contrast to the PbS-TCA UC film, the non-TCA UC film shows minimal in-growing UCPL signal (Figure 2e) or TES-ADT GSB (Figure SI 11a) due to inefficient triplet transfer. Further details are given in the Supporting Information, including the lifetimes extracted from exponential fits to the data.

| **Sample** | **$\tau_1$ / ns ($A_1$)** | **$\tau_2$ / ns ($A_2$)** | **$\tau_3$ / ns ($A_3$)** | **$\tau_4$ / ns ($A_4$)** |
|---|---|---|---|---|
| **PbS** | 25 ± 3 (24%) | 173 ± 9 (49%) | 940 ± 40 (27%) | - |
| **PbS-TCA** | 9 ± 1 (47%) | 58 ± 4 (39%) | 326 ± 31 (12%) | 12250 ± 2150 (2%) |

**Table 1:** PbS GSB (1035 nm) fit parameters for QD films. Multiexponential decays were convolved with a Gaussian instrument response function (1.5 ns). All PbS exciton lifetime components are shorter for the TCA-passivated QDs due to triplet energy transfer and surface recombination.

Efficiency Characterization of UC Films:

Next, UC metrics including IQE, external quantum efficiency (EQE), and threshold intensities were measured to characterize the performance of the TCA-enhanced BHJ films. IQE, the ratio of upconverted photons to absorbed photons, is widely reported in UC literature as $\phi_{UC}$—a key metric for the inherent efficiency of the UC system. It is primarily dependent on the efficiencies of triplet energy transfer from sensitizer to annihilator, TTA in the annihilator, and the fluorescence of the emitter (SI Equation 1). EQE, defined as the ratio of upconverted photons emitted to incident photons illuminating the film, is a useful metric when evaluating UC for applications. EQE is the product of the IQE and the film's optical absorption of NIR light (SI Equation 5). We measured EQE using the relative method (Equation 2, Methods), as frequently reported in thin-film TTA-UC studies.[22,23,27] Integrating sphere measurements were used to calculate the laser absorption of the films as reported by de Mello, et al.[52] (Methods) The IQE was then determined by normalizing the EQE to the film's absorption (SI, Equation 5). Here, IQE and EQE are reported out of a 50% maximum to account for the two-to-one nature of TTA-UC.[53]

The UC film performance was optimized further by varying the concentrations of TCA-enhanced PbS QDs, TES-ADT, and DBP (Figure SI 12). The optimal TES-ADT (300 mg/mL), 850 nm PbS QDs (75 mg/mL), and DBP (4 mg/mL) concentrations resulted in ~3.3 µm-thick films upon spin coating at 1600 r.p.m. for 30 seconds (Figure SI 14). To investigate this BHJ architecture under different NIR excitation wavelengths, UC films incorporating different PbS QD excitonic peaks (800, 900, 1020, and 1120 nm; Figure 1c) were fabricated. For each QD size, films with different QD concentrations were also prepared to tune the absorptivity (Figure SI 15). In these films, EQE increased with higher QD concentrations due to enhanced laser absorption (Figure 3a). Under 808 nm excitation, films with 800 nm QDs achieved a peak EQE of 0.15% (out of a 50% maximum) with approximately 12% absorption of the incident light. As the QD bandgap decreased (i.e., longer-wavelength absorption), EQE values dropped, reaching 0.0005% for films with 1120 nm QDs under 1208 nm excitation (Figure 3a). Figure 3e summarizes the dependence of EQE on PbS QD size and illustrates the variability in EQE across different QD batches and film fabrication conditions over eight independent experimental runs. For a given laser excitation, we observe that using the PbS QDs with a bluer excitonic peak (e.g., using 1120 nm QDs for 1208 nm excitation) allows optimal UC performance. We hypothesize that energy flow and UC efficiency is optimized when the energetic offset between the QD band edge and the TCA triplet energy is favorable.

To evaluate the intrinsic UC efficiency of the films, IQE was plotted for films with varying absorption (Figure 3b). IQE is a particularly relevant metric as it represents the maximum achievable performance once absorption is maximized, for example by utilizing plasmonic nanostructures[5] or cavities.[26,29] Interestingly, for all PbS QD sizes, IQE consistently decreased at high NIR absorption. This decline is attributed to the higher concentrations of PbS QDs, which act as acceptors in parasitic FRET-based back transfer of singlets and high-energy photons generated in the TES-ADT/DBP matrix.[27] This balance highlights the importance of controlling QD loading to maximize NIR absorption while minimizing back transfer of singlets. In contrast to the transient absorption kinetics obtained from PbS QD films (Figure 2d), kinetics from UC films show a prolonged PbS GSB lifetime upon addition of TCA (Figure SI 10) due to parasitic re-absorption of UCPL by PbS QDs. As with EQE, IQE decreased with decreasing QD bandgap, reaching champion IQE of 0.081% for 1120 nm QDs under 1208 nm excitation. Reproducibility statistics for IQE at various excitation wavelengths (808, 1130, and 1208 nm) are summarized as box-and-whisker plots and demonstrate that the system lies on the Pareto front of NIR UC performance at all tested wavelengths (Figure 1d). Champion IQE performance of 1.9%, 0.46%, and 0.081%, under 808, 1130, and 1208 nm excitation was observed (Table SI 1).

UC Performance with Decreasing QD Bandgap:

As seen above, UC performance decreases sharply with decreasing PbS QD bandgap. This decrease in IQE is attributed to two key factors: (1) the triplet energy transfer from the PbS QD to the TCA ligand ($\phi_{ET}$), and (2) parasitic back transfer of singlets from TES-ADT/DBP to PbS QDs ($\phi_{BT}$). As the QD bandgap reduces, the energy transfer from the PbS QD to the TCA ligand becomes less energetically favorable. To study this, we employed solution-state photoluminescence spectroscopy of the PbS QDs. Four different sizes of PbS QDs (800, 900, 1020, and 1120 nm PbS QDs) were ligand-exchanged with different TCA loadings (no TCA and 2, 3, 4, 5 volume/volume TCA equivalents). For these different QDs and TCA loadings, solutions in cuvettes were analyzed using absorbance and photoluminescence spectroscopy. Using the PbS QD PL normalized to absorbance, the quenching efficiency was calculated (SI Equation 10). As shown in Figure 3d, we observed an increase in quenching with increased TCA loading, highlighting the enhanced TET from PbS QDs to TCA. Additionally, we observe a decrease in the quenching efficiency with decreasing bandgap of the PbS QDs (Figure 3d). We attribute this decrease in exciton extraction from redder PbS QDs to the decreased energetic offset between the PbS QD bandgap and the TCA triplet.

This decrease in triplet energy transfer was further corroborated using UC threshold measurements. Due to the quadratic nature of the TTA process, UC efficiency exhibits a nonlinear dependence on photon intensity incident on the UC film. At low excitation intensities, triplet populations are too sparse for efficient annihilation, resulting in low UC efficiencies. As the excitation intensity increases, this relationship becomes linear, reflecting more efficient triplet–triplet interactions.[54,55] The TTA-UC threshold is defined as the incident power density at which UC emission transitions from quadratic to linear intensity dependence, marking the onset of efficient annihilation. Low threshold intensities are particularly attractive for enabling low-power applications such as photovoltaics and imaging. Figure 3c presents threshold intensities for UC films incorporating 800, 900, 1020, and 1120 nm PbS QDs under 808 nm excitation. As the QD bandgap decreases, the threshold intensity rises sharply—from 0.38 W/cm² for 800 nm QDs to 14 W/cm² for 1120 nm QDs for 808 nm excitation (SI Table 7).

Equation 1 shows the dependence of threshold intensity on absorption ($\alpha$), energy transfer from PbS to TES-ADT ($\phi_{TET}$), annihilator triplet diffusion constant ($D_T$), annihilation distance between the annihilators ($a_0$), and lifetime of triplet annihilators ($\tau_T$).[12] Given that the annihilator-side system is identical, the increase in threshold intensity can only be attributed to an increase in $\phi_{TET}$. In tandem with the PbS emission studies highlighted above, this data strongly implies that the triplet energy transfer from PbS QD to the TCA is reduced as the PbS QD bandgap is decreased. This limitation of the PbS/TCA/TES-ADT/DBP system could possibly be alleviated by using other ligands with lower triplet energy levels. However, the triplet energy level of TES-ADT (~1.08 eV) presents an energetic restriction to move beyond 1200 nm as well.

$$I_{th} = (8\pi\alpha\phi_{TET}D_T a_0)^{-1}(\tau_T)^{-2} \qquad \text{Equation 1}$$

FRET-based back transfer from the annihilator to the PbS QDs is a limitation of the BHJ architecture for UC. Here, the impact of decreasing PbS QD bandgap on parasitic back transfer of singlet excitons in TES-ADT/DBP was studied using photoluminescence quantum yield (PLQY) measurements with a 532 nm excitation. The 532 nm laser creates singlets in the TES-ADT/DBP through direct absorption. The resulting singlets can emit photons to relax or undergo non-radiative FRET to the PbS QDs. With higher PbS QD concentrations, FRET is increased, resulting in lower measured PLQYs. For the UC films shown in Fig 3a and 3b, the PLQY of the films were measured using an integrating sphere (Figure SI 17). Relative to a TES-ADT/DBP-only film, we observed significantly reduced PLQY for the UC films which possess PbS QDs. The PLQY showed a consist decrease with increasing amounts of PbS QDs. Additionally, the PLQY decreases more for redder QDs, corroborating that redder QDs may suffer from higher parasitic back transfer due to higher spectral overlap. We discuss a breakdown of the efficiency of each step of UC in the SI of this paper (SI Note 5, Table SI 6).

Mirror-Enhanced UC:

As reported by Wu, et al.,[29] introduction of mirrors onto UC films can enhance film absorption and UCPL. In a similar approach, after spin coating the UC film on glass, LiF (10 nm) and aluminum (100 nm) were sequentially thermally deposited onto the UC film to create an insulating buffer and mirror respectively. Under 1130 nm excitation, the mirrored films demonstrated a 4-fold improvement in collected UCPL (Figure 3f) and far brighter images of patterned 1200 nm LED light (Figure 3f insets), consistent with improved film absorption.

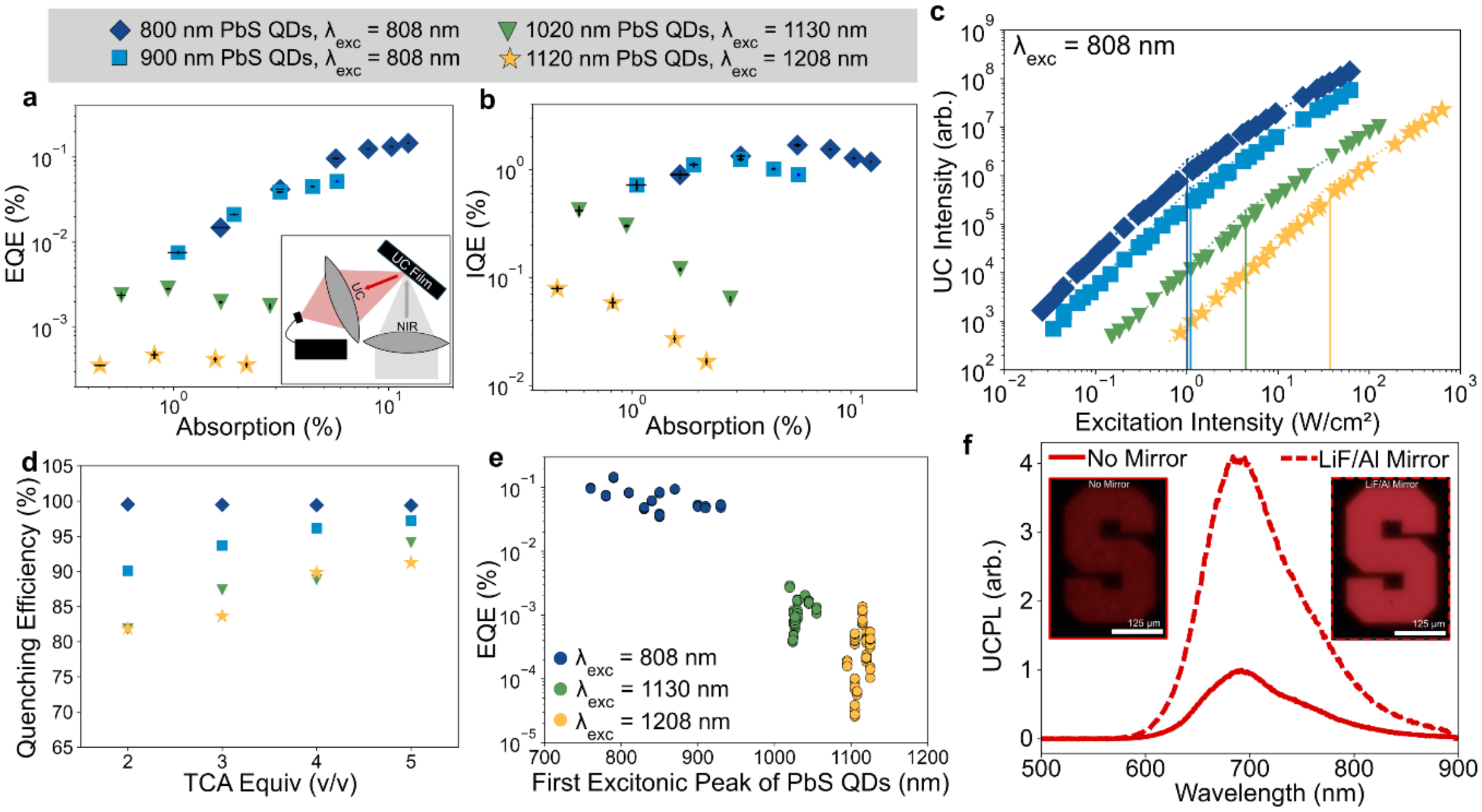

**Figure 3.** UC film performance with varying PbS QD first excitonic peaks: 800 nm (dark blue diamonds), 900 nm (light blue squares), 1020 nm (green triangles), 1120 nm (yellow stars). **(a)** EQE and **(b)** IQE are plotted as functions of laser absorption and QD size. The inset in panel 3a highlights the reflective mode used to measure EQE and IQE (Figure SI 20). **(c)** UC threshold intensities (measured using 808 nm excitation) show an increase with decreasing PbS QD bandgap. **(d)** PbS QD emission quenching upon binding of TCA can be used to approximate the energy

transfer efficiency from PbS QDs to TCA. **(e)** EQE distribution for different PbS QD sizes using data from 38 films across 12 experimental runs. **(f)** Incorporation of a mirror by evaporating LiF (10 nm) and aluminum (100 nm) on the UC film results in a 4× boost in UCPL under 1130 nm excitation. Mirror incorporation (right inset) results in brighter imaging of 1200 nm LED light at the same intensity and exposure in comparison to films without mirror incorporation (left inset).

Visible Imaging using 1200 nm Light:

While mechanisms like second harmonic generation, sum frequency generation, and lanthanide-doped nanoparticles have been demonstrated for NIR imaging and sensing, we believe TTA-UC is particularly interesting and can overcome prior limitations in bandwidth, conversion efficiency, and field of view (SI Note 9, SI Table 9). To evaluate the applicability of our UC thin films in low-intensity applications such as night vision, bioimaging, and photovoltaics, UC imaging was performed using an incoherent 1200 nm LED source with a FWHM of 65 nm (Figure 4b) which has been longpass filtered. A reflective imaging setup (Figure 4a) was constructed, where light passed through negative imaging masks—i.e., targets with transparent regions—to define the projected pattern. The transmitted NIR image was imaged to infinity using a 150 mm achromatic doublet lens and directed towards the UC film via a dichroic mirror positioned at 45°. This dichroic mirror reflected NIR light (>900 nm) into a 15× objective, which focused the image onto the UC film with a magnification of 0.0866. The power impinging on the UC film was on the order of ~0.1 mW across different mask configurations (Table SI 8), resulting in intensities on the film on the order of 0.4 $W/cm^2$. The upconverted visible emission was collected through the same objective and dichroic mirror, which now acted as a shortpass filter (<900 nm) to reject residual NIR light. The UC signal was then focused on a camera using a 200 mm tube lens. The upconverted image of the target mask was reimaged onto the camera sensor with a magnification of 1.33.

With an input of 1200 nm light (FWHM = 65 nm) and an output of 700 nm light (FWHM = 100 nm), this UC imaging achieves an anti-Stokes shift of roughly 500 nm, i.e., 0.73 eV (Figure 4b). Using this imaging setup, NIR objects can be imaged with a resolution of ~5-7 line pairs/mm (lp/mm) and upconverted images on the film can be imaged at ~80 lp/mm (Figure SI 25). The optical setup also enabled high-contrast UC imaging (contrast >20%) (Figure SI 25) with incident intensities at the imaging mask around 20 $mW/cm^2$ (SI Note 8). As demonstrated in Figures 4c and 4d, complex targets such as the Bucky Badger mascot and the University of Wisconsin–Madison logo (Figure SI 22) were successfully imaged under incoherent NIR illumination, with exposure times as short as 3 seconds, despite sub-threshold incident intensities at the UC film (<0.4 $W/cm^2$; SI Note 7).

Imaging Using Light Below the Silicon Bandgap:

A promising application of NIR-to-visible UC is in sensitizing silicon at photon energies below its bandgap (1.12 eV), enabling silicon-based photovoltaics to surpass the detailed-balance limit[56] and facilitating deeper NIR-sensitive photodetectors and cameras.[57] Gholizadeh, et al. previously demonstrated solution-phase TTA-UC below silicon's bandgap using 1140 nm excitation.[40] Here, we demonstrate thin-film TTA-UC at ~1200 nm (FWHM = 65 nm) while positioning a 350-µm-thick silicon wafer before entry into the imaging setup (Figure 4a). Despite the presence of the wafer, we observe bright UCPL (Figure 4e) and clear UC images (Figure 4f), with the reduced UCPL intensity and dimmer images resulting from reflections at the silicon–air interfaces (Figure SI 24, SI Note 6). This demonstration underscores the potential of this UC platform for photovoltaic enhancement and NIR-sensitive detection and imaging.

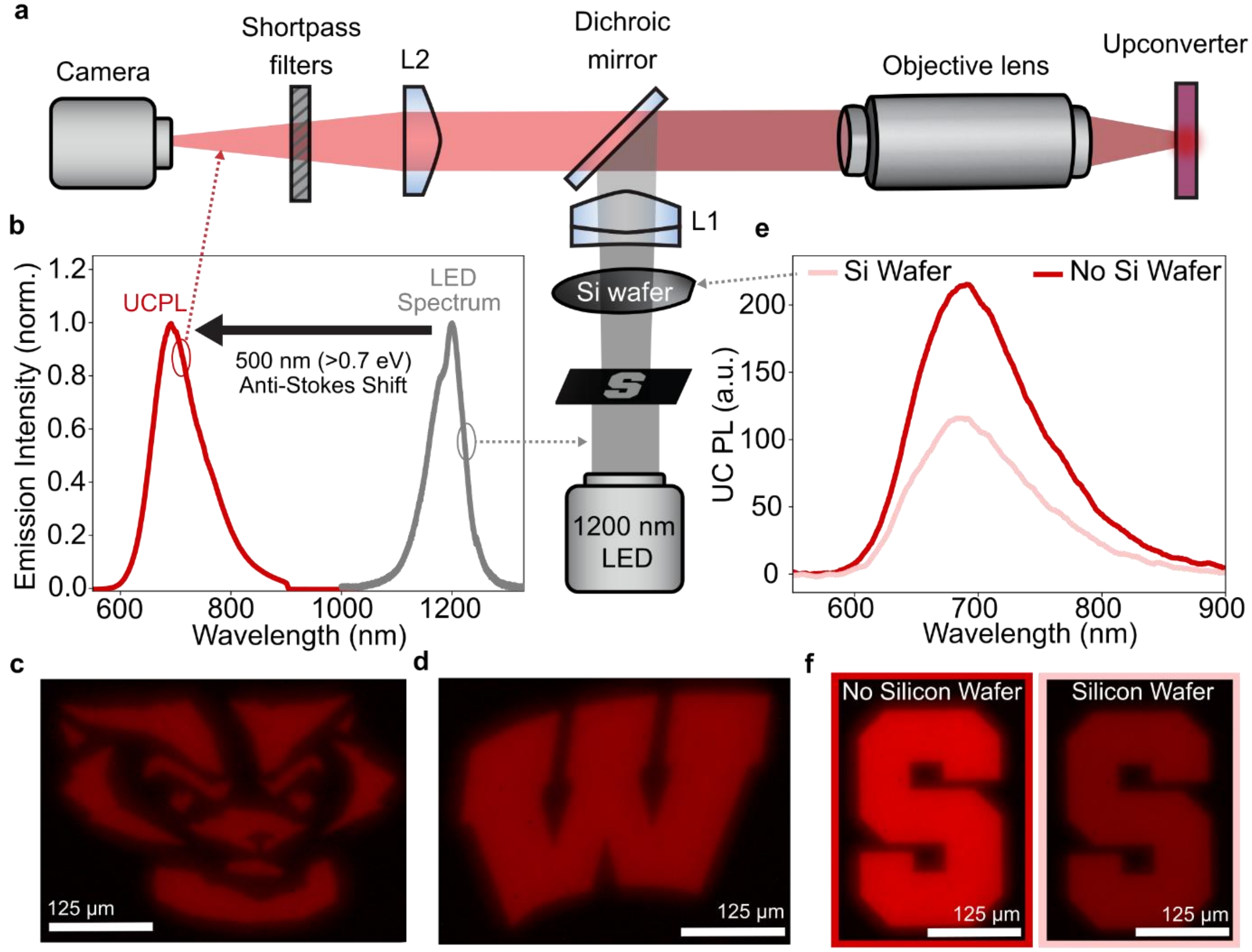

**Figure 4.** TCA-modified BHJ UC films enable reflective imaging using 1200 nm photons. **(a)** Schematic of the imaging setup. The target masks were backlit with a 1200 nm LED (with a 1050 nm long pass filter, not shown) and imaged onto the UC film through a 150 mm achromatic doublet (L1), dichroic filter, and objective (0.70 NA, 15×). The upconverted image was then imaged onto a camera using the same objective, dichroic filter, 200 mm tube lens (L2), and further short pass filters (Figure SI 21). **(b)** The LED spectrum (grey, raw data from Thorlabs datasheet) peaks at 1200 nm and the UC spectrum (red) peaks at 700 nm, resulting in an anti-Stokes shift >0.7 eV. This imaging setup using mirrored UC films was deployed to image masks in the shape of **(c)** Bucky Badger and **(d)** Wisconsin logo with 3 second imaging exposure time. By placing a silicon wafer (350 µm) in the path of the LED, sub-silicon imaging using UC is demonstrated. **(e)** Introducing the silicon wafer results in a reduction of UC PL ($\lambda_{exc}$ = 1208 nm) due to reflections

at the air/Si interfaces (Figure SI 24, SI Note 6)). **(f)** A Stanford logo was successfully imaged with a silicon wafer in the beam bath (mirrored BHJ UC film, 7.5 second exposure).

To conclude, we demonstrated a thin-film TTA-UC system that has high IQEs over the 800-1200 nm NIR-I and -II bands, including champion IQEs of 1.9%, 0.46%, and 0.101% under 808, 1130, and 1208 nm excitation, respectively. This was accomplished using BHJ films composed of PbS QDs, TES-ADT, and DBP, wherein the introduction of TCA ligands on the QD surface led to marked improvements in UC performance, attributed to enhanced film morphology and exciton extraction. Surface modification of the PbS QDs with TCA was studied via NMR spectroscopy and time-resolved spectroscopy was used to examine UC kinetics and corroborate improved exciton extraction by TCA. Steady-state emission spectroscopy and threshold measurements were used to demonstrate decreasing triplet energy transfer efficiency with deeper NIR-absorbing PbS QDs. Additionally, PLQY measurements were deployed to study the disadvantageous FRET from annihilator singlet to the PbS QDs. Further, the UC performance was further boosted by using reflective back mirrors to enhance absorption and light outcoupling. Leveraging these high-efficiency films, we demonstrated low-intensity UC imaging using incoherent 1200-nm LED excitation through a silicon wafer, underscoring the promise of UC for silicon-compatible photonic applications such as sub-bandgap photovoltaics and NIR detection.

METHODS.

**Materials:**

5-tetracene carboxylic acid (>95% by NMR) was purchased from HAARES ChemTech Inc. PbO (Puratronic$^{TM}$, 99.999%, metals basis), 1-octadecene (90%, technical grade), anhydrous toluene (99.85%, extra dry over molecular sieves, AcroSeal), anhydrous acetonitrile (99.9+%, extra dry, AcroSeal), and anhydrous chloroform (99.9%, Extra Dry over Molecular Sieve, Stabilized, AcroSeal™) were purchased from Thermo Scientific Chemicals. Oleic acid (technical grade, 90%), and hexamethyldisilathiane (synthesis grade) were purchased from Millipore Sigma. DBP (5,10,15,20-Tetraphenylbisbenz[5,6]indeno[1,2,3-cd:1′,2′,3′-lm]perylene, >99%, sublimed) and TES-ADT (5,11-Bis(triethylsilylethynyl)anthradithiophene (>99%), mixture of isomers) was purchased from Luminescence Technology Corp. Hexanes (ACS certified), isopropanol (ACS certified), and acetone (ACS certified) were purchased from Fisher Scientific. All chemicals were used as purchased unless specified. Continuous wave lasers were purchased from Dragon Lasers (532 nm, 735 nm, and 808 nm) and from Changchun New Industries Optoelectronics Co., Ltd. (1130 nm and 1208 nm). Imaging transparency masks were fabricated by Fineline Imaging, a division of EMS Thin Metal Parts, LLC.

**Synthesis of PbS quantum dots (850, 950, 1050, 1150 nm absorption peaks):**

PbS quantum dots were fabricated using a modified Hines hot-injection technique. To a two-neck 100 mL round bottom, 0.669 g of PbO, 30 mL of 1-octadecene, and 2 mL of oleic acid were added. The flask was heated under vacuum to degas and synthesize lead oleate (temperature was increased to 120 °C slowly by 30 °C every 10 minutes). The reaction flask was degassed at 120 °C for at least 2 hours resulting in a clear solution. The flask was refilled with nitrogen and set to the injection temperature. Depending on the desired size of QDs, the injection temperature was varied (Table 2). In a nitrogen atmosphere glovebox, 7.5 mL of dry toluene was charged with 350 µL of hexamethyldisilathiane (TMS-S). This solution was brought out of the glovebox and injected into the reaction flask with minimal exposure to air. Upon injection, the reaction flask was immediately removed from the heat plate and allowed to cool naturally. Upon reaching room temperature, the reaction was washed with acetone (4× volume of acetone was added, centrifuged at 8800 ×g for 2 mins). The resulting pellet was redispersed in hexanes (~30 mL) and washed with acetone (4× volume of acetone was added, centrifuged at 8800 ×g for 2 mins). The final pellet was redispersed in hexanes (~15 mL) and stored in the glovebox until use.

| **PbS first excitonic peak (nm)** | **Injection Temperature (°C)** |
|---|---|
| 800 | 70 |
| 900 | 85 |
| 1020 | 137 |
| 1120 | 170 |

**Table 2:** Injection temperatures variation was used to fabricate PbS QDs of varied sizes.

**Fabrication of upconversion devices:**

**Substrate treatment:** All samples, unless otherwise specified, were fabricated on 1 cm × 1 cm soda-lime glass of 1.1 mm thickness. All substrates were washed by sonicating for 5 minutes in a 1% hellmanex detergent solution in deionized water, followed by two 5-minute rounds of sonication in water, followed by two 5-minute rounds of sonication in acetone and two 5-minute rounds of sonication in isopropanol. The substrates were then dried under pressurized air to remove solvent and dust particles. Within 15 minutes before spin coating, the glass substrates were treated with UV ozone plasma for at least 15 minutes.

**TCA ligand exchange** (for 4 vol/vol equivalents): The PbS QDs stock solution was washed with acetone once more (4× acetone, 8800 ×g for 2 mins) to create a pellet. In a nitrogen-filled glovebox, this weighed pellet was dissolved in anhydrous toluene to make a 25 mg/mL PbS QD solution. To 1 mL of this PbS QD solution, 4 mL of a 10 mM solution of 5-tetracene carboxylic acid in acetonitrile was added while stirring. After 1 hour of stirring, the ligand exchange solution was removed and centrifuged (8800 ×g for 2 mins). The resulting pellet was dried to remove any residual solvents through evaporation. The dry pellet was weighed into a vial for further use. The TCA ligand density was modified by introducing different vol/vol equivalents of TCA solution (10 mM) to PbS QD solution (25 mg/mL).

**Note:** The 10 mM TCA solution in acetonitrile needs to be heated for 10 minutes at 80 °C to completely dissolve the TCA. It was allowed to cool to room temperature while stirring before use in the ligand exchange.

**UC solution preparation:** TES-ADT/DBP stock solutions were made by adding a DBP stock solution in chloroform (1 mg/mL) to a vial containing the required amount of TES-ADT to make a 300 mg/mL solution. This TES-ADT/DBP stock solution was then added to weighed ligand-exchanged PbS pellets to make a final PbS concentration of 75 mg/mL. To make lower concentration PbS UC solutions (50, 25, 10 mg/mL), the PbS/TES-ADT/DBP solution was diluted using the TES-ADT/DBP stock solution, thus maintaining the TES-ADT and DBP concentration across the PbS concentration sweep.

**Spin coating and encapsulation:** All spin coating was performed in a nitrogen-filled glovebox. The UC solutions in chloroform (70 µL) were spin-coated at 1500 rpm for 30 seconds with a 0.8 second ramp. The UC solution was loaded onto stationary samples before ramping. All samples were encapsulated with a 1.2 mm-thick microscope slide (Electron Microscopy Sciences) using UV-curing epoxy (purchased from Epoxy Technology) under a UV-lamp for 5-10 minutes.

**Thermal deposition of LiF and Aluminum:** The spin coated substrates were then loaded into an Angstrom Engineering thermal evaporation chamber. All evaporations were performed under a pressure of less than 1E-05 mbar. Both LiF and Aluminum were evaporated at a rate of ~1-4 Å/s.

**Fabrication of imaging films:** For imaging measurements, the UC solution was filtered through 0.1 µm PTFE membrane syringe filters (Whatman® Puradisc 13 syringe filters) before spin coating to ensure further uniformity.

**Note on UC film morphology and stability:** Over the span of a few hours, the films show crystalline grains on the micron-scale. Figure SI 26 shows an image of a backlit UC film, highlighting the crystalline phases. This crystallization results in the non-uniformity observed in Figure 2c. We note that the UC performance is retained despite crystallinity. We also observed that the evaporation of LiF/Aluminum onto the films slows the crystallization process. The images shown in Figure 4 were collected using films which were fabricated 3-4 weeks prior.

**UV-vis spectroscopy (absorption) measurement:** UV-vis spectroscopy was performed on PbS QD solutions using an Agilent Cary 6000i UV/vis/NIR spectrophotometer in transmission mode to measure the first excitonic peak. A 1 cm path length quartz cuvette was used, and absorption was measured in hexanes. Spectra of PbS QDs were taken to maintain optical density below 0.1 across 700-1300 nm.

**Nuclear magnetic resonance (NMR):** A Bruker Avance III 400MHz spectrometer equipped with a 5mm Z-Gradient iProbe and Avance Neo NanoBay console was used. The spectra were acquired using standard pulse sequences from the Bruker library, except for quantitative NMR spectra, where a relaxation delay (d1) of 30 seconds was used. All data was processed using Mestrenova. NMR studies were performed on $CDCl_3$ solutions of PbS QDs with the first excitonic peak at 1115 nm.

**Photoluminescence quantum yield (PLQY) measurement:** Photoluminescence quantum yields of TES-ADT/DBP films were measured using the de Mello method in an integrating sphere (from Labsphere).[52] The integrating sphere was calibrated using a radiometric light source (HL-3P-INT-CAL, Ocean Insight). A 532-nm-continuous-wave laser was used to excite the neat TES-ADT/DBP films. The sample was loaded into the integrating sphere such that the 532 nm laser had an approximately 8-degree incidence.

| **Sample** | **PLQY (%)** | **Absorption (%)** |
|---|---|---|
| 300 mg/mL TES-ADT, 1 mg/mL DBP in chloroform | ~2-9 | ~95 |
| 75 mg/mL TES-ADT, 0.25 mg/mL DBP in chloroform | ~25 | ~75 |

We note that spin coating TES-ADT/DBP in chloroform at high concentrations (>200 mg/mL) results in films which crystallize instantaneously, resulting in very low PLQY (2-9%). For similarly high concentrations of TES-ADT/DBP, UC films with PbS QDs are less crystalline. We believe the presence of PbS QDs disrupts TES-ADT crystallization and permits higher PLQY of TES-ADT/DBP in UC films.

**Steady-state photoluminescence (PL) and UC photoluminescence (UCPL) measurement:** All relative PL and UCPL measurements were conducted on the set up shown in SI Figure 20. The samples were illuminated with a 532, 808, 1130, or 1208 nm continuous-wave lasers at an angle of ~45 degrees. The emission from the samples was collected using a high numerical aperture lens and fiber-coupled into a spectrometer (QE pro high-performance, Ocean Insight). The fiber coupling to the detector included a filter holder with appropriate short pass filters (700 nm SP, 1000 nm SP from Thorlabs) to filter the laser out of the UCPL spectra. To calculate upconversion photoluminescence intensity, the emission was integrated from 550-850 nm.

**Upconversion external quantum efficiency (EQE) measurement:** We found that integrating sphere-based UC efficiency measurements were not accurate due to (1) signal-to-noise constraints, (2) laser signal saturation, and (3) UC emission reabsorption. Therefore, to measure the EQEs, we used the relative method introduced by Izawa et al. that has been widely adopted by the field.[22,23,27,34,58] The PL setup described above was used for these EQE measurements. A TES-ADT/DBP-only film of ~500 nm thickness was used as the standard with a known PLQY to match the UC spectral profile. This standard was used because its fabrication process, form factor, and emission profile most closely resemble the UC film itself; the only difference being that the standard film lacks the PbS QDs and TCA which are incorporated in UC films and thickness changes. The PLQY of this standard was ~25%. This PLQY measurement was done using a 532 nm laser *via* the de Mello integrating sphere methodology described above.

In PL measured ($I_{UC}$ and $I_{Std}$) for the relative EQE calculation (Equation 2), only the emission from the front face of the UC film and the standard film are collected, while waveguided light and backward emission are not collected. However, the PLQY measured for the standard film includes the waveguided emission, as the PLQY is measured in the integrating sphere. Therefore, based on Equation 2, this measurement calculates an EQE which includes the waveguided emission using the assumption that the fraction of light that is waveguided is the same for the UC films and the standard film, as is standardized in the literature.[22,23,27,58] We note that this method of calculating EQE does not account for topological and refractive index differences between films with and without PbS QDs.

The PL (integrated 550-850 nm, using 700 nm short pass or 950 nm short pass filter) from TES-ADT/DBP-only samples ($I_{std}$) was measured using a 532 nm continuous wave laser. With the same optical alignment, the UCPL (integrated 550-850 nm, using 700 nm short pass or 950 nm short pass filter) from upconversion films ($I_{UC}$) was measured using an 808 or 1130 or 1208 nm continuous-wave laser. For EQE calculations under 808 nm excitation, UCPL and PL were collected with 700 nm short pass filters to account for the spectral area cut off by the filter. All measurements were performed with appropriate short pass filter to remove the laser signal. The incident power of the 532 nm ($P_{std}$) and 808 nm ($P_{UC}$) lasers were measured using a Newport photodiode (Newport 818-SL). The incident power of the 1130 and 1208 nm laser ($P_{UC}$) were measured using a different Newport photodiode (Newport 818-IR). For each UC film, four measurements were performed on random positions across the film. The resulting averages and standard deviations were propagated using the Python uncertainties package. Equation 2 was used to calculate the EQEs of the different films.

For EQE measurements, laser excitation was maintained above the threshold intensities to collect optimal EQE performance in the linear regime. For the data reported in Figure 3a and 3b, laser intensities of 1.7, 13, 23 $W/cm^2$ for 808, 1130, and 1208 nm respectively, were used. The spot sizes were ~ 9.4E-03, 5.4E-03, and 5.8E-03 $cm^2$ respectively, for the 808, 1130, and 1208 nm lasers. The EQE and IQE numbers listed in this work are out of a 50% maximum.

$$\mathrm{EQE} = \frac{\mathrm{I_{UC}}}{\mathrm{I_{Std}}} \times \frac{\mathrm{P_{Std}}}{\mathrm{P_{UC}}} \times \frac{\lambda_{Std}}{\lambda_{UC}} \times (\%\mathrm{A_{Std}}) \times \mathrm{PLQY_{Std}} \qquad \text{Equation 2}$$

Where $I_{UC}$ = intensity of the collected UC spectrum, $I_{Std}$ = intensity of the collected PL spectrum from the standard film, $P_{UC}$ = power of the NIR laser, $P_{Std}$ = power of the 532 nm laser, $\lambda_{UC}$ = wavelength of the NIR laser, $\lambda_{Std}$ = 532, and $\%A_{Std}$ is the percent absorption of the standard film at 532 nm.

**UC-film absorption measurement:** The absorption of UC films was calculated using the de Mello method[52] (Equation 3) in an integrating sphere (from Labsphere). The integrating sphere was calibrated using a radiometric light source (HL-3P-INT-CAL, Ocean Insight). A 735 nm continuous wave laser was used to excite the UC films. The films were loaded into the sphere such that the laser had an approximately 8-degree incidence. The absorption and reflection losses of a bare glass substrate were subtracted from the film absorption to calculate the true absorption at 735 nm. This absorption was then scaled to calculate the absorption at the laser peak (808, 1130, or 1208 nm). This scaling was done using a correction factor based on the solution-state absorption spectrum of as-synthesized PbS QDs (Equation 4).

We note that scaling the thin film absorption at 735 nm to calculate the absorption at other wavelengths is an approximation. Since the thickness of the film is on the order of the wavelengths used, there may be changes in absorptivity and reflectivity of films with wavelength. The morphology of the film (like crystallization) may result in a wavelength and angle dependent reflection profile which may vary at 735, 808, 1130, and 1208. Therefore, the spectral shape of QD absorption in thin films may differ from that in solution-state. Solution-state absorbance measurements were performed on QD solutions with optical density between 0.04-0.10 at 735 nm. Solution-state absorbance spectra of PbS QDs were observed to redshift up to 25 nm with increasing QD concentration (Figure SI 13a). This redshift impacts the correction factor (the ratio of percent absorption at λ to the percent absorption at 735 nm) used in Equation 4 (Figure SI 13b).

$$\%\mathrm{Abs}_{735\mathrm{nm,film}} = \left(1 - \frac{\mathrm{I}_C}{\mathrm{I}_B}\right) \times 100\% \qquad \text{Equation 3}$$

Where $I_C$ = laser intensity measured in the integrating sphere when the laser is incident on the UC film and $I_B$ = laser intensity measured in the integrating sphere when the sample is in the sphere, but outside the laser beam path.

$$\%\mathrm{Abs}_{\lambda,\mathrm{film}} = \left(\frac{\%Abs_{\lambda,\mathrm{PbS\ Solution}}}{\%\mathrm{Abs}_{735\mathrm{nm,PbS\ Solution}}}\right) \times \%\mathrm{Abs}_{735\mathrm{nm,film}} \qquad \text{Equation 4}$$

**Upconversion internal quantum efficiency (IQE) measurement:** The IQE was calculated as the EQE divided by the UC film absorption at the laser excitation wavelength. To avoid inflated IQE values arising from measurement noise, IQEs for films with 735 nm absorption below 1% are not reported in this work.

**Upconversion threshold measurement:** Upconversion threshold was measured on the photoluminescence set up described above. The laser was set to the highest power needed for the measurement and the power was cut as needed using a neutral density filter wheel. The UC PL for the samples were measured across orders of magnitude of input power to span the linear and quadratic regimes in the log-log input-output power plots. The integration time was increased to achieve high signal at low input powers.

The input power for each point was measured using a Thorlabs power meter (PM100D meter with S120VC sensor, Thorlabs). The spot-size of the input laser was measured using a camera (CS165CU, Thorlabs) to calculate the intensity of the input light. The threshold intensity was calculated by fitting the log-log plot in the linear and quadratic regimes to find the abscissa of the point of intersection of those lines.

**Spot-size measurement:** Spot-sizes were calculated using a CS165CU Thorlabs camera. ND filters were used to prevent saturation of the camera. Since the measurements were performed at 45-degree incidence, an ellipsoidal area was calculated from the spot-size.

**Profilometry measurement:** A Dektak XT-S Stylus profilometer was used under 3 mg stylus force, with grooves scratched into the samples with a razor blade for measurement (Figure SI 14).

**NIR photoluminescence spectroscopy:** PL spectra of PbS QDs in solution state (Figure SI 16) were collected using a Horiba FluoroLog-3 system with a Xe arc lamp under 600 nm excitation.

**Visible imaging of 1200 nm LED:** For the illumination source, we used a 1200 nm LED (Thorlabs, M1200L4), collimated using an aspheric condenser lens (Thorlabs, ACL2520U), coupled with a 1050 nm long-pass filter (Thorlabs, FELH1050) to attenuate any shorter wavelengths. UC films fabricated using 1150 nm PbS QDs were used to perform visible imaging using NIR light. In Figure SI 27, the product of PbS QD absorption and 1200 nm LED emission is used to highlight the photons employed in these experiments (FWHM = 71 nm). This source was used to illuminate negative imaging masks, i.e., targets where the transparent regions are imaged. In Figures 2c, 3f, and 4d–f, we imaged various masks (Bucky Badger, Wisconsin logo, and Stanford logo) each approximately 4 mm × 4 mm in size.

An achromatic doublet (L1 in Figure 4a; Thorlabs, AC254-150-C-ML) with a focal length of 150 mm was used to image the mask into infinity space, which was then reflected towards the objective lens by a shortpass dichroic mirror (Thorlabs, DMSP900R) placed at 45 degrees. A 15× objective lens (Thorlabs, TL15X-2P) was used to focus the infrared image on the upconverter and collect the image of the resulting upconverted emission. This objective lens has a focal shift of ~5 microns across 600-1250 nm, which is a mechanism of resolution loss in this system. The power incident upon the upconverter was on the order of ∼0.2 mW, depending on the imaging mask used. The emitted light from the upconverter was captured by the same objective lens and transmitted through the dichroic mirror towards the camera. The image was then formed on the camera (Thorlabs, Kiralux camera) using a 200 mm tube lens (L2 in Figure 4a, Thorlabs, TTL200-S8) and additionally filtered using two 1000 nm filters and one 950 nm short-pass filter (Thorlabs, FESH1000, FESH0950) to remove any remaining NIR.

For the image given in Figure 4f (right) we also inserted a silicon wafer (350 μm double polished) between the imaging target and lens L1. In Figure SI 25, the imaging target was replaced with an Air Force resolution target (Thorlabs, R1DS1N). For the images shown in Figures 2c and 3f, the camera exposure was set to 8 seconds with a gain of 15 dB. For the images shown in Figure 4c and 4d, the camera exposure was set to 3 seconds with a gain of 15 dB. For the images shown in Figure 4f, the camera exposure was set to 7.5 seconds with a gain of 15 dB. For the image of the resolution target shown in Figure SI 25, a gain of 20 dB was used with camera exposure of 3 seconds.

**Quantum Chemical Calculations:**

All quantum chemical calculations were performed with the Q-Chem software package.[59] Calculations were performed in the gas phase for TCA and the complex between the TCA anion and $Pb^{2+}$ cation. Equilibrium geometries for the lowest energy singlet ($S_0$) and triplet ($T_1$) states for both species were optimized with the $\omega$B97M-V density functional,[60] using the using the def2-SVP basis set[61] (and the associated effective core potential for Pb[62]). CCSD(T)/def2-SVP single point calculations[63] on the resulting geometries were subsequently performed to obtain relative energies of the two spin states, which were largely in agreement with the $\omega$B97M-V results (reported in the supporting information).

**Transient absorption and time-resolved photoluminescence (TRPL) spectroscopy**

Films for transient absorption/TRPL in Figures 2d, 2e, and SI 10, 11a were fabricated using 1030 nm PbS QDs and concentrations of 300 mg/mL TES-ADT, 1 mg/mL DBP, 75 mg/mL PbS QDs. Film for TA in Figures SI 11b was fabricated using 1120 nm PbS QDs and concentrations of 300 mg/mL TES-ADT, 1 mg/mL DBP, 50 mg/mL PbS QDs.

Nanosecond transient absorption measurements were performed using a commercial setup (enVISion, Magnitude Instruments). Excitation was via an external pulsed $Nd:YVO_4$ laser (Picolo, Innolas, 1064 nm, $\tau \sim 800$ ps, $f = 2000$ Hz). Residual higher harmonics of the laser output (355/532 nm) were picked off by a 950 nm longpass filter (FELH950, Thorlabs) into a Si photodetector (DET10A2, Thorlabs), with the remaining 1064 nm light passing through an optical chopper modulated at 1000 Hz, before entering the sample region. Both the Si photodiode and chopper passed a TTL signal to the instrument for shot-to-shot noise reduction and pump on/off triggering, respectively. A xenon lamp was used as a continuous probe source, overlapped on the sample at 90° to the 1064 nm pump. Probe and PL wavelength selection was via a monochromator after the sample region, with time-resolution provided by a fast photodetector (Si for 620 nm and 700 nm; InGaAs for 1035 nm). A probe monochromator slit width of 10 nm was used. Excitation fluence was 1 mJ $cm^{-2}$ for all reported data.

Ultrafast transient absorption measurements were performed using a commercial setup (Harpia TA, Light Conversion) and a Nd:KGW regenerative amplifier (1030 nm, $\tau \sim 200$ fs, $f = 50$ kHz), which seeded an optical parametric amplifier (Orpheus-HP, Light Conversion) to generate the 500 nm pump, which was then modulated at $f/2$ repetition rate by an optical chopper. A separate portion of the 1030 nm seeded a sapphire crystal for white light probe generation after traversing a mechanical delay stage that controlled relative pump-probe time delay. Pump and probe were overlapped on a *d* ~ 200 μm spot on the sample. The probe transmission with and without the pump present was read out using a free-space spectrograph (Kymera 193i, Andor) to generate the differential absorption spectrum at each time delay.

TRPL measurements were performed using a home-built optical setup. Excitation was via a pulsed $Nd:YVO_4$ laser (Picolo, Innolas, 1064 nm, τ ~ 800 ps, *f* = 5000 Hz). Residual higher harmonics (355/532 nm) of the laser were rejected by a 950 nm longpass filter (FELH950, Thorlabs). The excitation beam was focused onto a *d* ~ 500 μm spot on the sample. PL was collected by a pair of off-axis parabolic mirrors, which focused the emission into a spectrograph (Acton SpectraPro SP-2150, Princeton Instruments) coupled to a high-speed intensified charge-coupled device camera (Pi-Max 4, Princeton Instruments, 300 g/mm, 500 nm blaze). Spectrum collection was electronically gated and delayed with respect to the laser source trigger to construct the time-axis (resolution ~1 ns). Excitation fluence was 300 μJ/cm$^2$. The kinetics were tracked by integrating the spectrum at 600-700 nm for each time point.

DATA AVAILABILITY

The authors declare that the main data supporting the findings of this study are available within the article and its Supplementary Information files. Additional data, including raw data files, are available from the corresponding author upon request. Source data are provided with this paper.

AUTHOR CONTRIBUTIONS

Conceptualization: PN, LP, OSL, MAK, DNC. Methodology: PN, LP, OSL, RH. Investigation and interpretation: PN, RH, LP, OSL, BPC, JS, JSE, ONA, DH. Supervision: TJM, TWS, MPN, MJYT, MAK, DNC. Writing - original draft: PN. Writing - review and editing: all authors. Funding acquisition: MAK, DNC.

FUNDING SOURCES

Funding from the Defense Advanced Research Projects Agency grant HR00112220010 (PN, RH, LP, OSL, JS, JSE, MAK, DNC) is acknowledged.

ACKNOWLEDGMENTS.

Part of this work was performed at the Stanford Nano Shared Facilities (SNSF), supported by the National Science Foundation under award ECCS-2026822. Part of this work was performed at the Stanford University Chemistry Department NMR Facility (RRID:SCR_023325). PN acknowledges the support of a Stanford Graduate Fellowship in Science & Engineering (SGF) as a Gabilan Fellow and the Chevron Fellowship in Energy. LP acknowledges support of a National Science Foundation Graduate Research Fellowship under grant DGE-2146755. OSL acknowledges the funding support of the Swiss National Science Foundation (SNSF), under grant number P500PT_222358. ONA acknowledges support from the Stanford University School of Engineering through the Stanford Undergraduate Research Fellowship (SURF) program. JS was a Wisconsin Distinguished Graduate Fellow for a portion of this work. JSE was supported by the Department of Defense (DoD) through the National Defense Science & Engineering Graduate (NDSEG) Fellowship Program. TJM and DH acknowledge support from the AMOS program of the US Department of Energy, Office of Science, Basic Energy Sciences, Chemical Sciences, and Biosciences Division. DH was a Stanford Science Fellow for the initial stages of this work. The Flatiron Institute is a division of the Simons Foundation. MPN recognizes the support of the UNSW Scientia Program and an ARC DECRA Fellowship (DE230100382). MJYT acknowledges support from an ARC Future Fellowship (FT230100002).

REFERENCES

1. Kinoshita, M. *et al.* Photon Upconverting Solid Films with Improved Efficiency for Endowing Perovskite Solar Cells with Near-Infrared Sensitivity. *ChemPhotoChem* **4**, 5271–5278 (2020).

2. Richards, B. S., Hudry, D., Busko, D., Turshatov, A. & Howard, I. A. Photon Upconversion for Photovoltaics and Photocatalysis: A Critical Review: Focus Review. *Chem. Rev.* **121**, 9165–9195 (2021).

3. Sanders, S. N. *et al.* Triplet fusion upconversion nanocapsules for volumetric 3D printing. *Nature* **604**, 474–478 (2022).

4. Limberg, D. K., Kang, J.-H. & Hayward, R. C. Triplet–Triplet Annihilation Photopolymerization for High-Resolution 3D Printing. *J. Am. Chem. Soc.* **144**, 5226–5232 (2022).

5. Hamid, R. *et al.* All-passive upconversion of incoherent near-infrared light at intensities down to 10$^{-7}$ W/cm$^2$. Preprint at https://doi.org/10.48550/ARXIV.2411.18707 (2024).

6. Jin, J. *et al.* Recent advances of triplet–triplet annihilation upconversion in photochemical transformations. *Curr. Opin. Green Sustain. Chem.* **43**, 100841 (2023).

7. Liu, Z. *et al.* Near-Infrared to Visible Photon Upconversion with Gold Quantum Rods and Aqueous Photo-Driven Polymerization. *J. Am. Chem. Soc.* jacs.5c08826 (2025) doi:10.1021/jacs.5c08826.

8. Castellanos-Soriano, J., Garnes-Portolés, F., Jiménez, M. C., Leyva-Pérez, A. & Pérez-Ruiz, R. In-Flow Heterogeneous Triplet–Triplet Annihilation Upconversion. *ACS Phys. Chem. Au* **4**, 242–246 (2024).

9. Szalkowski, M. *et al.* Advances in the photon avalanche luminescence of inorganic lanthanide-doped nanomaterials. *Chem. Soc. Rev.* **54**, 983–1026 (2025).

10. Zhou, J., Liu, Q., Feng, W., Sun, Y. & Li, F. Upconversion Luminescent Materials: Advances and Applications. *Chem. Rev.* **115**, 395–465 (2015).

11. Malhotra, K. *et al.* Lanthanide-Doped Upconversion Nanoparticles: Exploring A Treasure Trove of NIR-Mediated Emerging Applications. *ACS Appl. Mater. Interfaces* **15**, 2499–2528 (2023).

12. Bharmoria, P., Bildirir, H. & Moth-Poulsen, K. Triplet–triplet annihilation based near infrared to visible molecular photon upconversion. *Chem. Soc. Rev.* **49**, 6529–6554 (2020).

13. Schloemer, T. *et al.* Nanoengineering Triplet–Triplet Annihilation Upconversion: From Materials to Real-World Applications. *ACS Nano* **17**, 3259–3288 (2023).

14. Chen, K., Luan, Q., Liu, T., Albinsson, B. & Hou, L. Semiconductor nanocrystals-based triplet-triplet annihilation photon-upconversion: Mechanism, materials and applications. *Responsive Mater.* **3**, e20240030 (2025).

15. Bossanyi, D. G. *et al.* In optimized rubrene-based nanoparticle blends for photon upconversion, singlet energy collection outcompetes triplet-pair separation, not singlet fission. *J. Mater. Chem. C* **10**, 4684–4696 (2022).

16. Niihori, Y. & Mitsui, M. Harnessing metal cluster sensitizers for triplet–triplet annihilation photon upconversion: Strategies for performance enhancement. *Chem. Phys. Rev.* **6**, 031301 (2025).

17. Bossanyi, D. G. *et al.* Spin Statistics for Triplet–Triplet Annihilation Upconversion: Exchange Coupling, Intermolecular Orientation, and Reverse Intersystem Crossing. *JACS Au* **1**, 2188–2201 (2021).

18. Millington, O. *et al.* The Interplay of Strongly and Weakly Exchange-Coupled Triplet Pairs in Intramolecular Singlet Fission. *J. Am. Chem. Soc.* **146**, 29664–29674 (2024).

19. Huang, Z. *et al.* Enhanced Near-Infrared-to-Visible Upconversion by Synthetic Control of PbS Nanocrystal Triplet Photosensitizers. *J. Am. Chem. Soc.* **141**, 9769–9772 (2019).

20. Wu, M. *et al.* Solid-state infrared-to-visible upconversion sensitized by colloidal nanocrystals. *Nat. Photonics* **10**, 31–34 (2016).

21. Bi, P. *et al.* Donor-acceptor bulk-heterojunction sensitizer for efficient solid-state infrared-to-visible photon up-conversion. *Nat. Commun.* **15**, 5719 (2024).

22. Hu, M. *et al.* Bulk Heterojunction Upconversion Thin Films Fabricated via One-Step Solution Deposition. *ACS Nano* **17**, 22642–22655 (2023).

23. Izawa, S. & Hiramoto, M. Efficient solid-state photon upconversion enabled by triplet formation at an organic semiconductor interface. *Nat. Photonics* **15**, 895–900 (2021).

24. Geva, N. *et al.* A Heterogeneous Kinetics Model for Triplet Exciton Transfer in Solid-State Upconversion. *J. Phys. Chem. Lett.* **10**, 3147–3152 (2019).

25. Gray, V. *et al.* Ligand-Directed Self-Assembly of Organic-Semiconductor/Quantum-Dot Blend Films Enables Efficient Triplet Exciton-Photon Conversion. *J. Am. Chem. Soc.* **146**, 7763–7770 (2024).

26. Wu, M., Lin, T.-A., Tiepelt, J. O., Bulović, V. & Baldo, M. A. Nanocrystal-Sensitized Infrared-to-Visible Upconversion in a Microcavity under Subsolar Flux. *Nano Lett.* **21**, 1011–1016 (2021).

27. Narayanan, P. *et al.* Alleviating Parasitic Back Energy Transfer Enhances Thin Film Upconversion. *Adv. Opt. Mater.* **13**, 2500252 (2025).

28. Nienhaus, L. *et al.* Speed Limit for Triplet-Exciton Transfer in Solid-State PbS Nanocrystal-Sensitized Photon Upconversion. *ACS Nano* **11**, 7848–7857 (2017).

29. Wu, M., Jean, J., Bulović, V. & Baldo, M. A. Interference-enhanced infrared-to-visible upconversion in solid-state thin films sensitized by colloidal nanocrystals. *Appl. Phys. Lett.* **110**, 211101 (2017).

30. Nienhaus, L. *et al.* Solid-state infrared-to-visible upconversion for sub-bandgap sensitization of photovoltaics. in *2018 IEEE 7th World Conference on Photovoltaic Energy Conversion (WCPEC) (A Joint Conference of 45th IEEE PVSC, 28th PVSEC & 34th EU PVSEC)* 3698–3702 (IEEE, Waikoloa, HI, USA, 2018). doi:10.1109/PVSC.2018.8547686.

31. Dziobek-Garrett, R., Imperiale, C. J., Wilson, M. W. B. & Kempa, T. J. Photon Upconversion in a Vapor Deposited 2D Inorganic–Organic Semiconductor Heterostructure. *Nano Lett.* **23**, 4837–4843 (2023).

32. Duan, J. *et al.* Efficient solid-state infrared-to-visible photon upconversion on atomically thin monolayer semiconductors. *Sci. Adv.* **8**, eabq4935 (2022).

33. Radiunas, E. *et al.* CN-Tuning: A Pathway to Suppress Singlet Fission and Amplify Triplet-Triplet Annihilation Upconversion in Rubrene. *Adv. Opt. Mater.* **13**, 2403032 (2025).

34. Amemori, S., Sasaki, Y., Yanai, N. & Kimizuka, N. Near-Infrared-to-Visible Photon Upconversion Sensitized by a Metal Complex with Spin-Forbidden yet Strong $S_0$ –$T_1$ Absorption. *J. Am. Chem. Soc.* **138**, 8702–8705 (2016).

35. Nienhaus, L. *et al.* Triplet-Sensitization by Lead Halide Perovskite Thin Films for Near-Infrared-to-Visible Upconversion. *ACS Energy Lett.* **4**, 888–895 (2019).

36. Zhang, J. *et al.* Synthetic Conditions for High-Accuracy Size Control of PbS Quantum Dots. *J. Phys. Chem. Lett.* **6**, 1830–1833 (2015).

37. Green, P. B. *et al.* Controlling Cluster Intermediates Enables the Synthesis of Small PbS Nanocrystals with Narrow Ensemble Line Widths. *Chem. Mater.* **32**, 4083–4094 (2020).

38. Song, F. *et al.* Advances of Fluorescence Molecular Imaging: NIR-II Window, Probes, and Tomography. *Laser Photonics Rev.* **19**, 2400275 (2025).

39. Zhao, M. & Chen, X. Recent Advances in NIR-II Materials for Biomedical Applications. *Acc. Mater. Res.* **5**, 600–613 (2024).

40. Gholizadeh, E. M. *et al.* Photochemical upconversion of near-infrared light from below the silicon bandgap. *Nat. Photonics* **14**, 585–590 (2020).

41. Tripathi, N. & Kamada, K. Enhanced Near-Infrared-to-Visible Upconversion by a Singlet Sink Approach in a Quantum-Dot-Sensitized Triplet–Triplet Annihilation System. *ACS Appl. Nano Mater.* **7**, 2950–2955 (2024).

42. Tripathi, N., Ando, M., Akai, T. & Kamada, K. Near-infrared-to-visible upconversion from 980 nm excitation band by binary solid of PbS quantum dot with directly attached emitter. *J. Mater. Chem. C* **10**, 4563–4567 (2022).

43. Tripathi, N., Ando, M., Akai, T. & Kamada, K. Efficient NIR-to-Visible Upconversion of Surface-Modified PbS Quantum Dots for Photovoltaic Devices. *ACS Appl. Nano Mater.* **4**, 9680–9688 (2021).

44. Nishimura, N. *et al.* Photon upconversion utilizing energy beyond the band gap of crystalline silicon with a hybrid TES-ADT/PbS quantum dots system. *Chem. Sci.* **10**, 4750–4760 (2019).

45. Lekavičius, J. *et al.* Aggregation favors singlet formation in TES-ADT triplet annihilator for photon upconversion. *Chem. Sci.* **17**, 6230–6237 (2026).

46. Ho, E. A., Soni, A., Zhai, F. & Wang, L. Pseudo-Solid-State Polymer Materials for QD-Sensitized NIR-I and NIR-II Upconversion Beyond the Silicon Bandgap. *Adv. Mater.* e12741 (2025) doi:10.1002/adma.202512741.

47. Zhao, S. *et al.* Efficient Near-Infrared to Blue Photon Upconversion by Ultrafast Spin Flip and Triplet Energy Transfer at Organic/2D Semiconductor Interface. *Angew. Chem.* **137**, e202420070 (2025).

48. Wang, L. *et al.* Interfacial Trap-Assisted Triplet Generation in Lead Halide Perovskite Sensitized Solid-State Upconversion. *Adv. Mater.* **33**, 2100854 (2021).

49. Huang, Z. *et al.* Hybrid Molecule–Nanocrystal Photon Upconversion Across the Visible and Near-Infrared. *Nano Lett.* **15**, 5552–5557 (2015).

50. Allardice, J. R. *et al.* Engineering Molecular Ligand Shells on Quantum Dots for Quantitative Harvesting of Triplet Excitons Generated by Singlet Fission. *J. Am. Chem. Soc.* **141**, 12907–12915 (2019).

51. Huang, Z. & Tang, M. L. Designing Transmitter Ligands That Mediate Energy Transfer between Semiconductor Nanocrystals and Molecules. *J. Am. Chem. Soc.* **139**, 9412–9418 (2017).

52. De Mello, J. C., Wittmann, H. F. & Friend, R. H. An improved experimental determination of external photoluminescence quantum efficiency. *Adv. Mater.* **9**, 230–232 (1997).

53. Zhou, Y., Castellano, F. N., Schmidt, T. W. & Hanson, K. On the Quantum Yield of Photon Upconversion via Triplet–Triplet Annihilation. *ACS Energy Lett.* **5**, 2322–2326 (2020).

54. Monguzzi, A., Tubino, R., Hoseinkhani, S., Campione, M. & Meinardi, F. Low power, non-coherent sensitized photon up-conversion: modelling and perspectives. *Phys. Chem. Chem. Phys.* **14**, 4322 (2012).

55. Monguzzi, A., Mezyk, J., Scotognella, F., Tubino, R. & Meinardi, F. Upconversion-induced fluorescence in multicomponent systems: Steady-state excitation power threshold. *Phys. Rev. B* **78**, 195112 (2008).

56. Shockley, W. & Queisser, H. J. Detailed Balance Limit of Efficiency of *p-n* Junction Solar Cells. *J. Appl. Phys.* **32**, 510–519 (1961).

57. Xiang, H. *et al.* Upconversion nanoparticles extending the spectral sensitivity of silicon photodetectors to λ = 1.5 µm. *Nanotechnology* **31**, 495201 (2020).

58. Sakamoto, Y., Izawa, S., Ohkita, H., Hiramoto, M. & Tamai, Y. Triplet sensitization via charge recombination at organic heterojunction for efficient near-infrared to visible solid-state photon upconversion. *Commun. Mater.* **3**, 76 (2022).

59. Epifanovsky, E. *et al.* Software for the frontiers of quantum chemistry: An overview of developments in the Q-Chem 5 package. *J. Chem. Phys.* **155**, 084801 (2021).

60. Mardirossian, N. & Head-Gordon, M. *ω* B97M-V: A combinatorially optimized, range-separated hybrid, meta-GGA density functional with VV10 nonlocal correlation. *J. Chem. Phys.* **144**, 214110 (2016).

61. Weigend, F. & Ahlrichs, R. Balanced basis sets of split valence, triple zeta valence and quadruple zeta valence quality for H to Rn: Design and assessment of accuracy. *Phys. Chem. Chem. Phys.* **7**, 3297 (2005).

62. Metz, B., Stoll, H. & Dolg, M. Small-core multiconfiguration-Dirac–Hartree–Fock-adjusted pseudopotentials for post- ***d*** main group elements: Application to PbH and PbO. *J. Chem. Phys.* **113**, 2563–2569 (2000).

63. Raghavachari, K., Trucks, G. W., Pople, J. A. & Head-Gordon, M. A fifth-order perturbation comparison of electron correlation theories. *Chem. Phys. Lett.* **157**, 479–483 (1989).

**Supporting Information**

**Figures:**

**Tables:**

**Notes:**

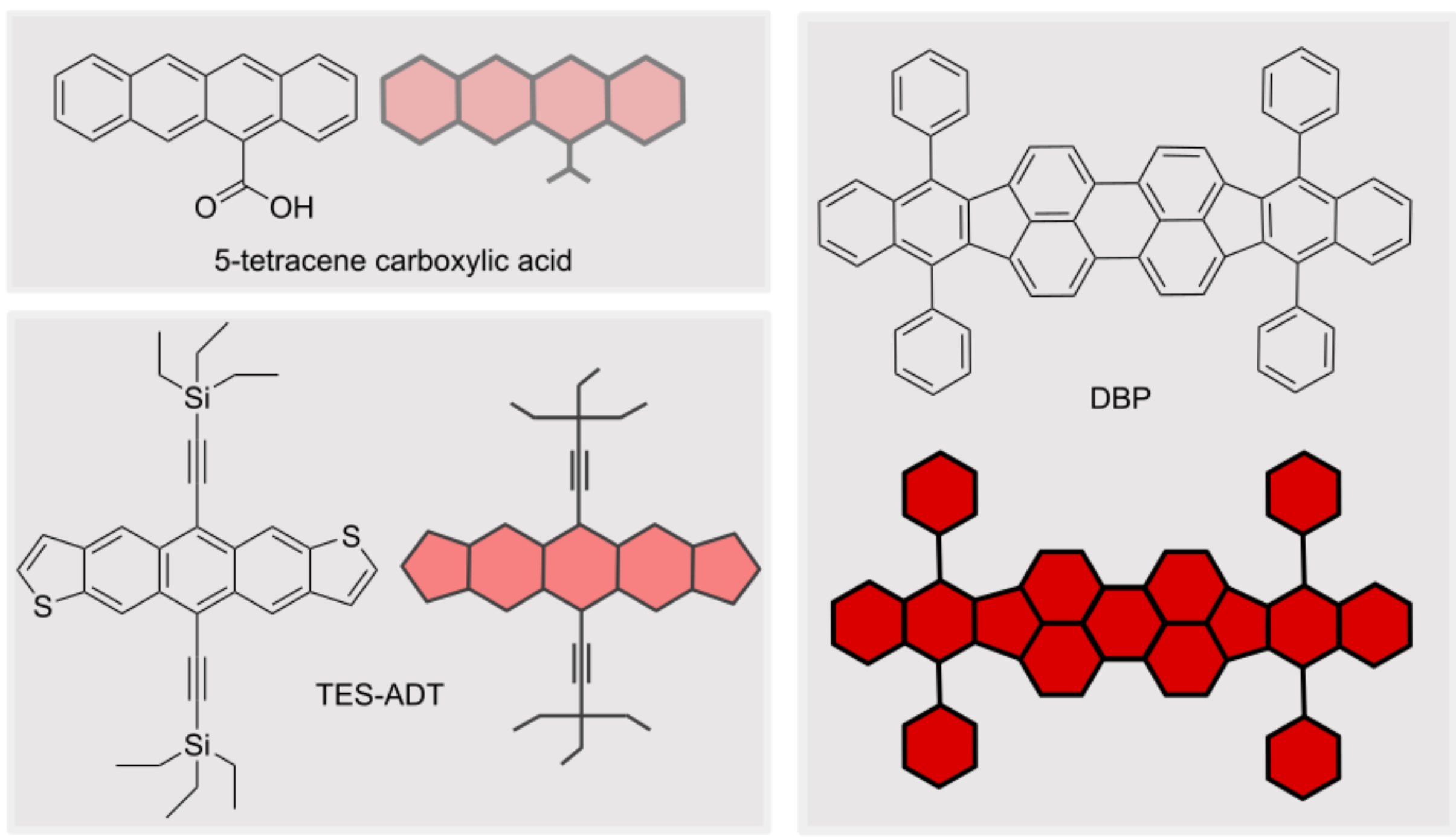

**Figure SI 1:** Chemical structure of TCA (top left), TES-ADT (bottom left), and DBP (right).

**SI Note 1:** To better facilitate communication and comparisons in the TTA community, Zhou, et al. have proposed a unified labeling scheme to describe quantum yield in TTA-UC.[1] The table below summarizes the equations to define various efficiency metrics. In this paper, we use IQE and EQE as metrics which are defined below.

**$\Phi_{UC}$** is the standard TTA-UC emission quantum yield (maximum of 50%)

**By the definition used in this work, $\Phi_{UC}$ = IQE**

**$\eta_{UC}$** is the normalized TTA-UC emission quantum yield (maximum of 100%)

**$\Phi_{UCs}$** is the singlet generation quantum yield

**$\Phi_{UCg}$** is the photon generation quantum yield

**$\Phi_{TTA}$** is the triplet-triplet annihilation efficiency (i.e., conversion of two triplets to one singlet) (maximum of 50%)

**$\Phi_{ISC}$** is the sensitizer intersystem crossing efficiency

**$\Phi_{TET}$** is the triplet energy transfer efficiency

**$\Phi_{f}$** is the fluorescence quantum yield (PLQY) of the annihilator/emitter

**$\Phi_{out}$** is the outcoupling efficiency of upconverted photons (i.e., losses from annihilator reabsorption)

**$\Phi_{BT}$** is the back energy transfer efficiency from annihilator singlet to sensitizer

In the measurements reported in this work, an integrating sphere is used to include waveguided upconverted photons in the calculation for IQE.

$$\phi_{UC} = \frac{\#observed\ UC\ photons}{\#absorbed\ photons} = \phi_{ISC} \cdot \phi_{TET} \cdot \phi_{TTA} \cdot \phi_f \cdot \phi_{out} \cdot (1 - \phi_{BT}) = \boldsymbol{IQE} \quad \text{Equation 1}$$

$$\eta_{UC} = \phi_{UC} \times 2 \quad \text{Equation 2}$$

$$\phi_{UC_g} = \phi_{ISC} \cdot \phi_{TET} \cdot \phi_{TTA} \cdot \phi_f \cdot (1 - \phi_{ET}) = \frac{\phi_{UC}}{\phi_{out}} \quad \text{Equation 3}$$

$$\phi_{UC_s} = \phi_{ISC} \cdot \phi_{TET} \cdot \phi_{TTA} = \frac{\phi_{UC}}{\phi_f \cdot \phi_{out} \cdot (1 - \phi_{BT})} \quad \text{Equation 4}$$

$$\boldsymbol{EQE} = \%absorption\ \times IQE \quad \text{Equation 5}$$

| **UC System** | **Laser Exc (nm)** | **UC Emission (nm)** | **IQE (%, 50% max)** | **Reported values and any further calculations** |
|---|---|---|---|---|
| PdPc/CN-rubrene[2] | 730 | 610 | 3.00 | $\Phi_{UC}$ = 3.0% |
| $WSe_2$/TIPS-anthracene[3] | 730 | 450 | 1.2 | $\Phi_{UC}$ = 1.2% |
| ITIC/rubrene[4] | 750 | 565 | 1.42 | $\Phi_{UC}$ = 1.42% |
| ITIC/rubrene/DBP[4] | 750 | 610 | 2.53 | $\Phi_{UC}$ = 2.53% |
| $MoSe_2$/rubrene/DBP[5] | 772 | 610 | 1.1 | $\Phi_{UC}$ = 1.1% |
| PbS/TCA/TES-ADT[6] | 785 | 610 | 0.36 | $\eta_{UC}$ = 0.72%<br>In this paper, $\eta_{UC}$ is defined to be (2* $\Phi_{UC}$)<br>Therefore, $\Phi_{UC} = \eta_{UC}/2 = 0.72/2 = 0.36\%$ |
| Perovskite/rubrene/DBP[7] | 785 | 610 | 0.15 | Reported: $\Phi_f$ = 9.3% and "$\eta_{UC}$" = 3.1%<br>In this paper, $\eta_{UC}$ is defined to be (2* $\Phi_{UC}/\Phi_f$)<br>Therefore, $\Phi_{UC} = \eta_{UC}*\Phi_f/2 = 3.1 * 0.093/2 = 0.15\%$ |
| Perovskite/rubrene/DBP[8] | 785 | 610 | 0.11 | Reported: $\Phi_f$ = 45.6% and "$\Phi_{UC}$" = 0.489%<br>In this paper, $\Phi_{UC}$ is defined to be (2*"$\Phi_{UC}$"/$\Phi_f$)<br>Therefore, $\Phi_{UC}$ = "$\Phi_{UC}$"*$\Phi_f$/2 = 0.489 * 0.456/2 = 0.11% |
| PbS/rubrene/DBP[9] | 808 | 610 | 1.6 | Reported: "IQE" = 7%<br>In this paper, IQE is defined to be (2* $\Phi_{UC}/\Phi_f$)<br>We use $\Phi_f$ = 45% PLQY of DBP/rubrene based discussion with the authors<br>Therefore, $\Phi_{UC}$ = "IQE"*$\Phi_f$/2 = 7 * 0.45/2 = 1.6% |
| PbS/rubrene/DBP[10] | 808 | 610 | 0.28 | Reported: $\Phi_f$ = ~46% and "$\eta_{UC850}$" = 1.23%<br>In this paper, $\eta_{UC}$ is defined to be (2* $\Phi_{UC}/\Phi_f$)<br>Therefore, $\Phi_{UC} = \eta_{UC}*\Phi_f/2 = 1.23 * 0.46/2 = 0.28\%$ |
| PbS/rubrene/DBP[10] | 808 | 610 | 0.11 | Reported: $\Phi_f$ = ~44% and "$\eta_{UC960}$" = 0.51%<br>In this paper, $\eta_{UC}$ is defined to be (2* $\Phi_{UC}/\Phi_f$)<br>Therefore, $\Phi_{UC} = \eta_{UC}*\Phi_f/2 = 0.51 * 0.44/2 = 0.11\%$ |
| PbS/rubrene/DBP[10] | 808 | 610 | 0.049 | Reported: $\Phi_f$ = ~47% and "$\eta_{UC1010}$" = 0.21%<br>In this paper, $\eta_{UC}$ is defined to be (2* $\Phi_{UC}/\Phi_f$)<br>Therefore, $\Phi_{UC} = \eta_{UC}*\Phi_f/2 = 0.21 * 0.47/2 = 0.049\%$ |
| | | | | An excitation UC spectrum for these films was collected, showing UC from ~930-1130 nm. |
| PbS/rubrene/DBP[11] | 808 | 610 | 1.6 | Reported: “$\eta_{UC}$” = 7%<br>In this paper, $\eta_{UC}$ is defined to be (2* $\Phi_{UC}/\Phi_f$)<br>We use $\Phi_f$ = 45% PLQY of DBP/rubrene based discussion with the authors<br>Therefore, $\Phi_{UC}$ = “IQE”*$\Phi_f$/2 = 7 * 0.45/2 = 1.6% |
| PbS/rubrene/DBP[12] | 980 | 610 | 0.36 | Reported: “$\eta_{UC}$” = 1.6%<br>In this paper, $\eta_{UC}$ is defined to be (2* $\Phi_{UC}/\Phi_f$)<br>We use $\Phi_f$ = 45% PLQY of DBP/rubrene based discussion with the authors<br>Therefore, $\Phi_{UC}$ = “IQE”*$\Phi_f$/2 = 1.6 * 0.45/2 = 0.36% |
| Y6/rubrene/DBP[13] | 808 | 610 | 0.51 | Reported: EQE = 0.19%, absorption = 37.1%<br>Therefore, IQE = $\Phi_{UC}$ = 0.19/0.371 = 0.51% |
| PbS/TCA/rubrene/DBP[14] | 808 | 610 | 0.78 | Reported: EQE = 0.0045%, absorption = 0.58%<br>Therefore, IQE = $\Phi_{UC}$ = 0.0045/0.0058 = 0.78% |
| PbS/TCA/rubrene/DBP[14] | 808 | 610 | 0.17 | Reported: EQE = 0.01%, absorption = 6.0%<br>Therefore, IQE = $\Phi_{UC}$ = 0.01/0.06 = 0.17% |
| PBQx-TCI/PYIT1/rubrene[15] | 808 | 610 | 2.20 | $\Phi_{UC}$ = 2.2% |
| Y6/rubrene[4] | 850 | 565 | 0.52 | IQE = $\Phi_{UC}$ = 0.515% |
| Os complex/rubrene[16] | 938 | 610 | 0.22 | $\Phi_{UC}$ = 0.22% |
| Os complex/rubrene/DBP[17] | 730, 938 | 610 | 2.05 | Reported: $\Phi'_{UC}$ = 4.1%<br>In this paper, $\Phi'_{UC}$ is defined to be (2* $\Phi_{UC}$)<br>Therefore, IQE = $\Phi_{UC}$ = 4.1/2 = 2.05%<br><br>IQE in this work was measured at 730 nm due to insufficient detector sensitivity at 938 nm. |
| PbS/TES-ADT[18] | 975 | 660 | 0.34 | $\Phi_{UC}$ = 0.34% |
| PbS/TES-ADT/DBP[19] | 980 | 700 | 0.75 | Reported: $\eta_{UC}$ = 1.5%<br>In this paper, $\eta_{UC}$ is defined to be (2* $\Phi_{UC}$)<br>Therefore, IQE = $\Phi_{UC}$ = 1.5/2 = 0.75% |
| PbS/rubrene/DBP/reflector[20] | 980 | 610 | 0.80 | Reported: $\Phi_{UCs}$ = 0.90% and $\Phi_f$ = ~89%<br>In this paper, $\Phi_{UCs}$ is defined to be $(\Phi_{UC}/\Phi_f)$<br>Therefore, $\Phi_{UC} = \eta_{UC} * \Phi_f = 0.90 * 0.89 =$ 0.80% |
| PbS/TCA/TES-ADT[6] | 1064 | 610 | 0.011 | $\eta_{UC}$ = 0.022%<br>In this paper, $\eta_{UC}$ is defined to be $(2 * \Phi_{UC})$<br>Therefore, $\Phi_{UC} = \eta_{UC}/2 = 0.022/2 =$ 0.011% |
| PbS-TCA/TES-ADT/DBP | 808 | 700 | 1.9 | This work (champion) |
| PbS-TCA/TES-ADT/DBP | 1130 | 700 | 0.46 | This work (champion) |
| PbS-TCA/TES-ADT/DBP | 1208 | 700 | 0.101 | This work (champion) |

**Table SI 1:** Summary of IQE values reported in literature for various NIR-to-visible UC systems.

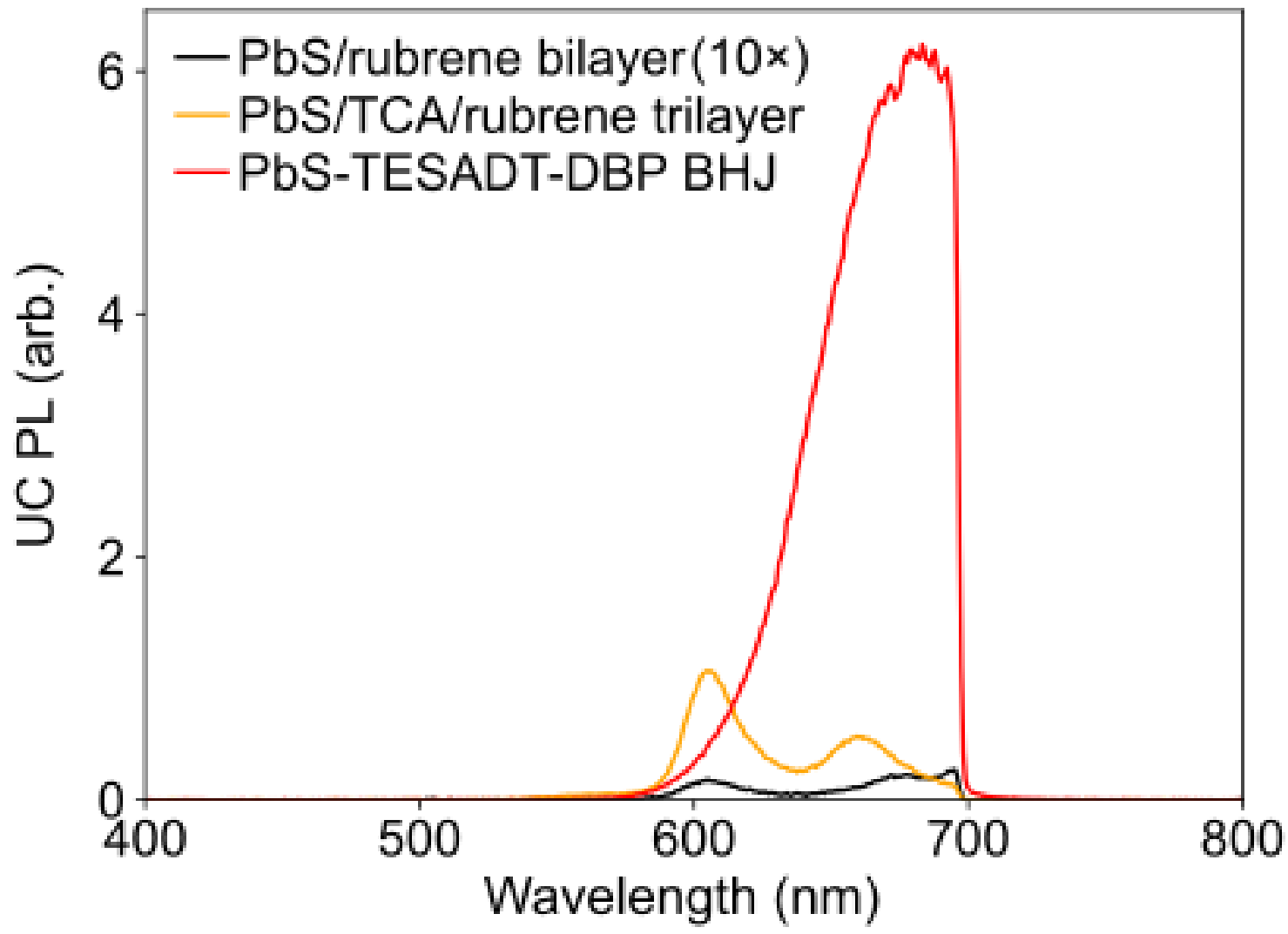

**Figure SI 2:** Comparison of UC PL from 960 nm QDs in different UC systems under identical 808 nm laser excitation. The PbS/rubrene bilayer UC (black trace) is magnified 10-fold to fit within the same y-axis range.

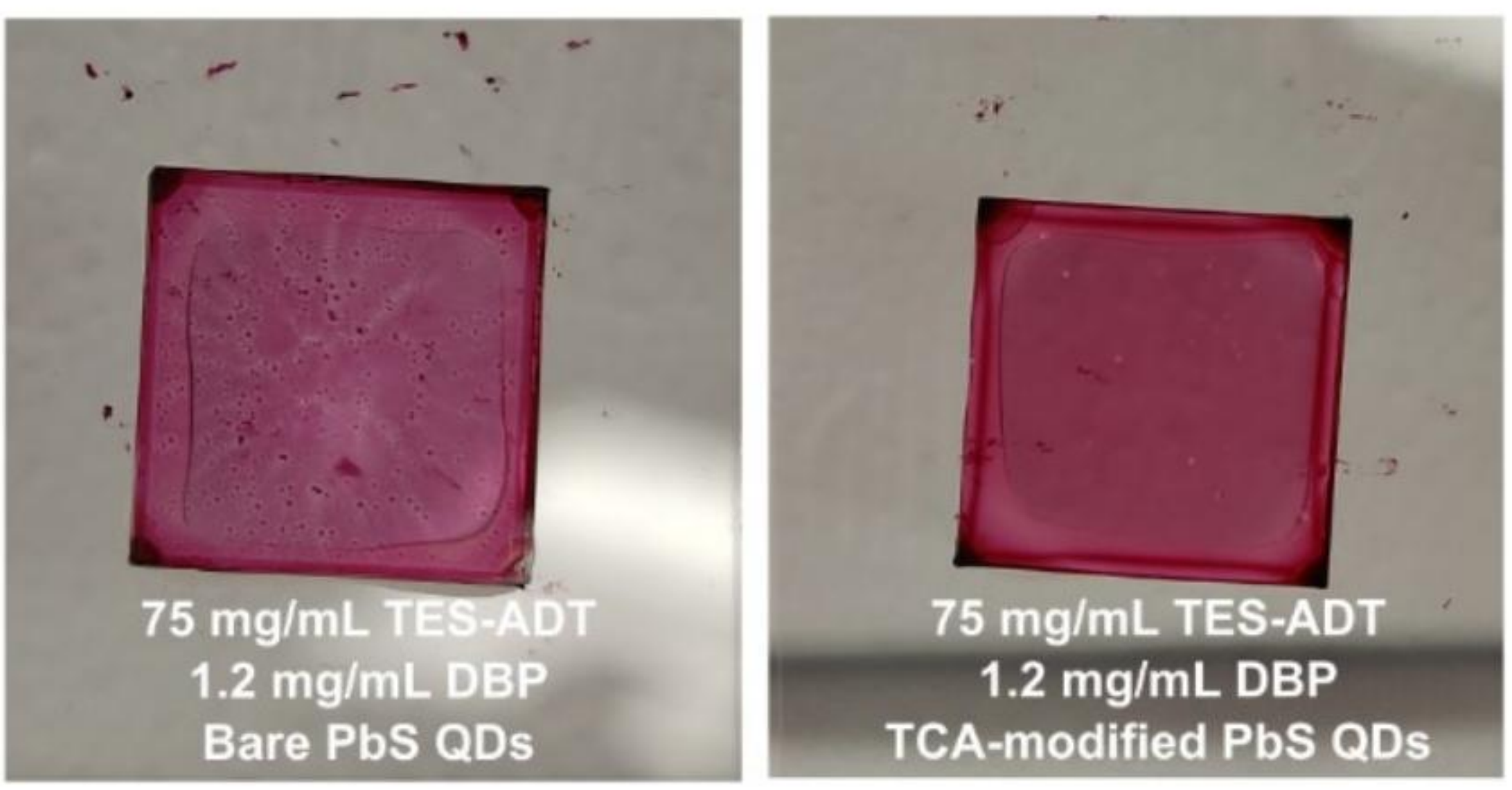

**Figure SI 3:** TCA-modified PbS QDs show better solubility in the TES-ADT/DBP matrix, resulting in more uniform films.

| | | TCA | | | Complex of TCA anion with $Pb^{2+}$ | | |
|---|---|---|---|---|---|---|---|
| **Geometry** | **Electronic State** | **$\omega$B97M-V/ def2-SVP** | **CCSD(T) / def2-SVP** | **$\omega$B97M-V/ def2-TZVPPD** | **$\omega$B97M-V/ def2-SVP** | **CCSD(T) / def2-SVP** | **$\omega$B97M-V/ def2-TZVPPD** |
| $S_0$ | $S_0$ | 0.00 | 0.00 | 0.00 | 0.00 | 0.00 | 0.00 |
| | $T_1$ | 1.77 | 1.82 | 1.74 | 1.16 | 1.16 | 1.20 |
| $T_1$ | $S_0$ | 0.43 | 0.32 | 0.47 | 0.61 | 0.36 | 0.69 |
| | $T_1$ | 1.31 | 1.41 | 1.33 | 0.49 | 0.49 | 0.58 |

**Table SI 2:** Energies of the lowest energy singlet ($S_0$) and triplet ($T_1$) states of TCA and the complex of TCA anion with $Pb^{2+}$, as computed with density functional theory ($\omega$B97M-V) and coupled cluster theory [CCSD(T)]. All energies are in eV, and are shown relative to the $S_0$ minimum (Franck-Condon point). The $S_0$ and $T_1$ geometries were optimized with $\omega$B97M-V/def2-SVP. The CCSD(T) calculations for the triplet were carried out with restricted open-shell Hartree-Fock orbitals as unrestricted Hartree-Fock calculations of the triplet had very large amounts of spin-contamination ($\langle S^2 \rangle \sim 3$ vs the ideal value of 2). We note that $\omega$B97M-V predicts $T_1$ to be more stable than $S_0$ at the minimum energy geometry of the $T_1$ state for TCA-Pb complex, but this is likely on account of overestimation of the $S_0$ state energy due to use of spin-restricted orbitals (as the $S_0$ state at this geometry is likely to have significant biradicaloid character). The much more accurate CCSD(T) method predicts a small but positive energy gap between the $S_0$ and $T_1$ states at this geometry.

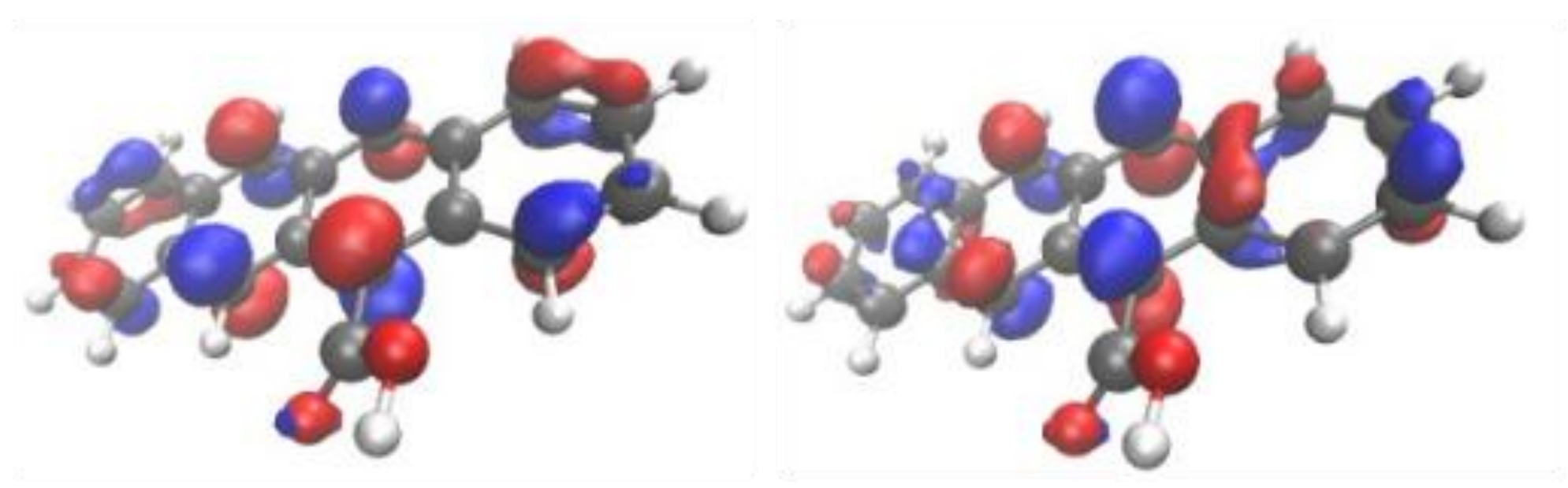

**Figure SI 4:** Singly occupied molecular orbitals (from $\omega$B97M-V/def2-TZVPPD) for the $T_1$ state of TCA at the $T_1$ optimized geometry. Both unpaired electrons are delocalized over the acene ring.

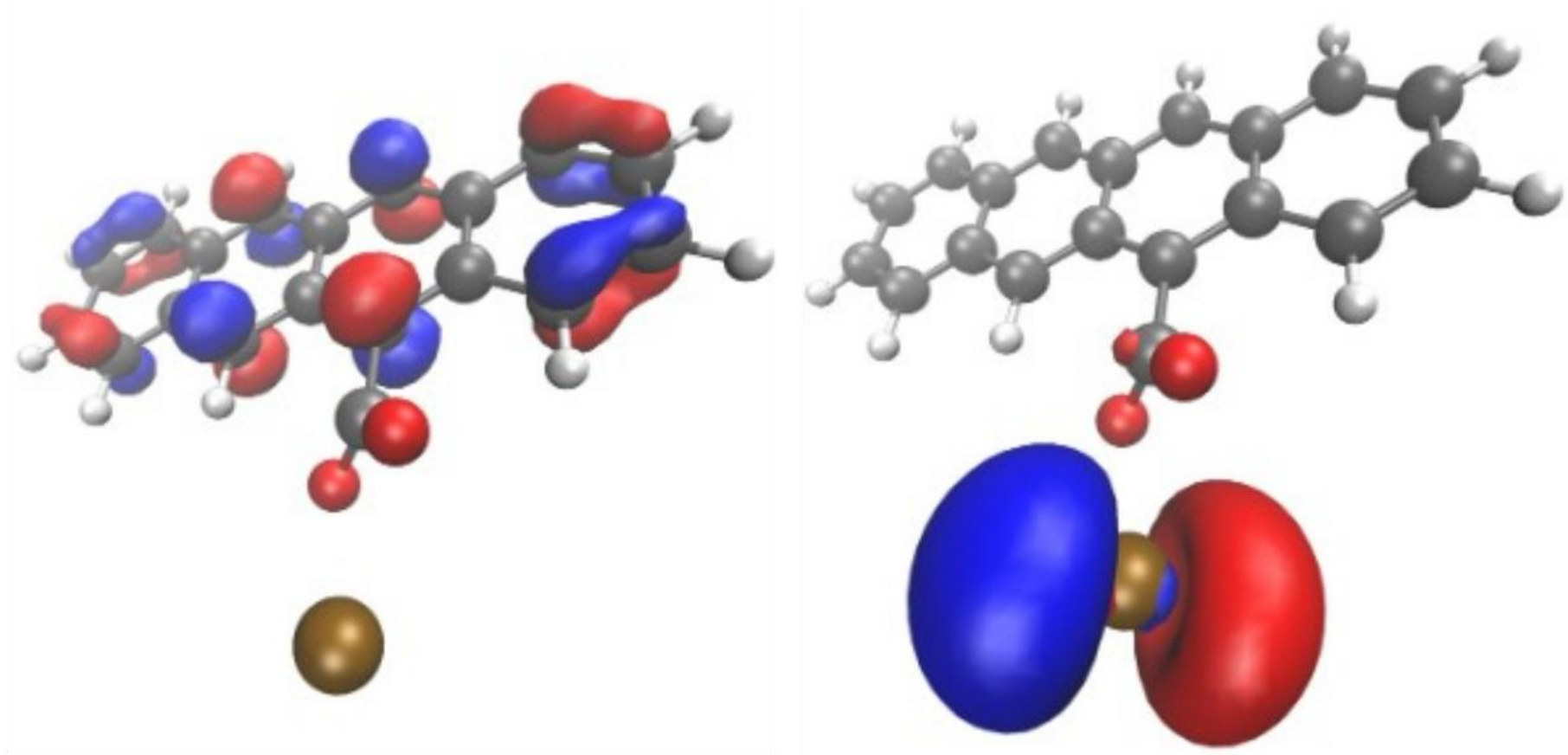

**Figure SI 5:** Singly occupied molecular orbitals (from $\omega$B97M-V/def2-TZVPPD) for the $T_1$ state of TCA anion complexed with $Pb^{2+}$ at the $T_1$ optimized geometry. One unpaired electron is localized on the acene ring while the other occupies a 6p orbital of Pb, demonstrating charge-transfer character of this electronic state.

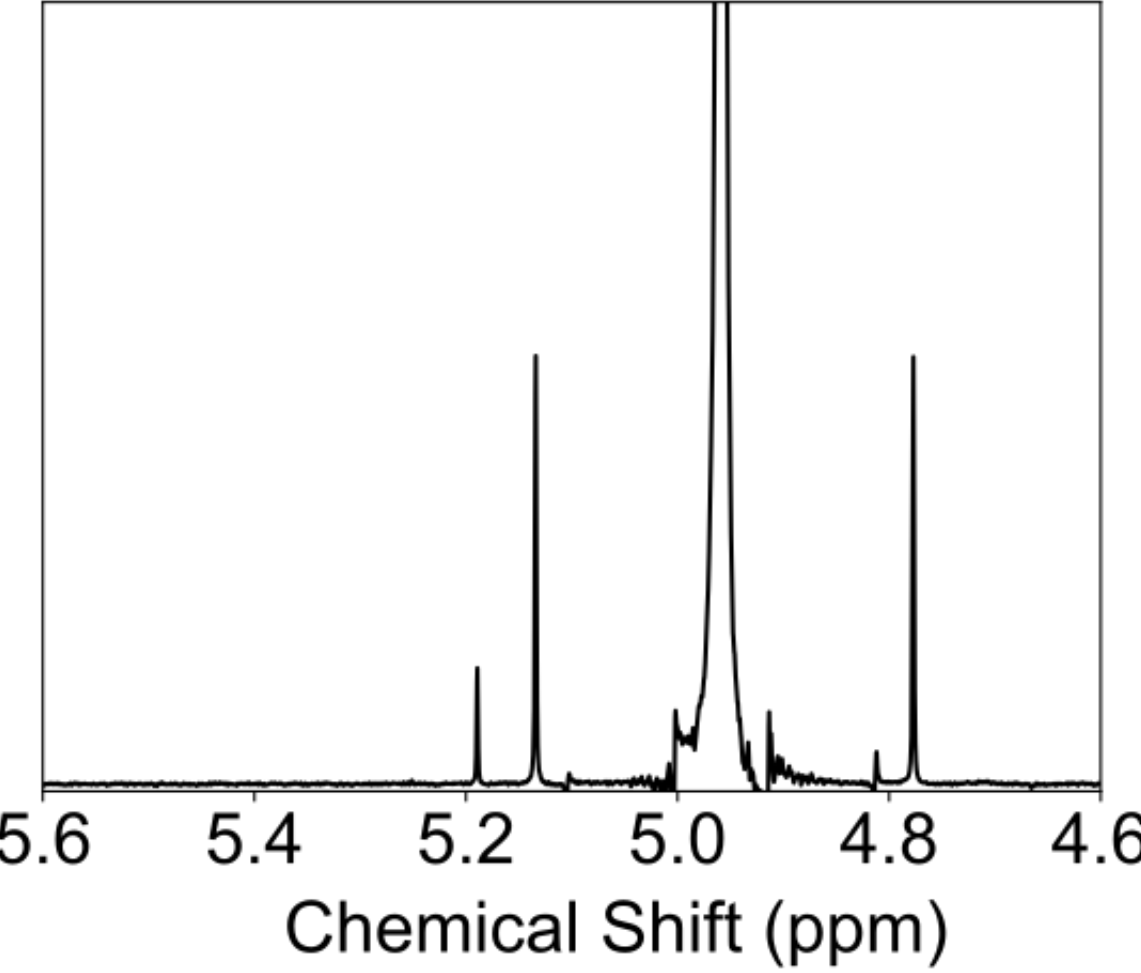

**Figure SI 6:** NMR spectrum of dibromomethane (DBM, $CH_2Br_2$, the standard used for quantification of ligand density on PbS QDs).

**SI Note 2:** The NMR spectra in Figure 2a and Figure SI 3, 4 show two unchanging peaks between 5.0 and 5.2 ppm. These peaks are confirmed to arise from the standard (DBM) used to perform quantitative NMR. The two symmetric peaks (4.75 ppm and 5.10 ppm) are C-13 satellite peaks. The signal at 5.2 ppm is an unidentified, unchanging peak present in the DBM standard.

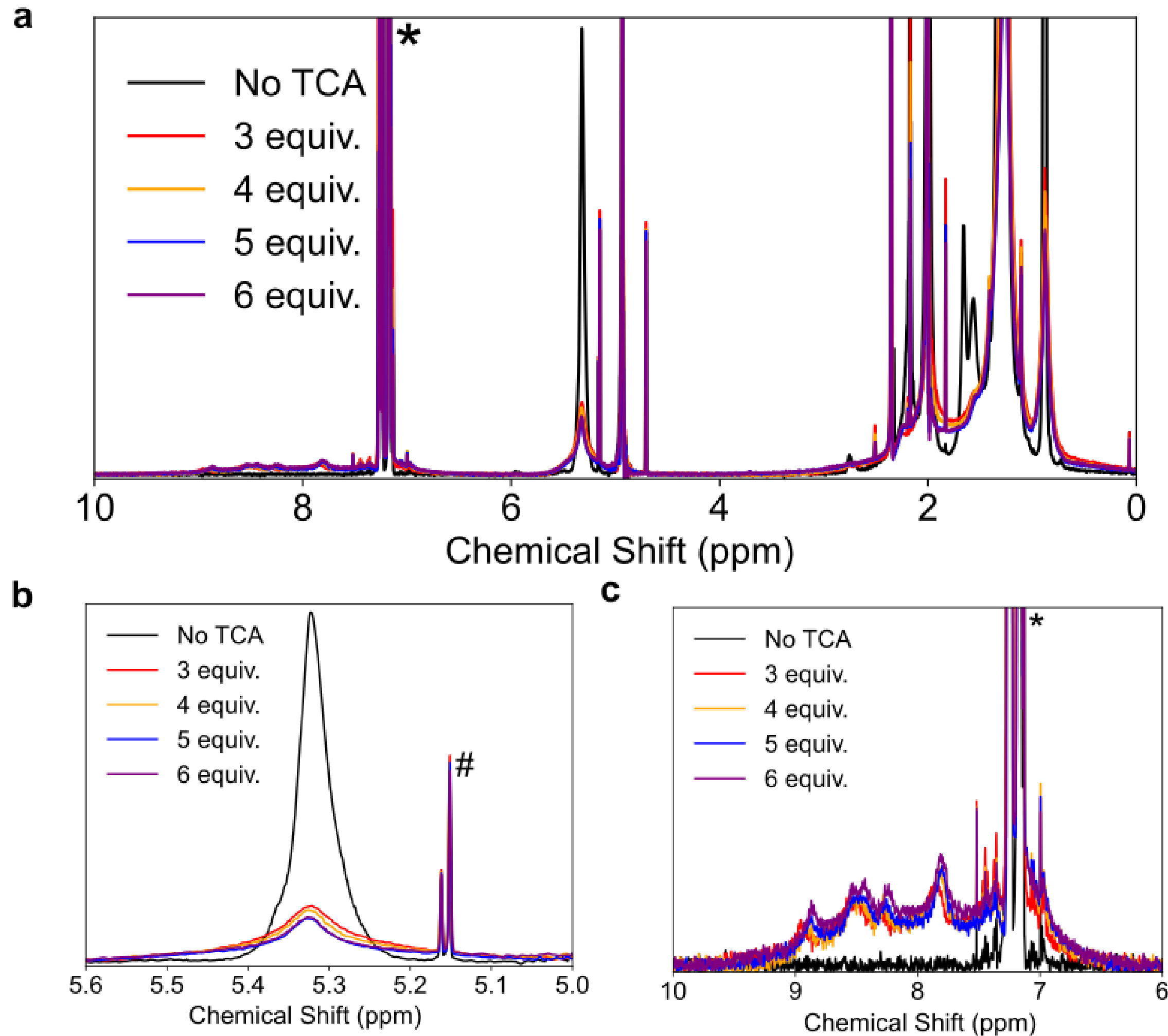

**Figure SI 7: (a)** Full NMR spectra of PbS QDs with varying TCA ligand exchange equivalents (vol/vol) normalized to the DBM peak at 4.94 ppm (* denotes residual toluene from sample preparation and non-deuterated chloroform solvent peaks). **(b)** Oleic acid peak reduces with increasing equivalents of TCA (# denotes impurity peaks in the DBM standard; see SI Note 2). **(c)** TCA peak increases with increasing equivalents of TCA (* denotes toluene and chloroform solvent peaks).

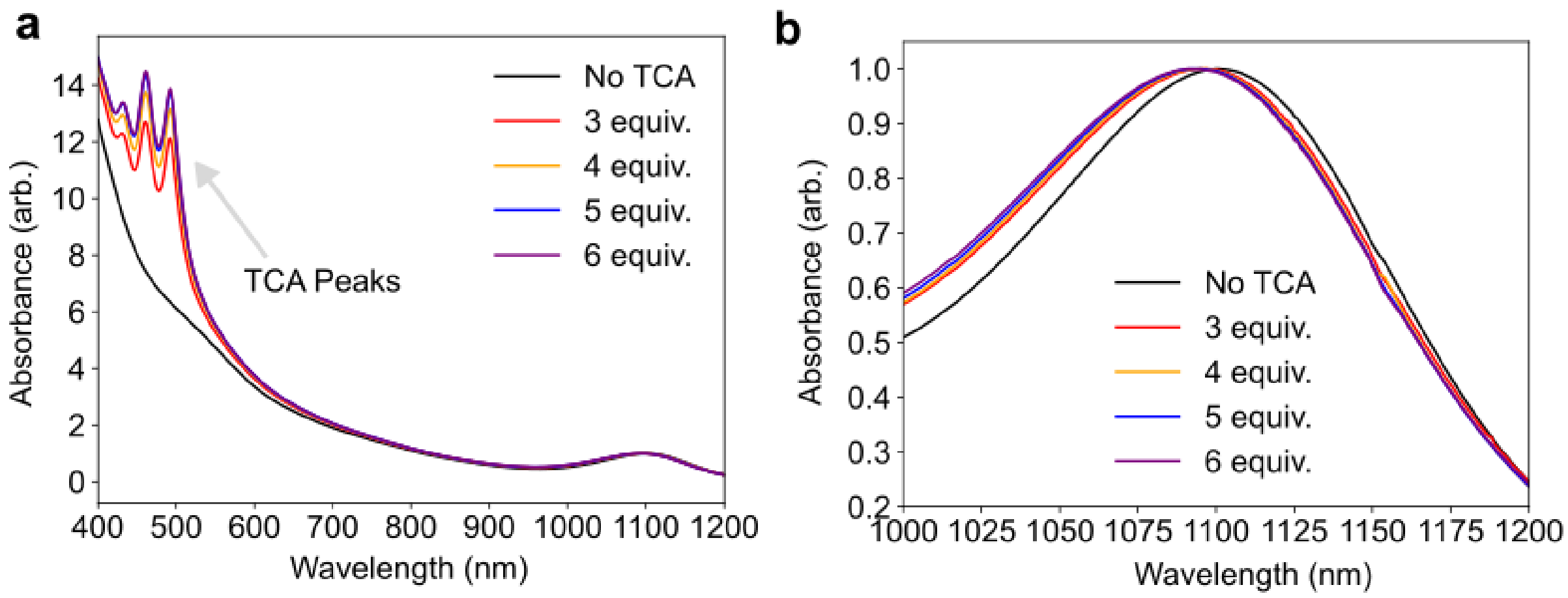

**Figure SI 8: (a)** Absorption spectra of PbS QDs with increasing TCA ligand exchange equivalents (vol/vol) show increasing TCA absorption intensity. **(b)** Increasing TCA density results in a mild blue-shift of the PbS QD first excitonic peak.

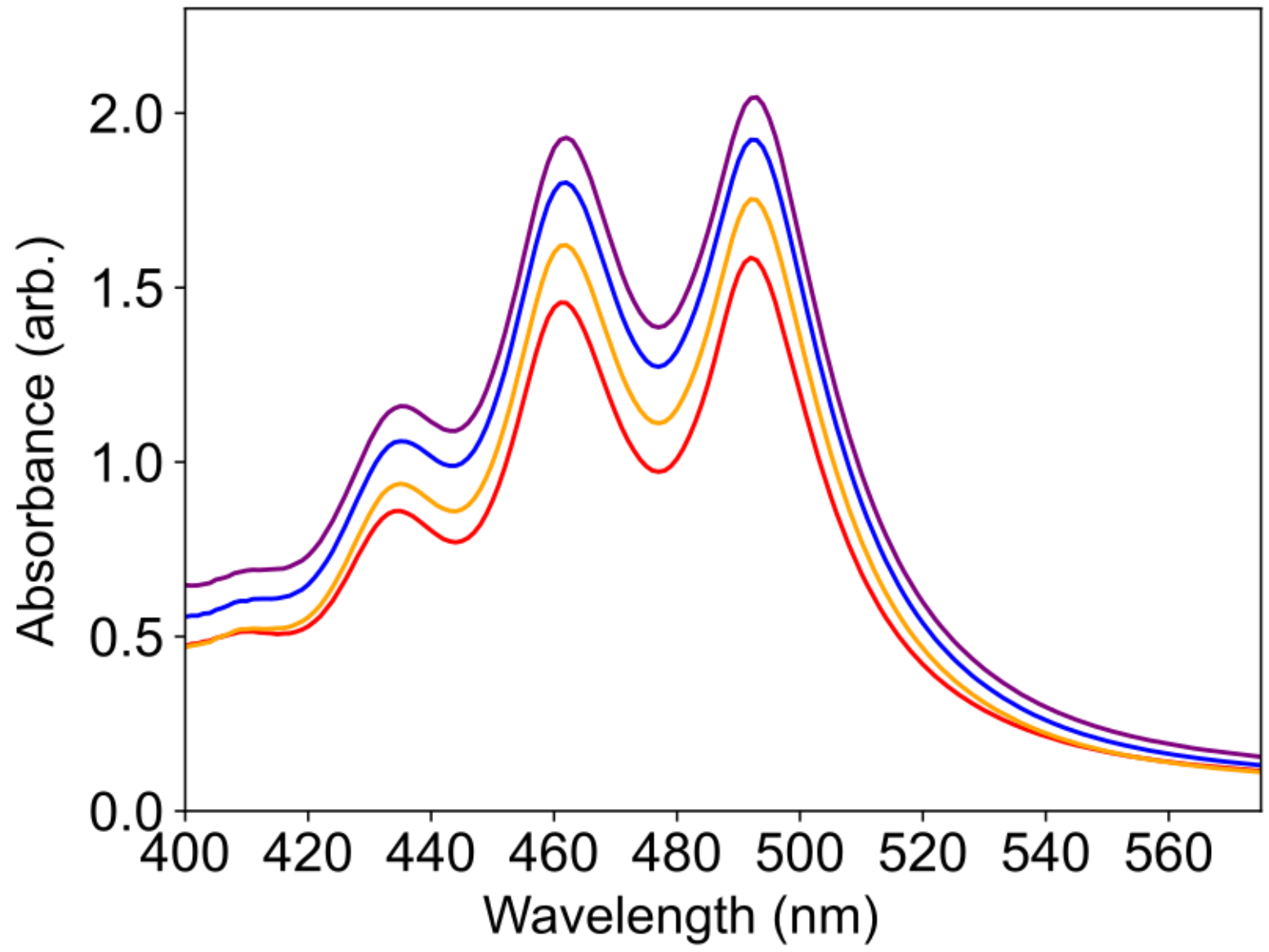

**Figure SI 9:** The scaled PbS QD absorption was subtracted from the absorption spectra for TCA-modified QDs to isolate the TCA peaks. An increase in TCA concentration with increasing ligand equivalents was observed.

| Sample | Oleic acid density (per $nm^2$) using NMR | TCA density (per $nm^2$) using NMR | TCA density (per $nm^2$) using UV-vis |
|---|---|---|---|
| PbS QDs (No TCA) | 5.79 | NA | NA |
| PbS QDs + 3 equiv. TCA | 4.11 | 0.99 | 3.99 |
| PbS QDs + 4 equiv. TCA | 4.08 | 1.00 | 4.44 |
| PbS QDs + 5 equiv. TCA | 3.25 | 1.14 | 4.74 |
| PbS QDs + 6 equiv. TCA | 3.32 | 1.26 | 4.88 |

**Table SI 3:** Calculated ligand densities of oleic acid and TCA using NMR and UV-vis absorption spectroscopy.

**SI Note 3:** The concentration and diameter of PbS QDs was estimated as reported by Moreels, et al.[21] The ligand densities of TCA and oleic acid (per QD and per $nm^2$) were calculated using NMR[22] or by UV-vis absorption spectroscopy.

PbS QDs (with or without TCA exchange) were dissolved in a $CDCl_3$ solution with known concentration of the DBM standard (0.144 M). By integrating the NMR peaks of TCA (7.30-9.50 ppm) and oleic acid (5.10-5.70 ppm) relative to the peaks of DBM (4.85-5.00 ppm), the concentrations of oleic acid and TCA were calculated (Table SI 3) using the following formula:

$$C_{ligand}\ (M) = \frac{I_{ligand}}{I_{DBM}} \times \frac{N_{DBM}}{N_{ligand}} \times C_{DBM} \qquad \text{Equation 6}$$

where I, N, and C are the NMR integral area, number of nuclei, and concentration of the species of interest. We note that $N_{DBM} = 2$, $N_{oleic\ acid} = 2$, and $N_{TCA} = 7$. Despite TCA having 11 protons, we only integrate regions corresponding to 7 protons as the signal from the remaining 4 protons overlaps with residual toluene and chloroform peaks. The resulting ligand concentrations were used to calculate ligand density using equation 7.

$$\text{Ligand density (nm}^{-2}\text{)} = \frac{\text{Ligand Concentration } C_{ligand} \text{ (M)}}{\text{Surface Area of QD (nm}^2\text{)} \times \text{PbS QDs Concentration (M)}} \qquad \text{Equation 7}$$

The ligand densities of TCA were also calculated purely based on absorption spectra. For absorption measurements, the NMR solution was diluted in chloroform. As shown in Figure SI 9, the pure PbS QD absorption can be subtracted from the spectra of TCA-modified QDs to isolate the signal from TCA alone. The resulting absorption signal was normalized by the measured extinction coefficient of TCA in chloroform at 480 nm (4.237 $mM^{-1}cm^{-1}$) to calculate ligand density (Equations 8 and 7).

$$C_{TCA}\ (M) = \frac{\text{Corrected absorbance at 480 nm (arb.)}}{\varepsilon_{480}\ (\text{mM}^{-1}\text{cm}^{-1}) \times 1\ \text{cm}} \qquad \text{Equation 8}$$

We observe larger ligand densities as estimated using UV-vis spectroscopy. We assign this discrepancy to (1) an inaccurate extinction coefficient for TCA due to changes in absorption spectral shapes for bound- versus free-TCA and (2) imperfect subtraction of PbS QD absorption from the absorption spectra of PbS-bound TCA.[23,24]

| Sample | Rel. Int. oleic acid (2H) | Rel. Int. TCA (7H) | PbS QD concentration (M) | TCA Ligand density (per QD) |
|---|---|---|---|---|
| PbS QDs (No TCA) | 0.162 | NA | 9.09E-05 | NA |
| PbS QDs + 3 equiv. TCA | 0.089 | 0.0748 | 7.06E-05 | 27.7 |
| PbS QDs + 4 equiv. TCA | 0.088 | 0.0753 | 7.03E-05 | 28.0 |
| PbS QDs + 5 equiv. TCA | 0.072 | 0.0881 | 7.22E-05 | 31.9 |
| PbS QDs + 6 equiv. TCA | 0.076 | 0.1011 | 7.46E-05 | 35.5 |

**Table SI 4:** NMR peak integrals for oleic acid and TCA (normalized to DBM peak integrals) across different ligand loadings were used to calculate the ligand density per QD.

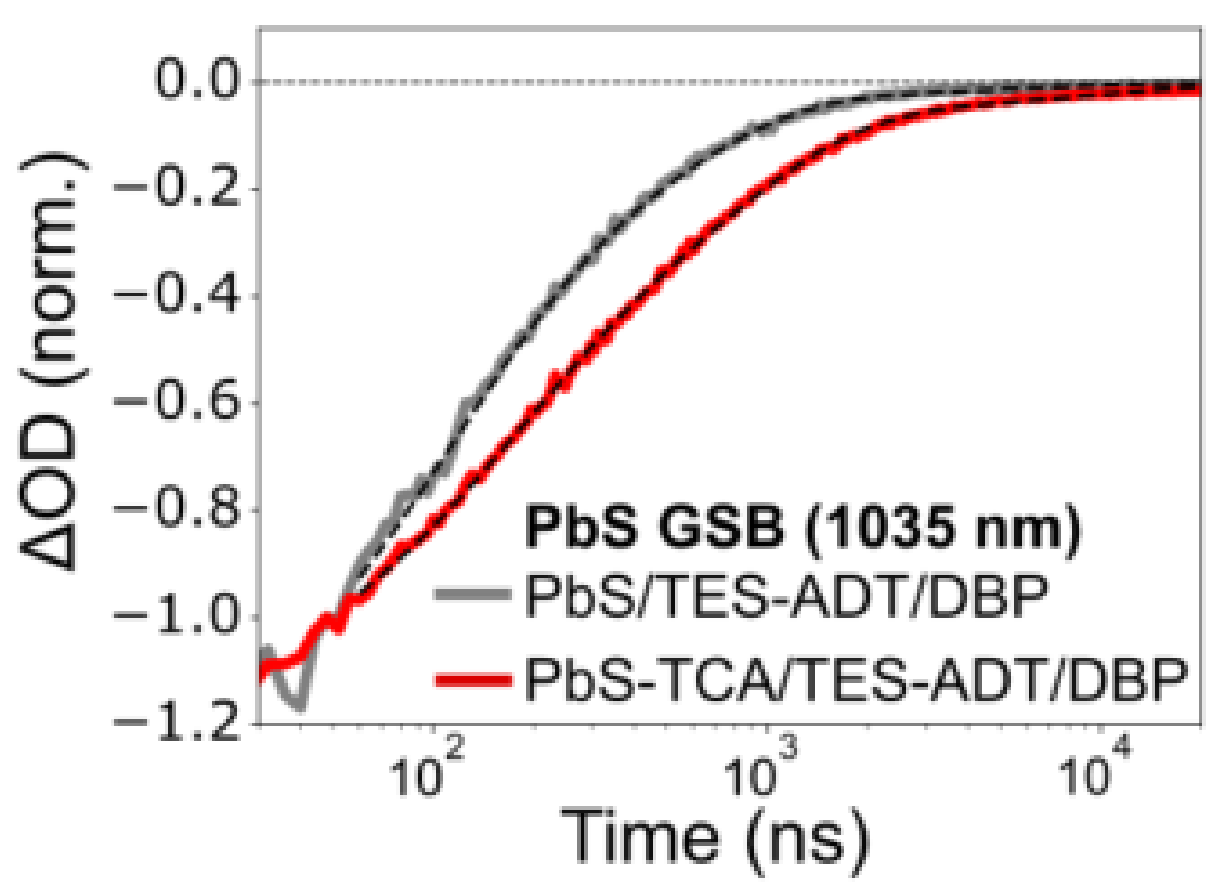

**Figure SI 10:** Transient absorption kinetics of PbS GSB (with and without TCA) in UC films measured at ~1035 nm. The increase in the lifetime of the PbS GSB upon the incorporation of TCA-modified PbS in UC films is assigned to parasitic re-absorption of UCPL by PbS QDs via FRET.

| Sample | $\tau_1$ / ns ($A_1$) | $\tau_2$ / ns ($A_2$) | $\tau_3$ / ns ($A_3$) | $\tau_4$ / ns ($A_4$) |
|---|---|---|---|---|
| **PbS** | 96 ± 4 (60%) | 445 ± 19 (37%) | 5440 ± 870 (3%) | - |
| **PbS-TCA** | 132 ± 4 (47%) | 709 ± 17 (47%) | 9710 ± 680 (6%) | - |

**Table SI 5:** PbS GSB (1035 nm) fit parameters for UC films. Fits were applied at >40 ns due to early-time detection artifact. All PbS exciton lifetime components are longer for the TCA-passivated QDs due to parasitic re-absorption.

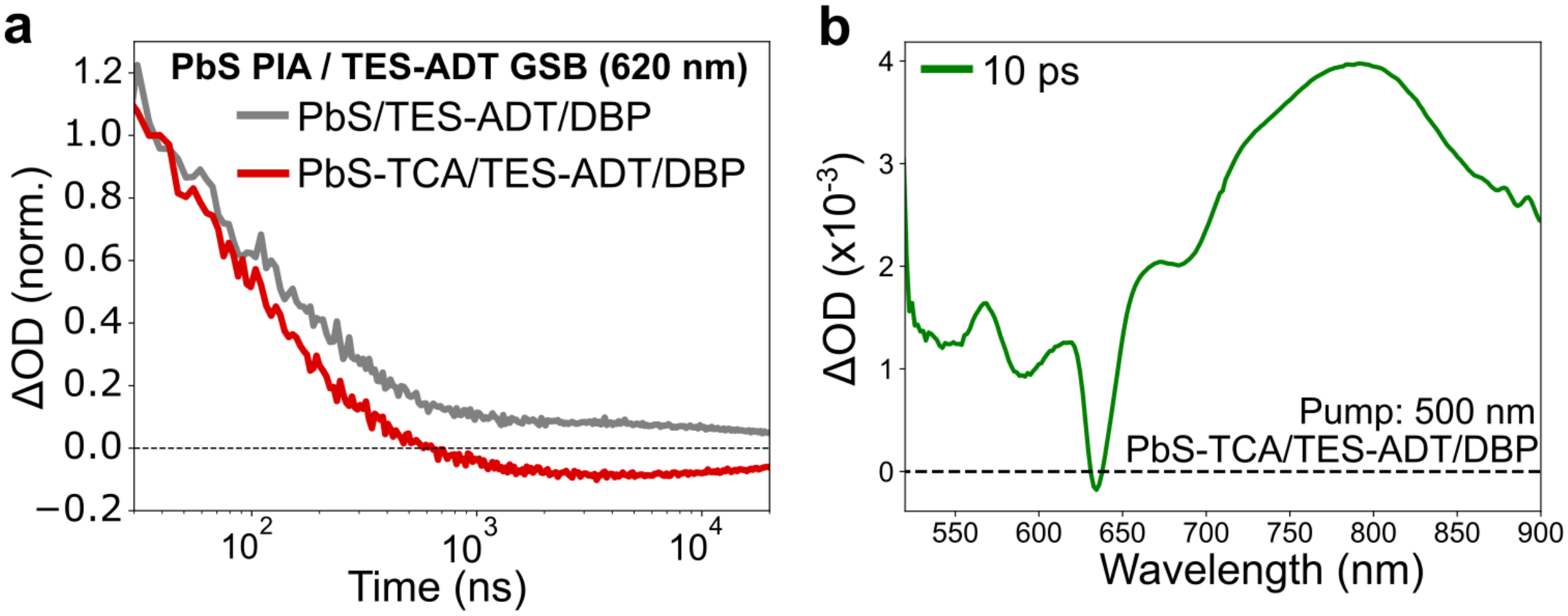

**Figure SI 11: (a)** Nanosecond transient absorption kinetics measured at 620 nm probe after 1064 nm excitation of PbS QDs in UC films (with and without TCA). ΔOD is initially positive due to photoinduced absorption (PIA) by excited PbS QDs. In the PbS-TCA system, triplet energy transfer to TES-ADT is efficient, leading to considerable build-up at later times of TES-ADT ground-state bleach (GSB, ΔOD < 0, as identified by fsTA spectrum). **(b)** Femtosecond transient absorption spectrum of PbS-TCA UC thin film 10 ps after excitation at 500 nm. At this excitation wavelength, both the TES-ADT and PbS QDs are excited and so this spectral window includes contributions from both the TES-ADT GSB and PbS PIA. On this timescale, the TES-ADT GSB is contributing most strongly at 600-630 nm, and used to inform the choice of probe wavelength for nanosecond transient absorption kinetics.

**SI Note 4:** The PL kinetics of PbS-TCA UC film after 1064 nm excitation, as presented in Figure 2e of the main text, are fitted with a model function of the form:

$$\mathrm{PL}(t) = A_{\mathrm{UC}}\left[\exp\left(-\frac{t}{\tau_{\mathrm{T}}}\right) - \exp\left(-\frac{t}{\tau_{\mathrm{TET}}}\right)\right]^2 + A_1 \exp\left(-\frac{t}{\tau_{\mathrm{F}}}\right) + A_2 \exp\left(-\frac{t}{\tau_{\mathrm{T}}}\right) \qquad \text{Equation 9}$$

where the first set of terms describes the UC kinetics as the squared difference of two exponentials.[25] Here, $\tau_{\mathrm{T}}$ corresponds to the TES-ADT triplet decay, and $\tau_{\mathrm{TET}}$ is the timescale of the ingrowing PL component due to triplet energy transfer. Due to the high peak light intensities achieved by the pulsed laser used to excite the sample, a small contribution of direct PL following multiphoton absorption is also observed, which creates an offset. This offset is modelled as a biexponential function with lifetimes corresponding to prompt ($\tau_{\mathrm{F}}$) and delayed fluorescence, where the latter follows the same triplet lifetime as for UC ($\tau_{\mathrm{T}}$). The fit commences at 1 ns, yielding extracted fit parameters of:

$\tau_{\mathrm{TET}}$ = 51 ± 1 ns, $\tau_{\mathrm{T}}$ = 1145 ± 5 ns, $\tau_{\mathrm{F}}$ < 1 ns;

$A_{\mathrm{UC}}$ = 15.2 arb. u., $A_1$ = 29.6 ± 0.9 arb. u., $A_2$ = 1.8 arb. u.

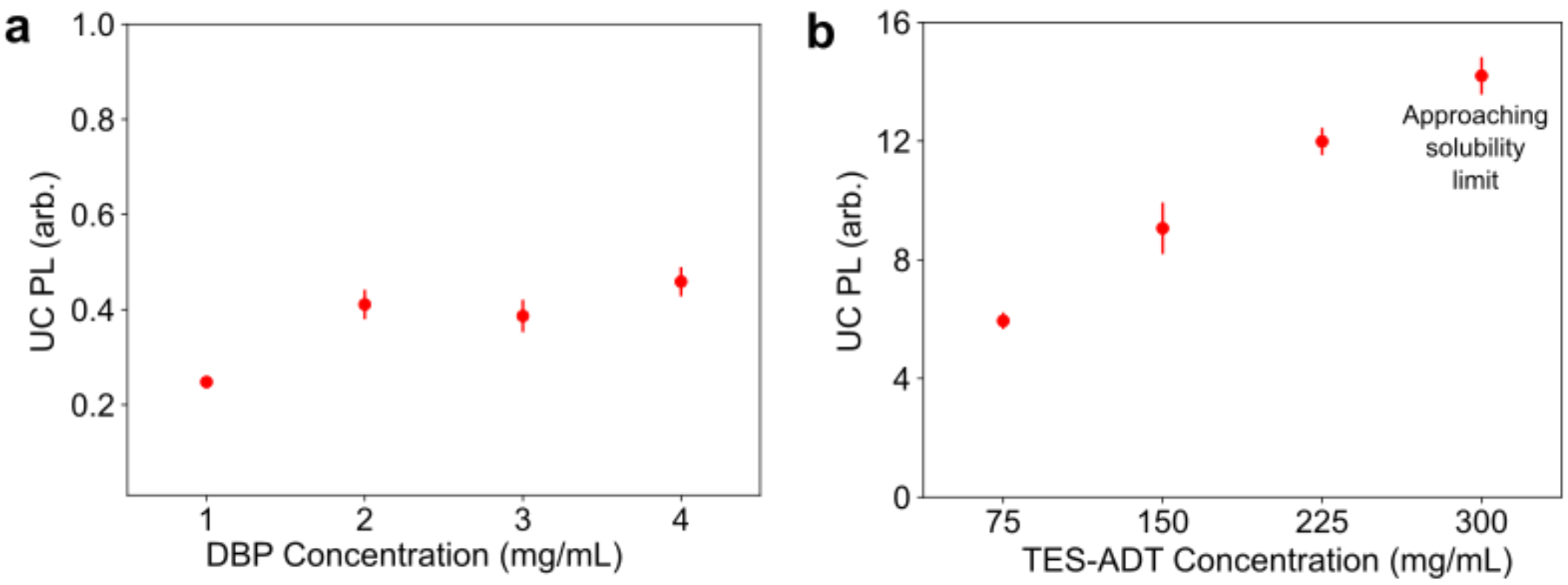

**Figure SI 12:** Optimization of **(a)** DBP, **(b)** TES-ADT concentrations in UC films based on integrated UCPL of the films under 808 nm laser excitation.

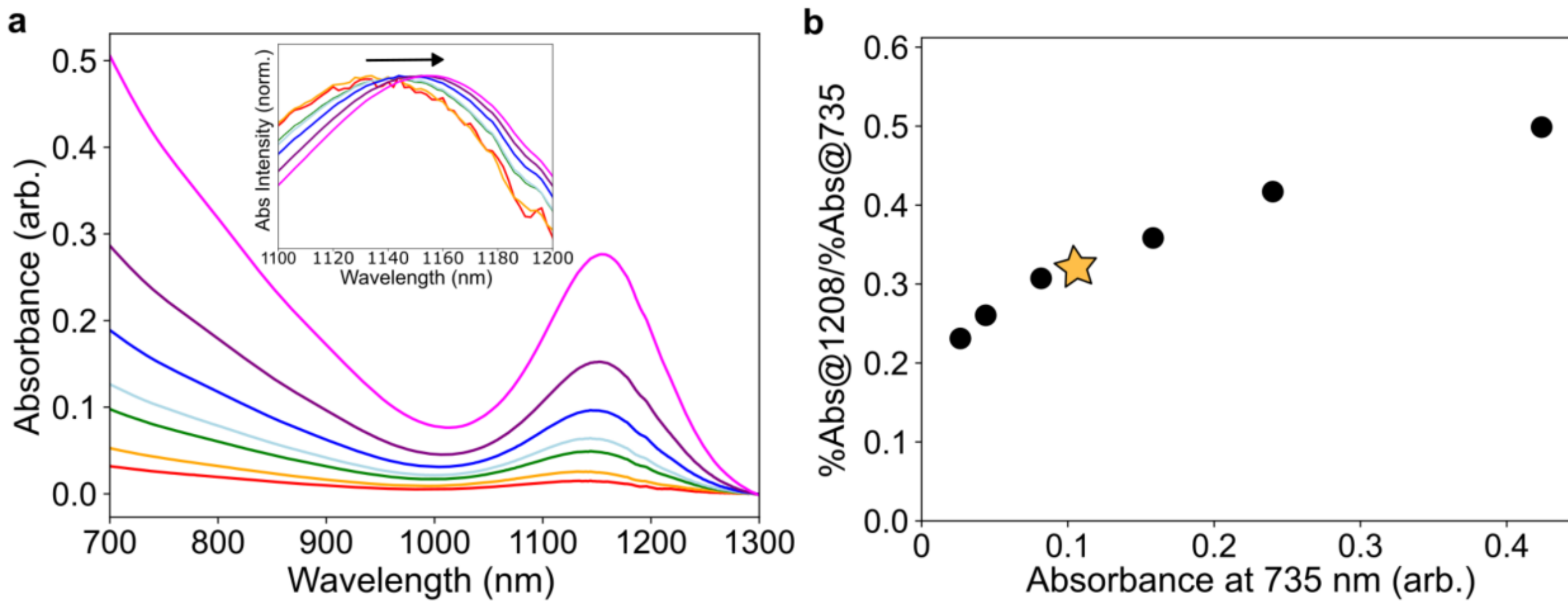

**Figure SI 13: (a)** Absorbance spectra of PbS QDs in hexanes with increasing concentration show a redshift in the first excitonic peak (highlighted through normalized spectra in the inset). **(b)** The ratio of percent absorption at 1208 nm to percent absorption at 735 nm for ~1150 nm PbS QDs is observed to increase with QD concentration in the solution. Ratios corresponding to ~0.1 absorbance at 735 nm were used in this work to convert film absorption at 735 nm to film absorption at 1208 nm (Main Text Equation 3).

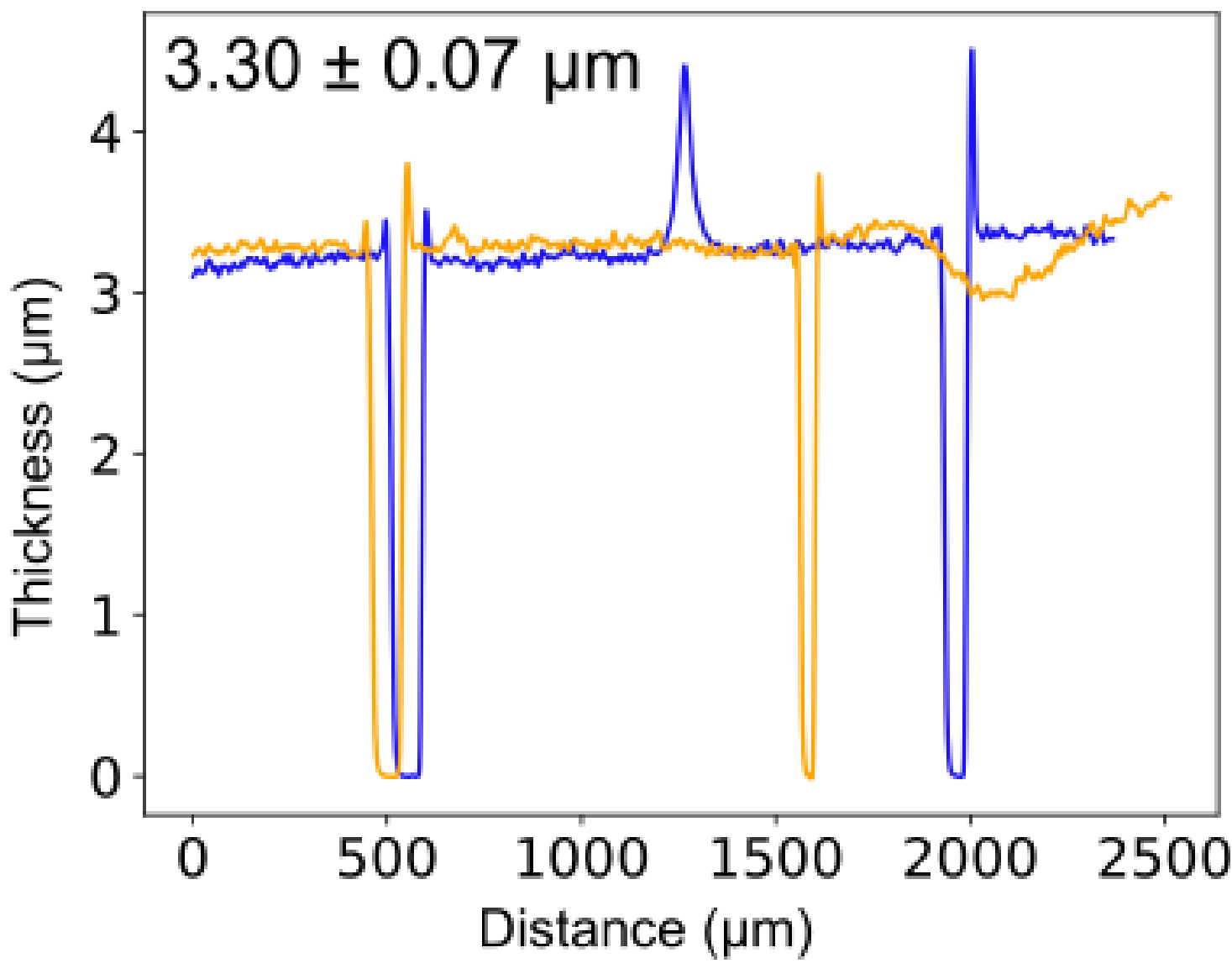

**Figure SI 14:** Thickness of UC films (300 mg/mL TES-ADT, 1 mg/mL DBP, 75 mg/mL PbS) were measured using profilometry across cuts on the film made using a razor blade. Yellow and blue traces and cuts across different parts of an UC film.

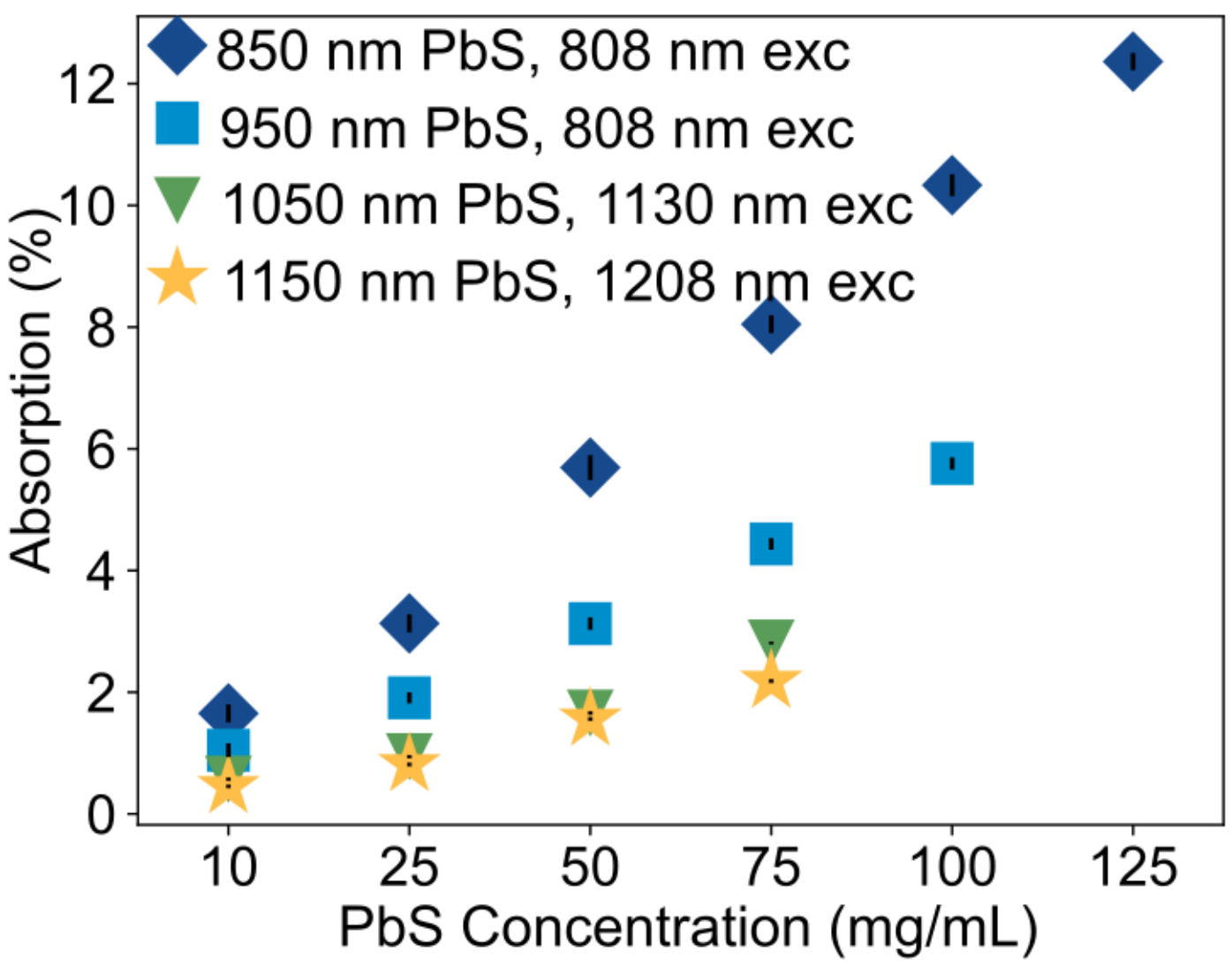

**Figure SI 15:** Dependence of laser absorption on PbS concentration used to make UC films.

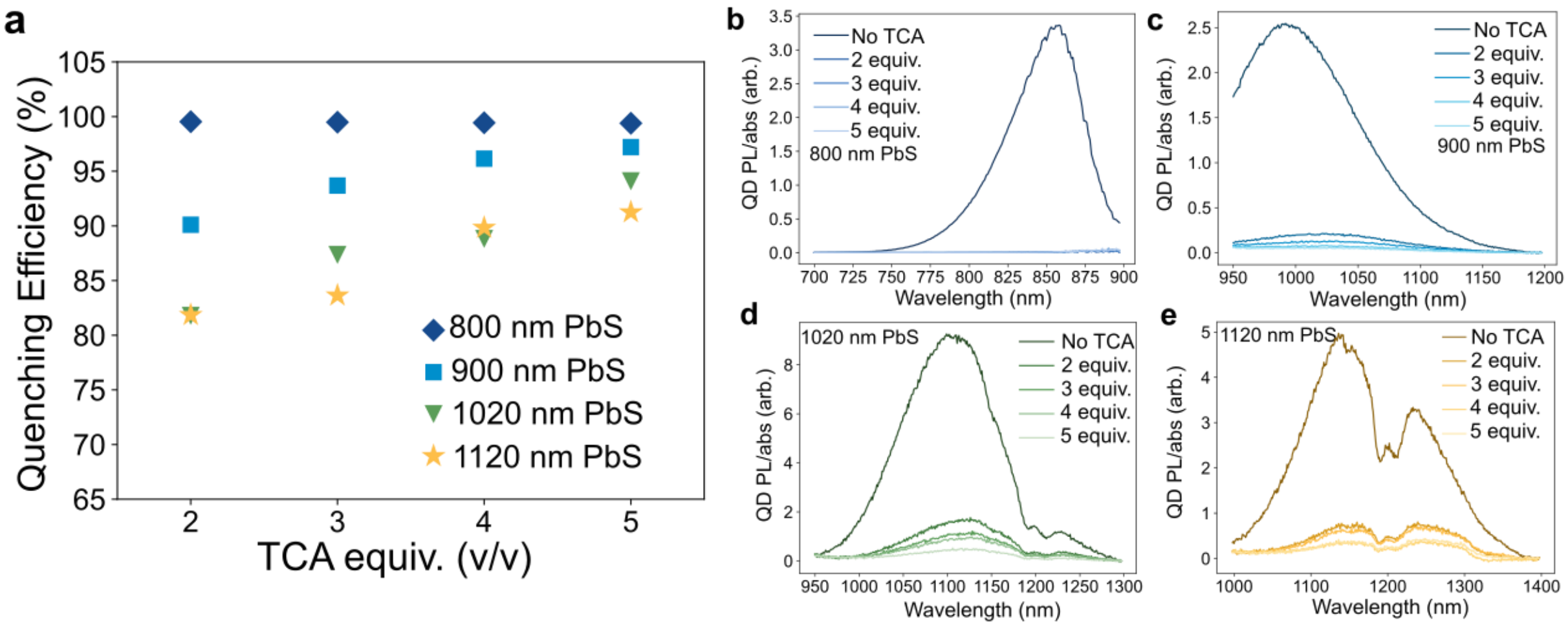

**Figure SI 16: (a)** Quenching efficiency (as calculated using Equation 11) for different PbS QD sizes and TCA loadings. **(b, c, d, e)** PbS QD PL (under 600 nm excitation) normalized to absorbance at 600 nm (as calculated using Equation 10). We note that the near infrared detector possesses an artifact around 1200 nm resulting in non-Gaussian emission profiles. Importantly, we also highlight that these measurements were performed with colloidal solutions of PbS QDs. Therefore, polarity-induced changes in energy levels of the QDs and the TCA may affect the accuracy of the measured quenching efficiencies.

$$PL\ normalized\ to\ absorption = PL_A = \frac{Integrated\ PL\ intensity}{Absorbance\ @\ 600\ nm} \qquad \text{Equation 10}$$

$$Quenching\ efficiency = \frac{PL_A_{No\ TCA} - PL_A_{x\ equiv.}}{PL_A_{No\ TCA}} \qquad \text{Equation 11}$$

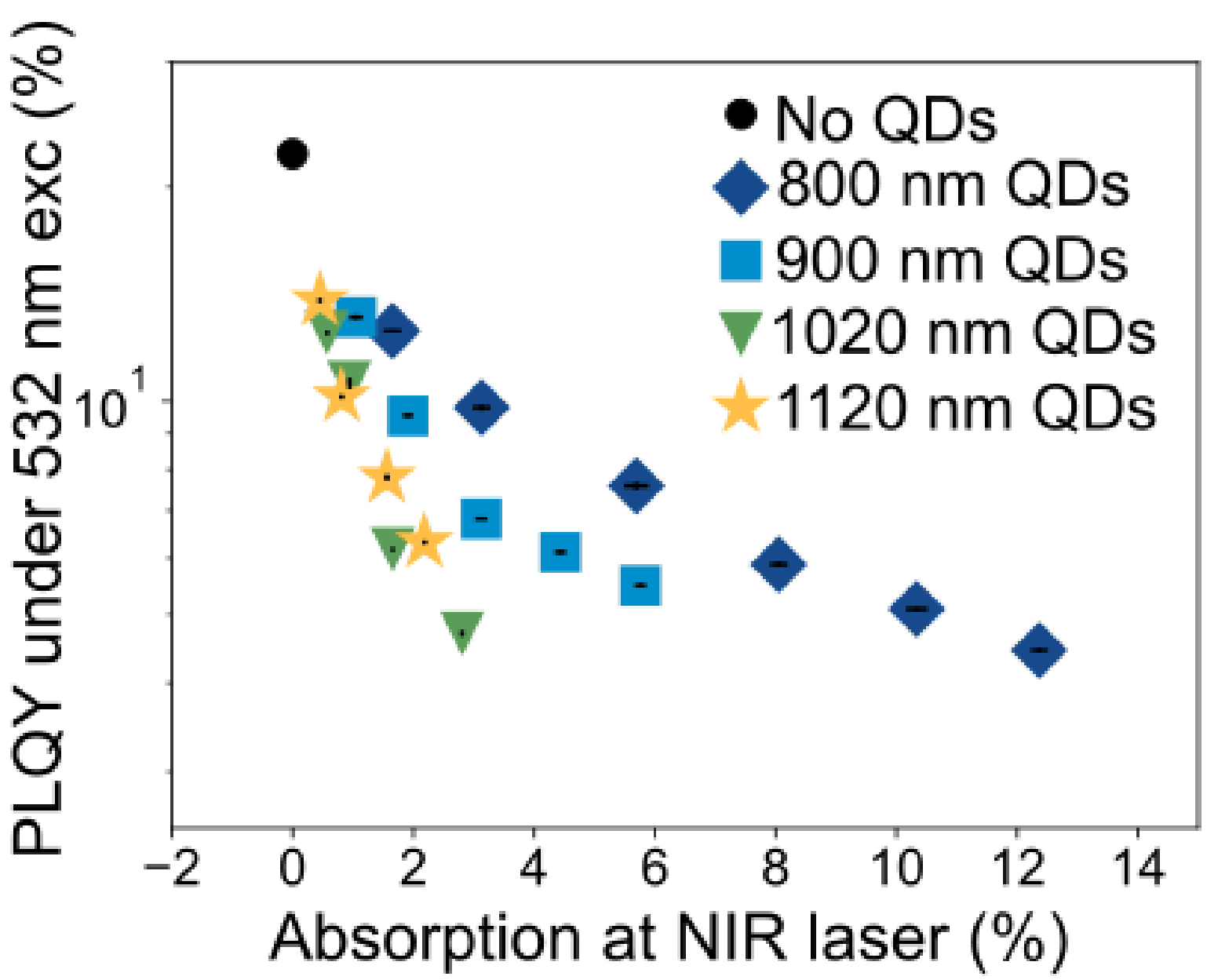

**Figure SI 17:** PLQY of the UC films shown in Figure 3a, 3b under 532 nm excitation. We note that PbS QDs in these films do possess 532 nm absorption, but the PbS QD contribution is estimated to be less than 10% of the total 532 nm absorption, leading to a minimal impact on the measured PLQY.

**SI Note 5:** The total UC quantum efficiency is given by:

$$IQE = \Phi_{\mathrm{UC}} = \phi_{\mathrm{ET}}\phi_{\mathrm{TET}}\phi_{\mathrm{TTA}}\phi_f(1-\phi_{\mathrm{BT}}) \qquad \text{Equation 12}$$

This accounts for the efficiencies of energy transfer (ET) from PbS to TCA; triplet exciton transfer (TET) from TCA to TES-ADT; triplet–triplet annihilation (TTA) in TES-ADT; photoluminescence of TES-ADT/DBP (F); and back transfer (BT) of UC emission, either via reabsorption or reverse-FRET. These are summarized schematically in Figure SI 18. Below, we quantify the efficiencies of individual steps to highlight the key shortcomings of the system.

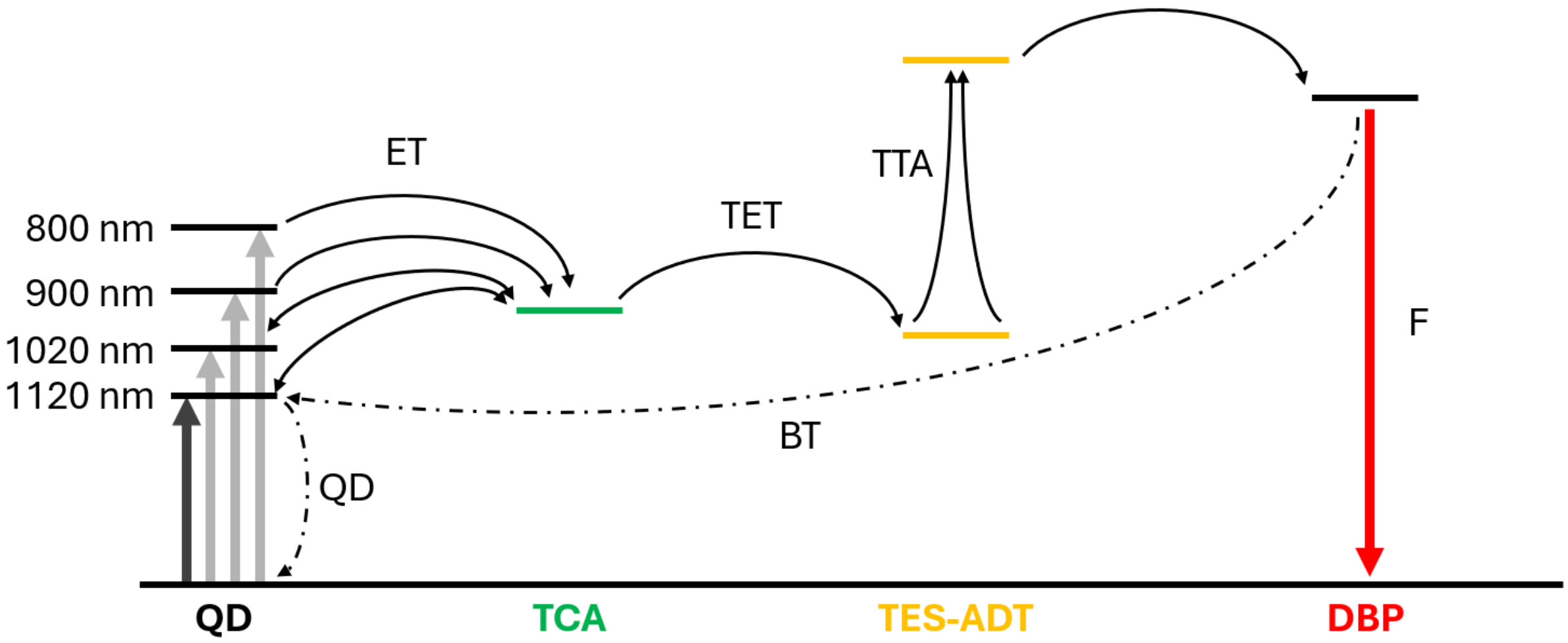

**Figure SI 18:** Photocycle for PbS QD-sensitized upconversion; terms correspond to dynamic steps outlined in Equation 12.

Experimentally, we find that the photoluminescence (PL) yield ($\phi_{\mathrm{PL}}$) of the TES-ADT/DBP composite is ~25%. As such, these devices would exhibit a maximum $\Phi_{\mathrm{UC}}$ of ~12.5%. The quantum yields observed are lower than this, implying less-than-unity yields of other processes.

As discussed above, we observe the ET from QDs to TCA drop using steady-state measurements of QD solutions with TCA (Figure SI 16). However, to accurately quantify ET in thin film, we extract a measure of the rate constant by integrating the normalized PL decay. For a single exponential,

$$A = \int_0^\infty \exp(-k_{QD}t)dt = \frac{1}{k_{QD}}$$

Where the intrinsic decay of the QD competes with energy transfer we add the ET rate constant and obtain

$$B = \int_0^\infty \exp(-((k_{ET} + k_{QD})t)dt = \frac{1}{k_{ET} + k_{QD}}$$

The ET yield can be expected to be

$$\phi_{\mathrm{ET}} = \frac{k_{ET}}{k_{ET} + k_{QD}} = 1 - \frac{B}{A}$$

Therefore, to gauge this quantity from the experimental traces we integrate the normalized PL decay (Figure SI 19) of 1130 nm QDs, both without and with the TCA ligands to obtain integrals $A$ and $B$, respectively. This procedure calculates a $\phi_{\mathrm{ET}}$ of 50.2% for 1130 nm QDs.

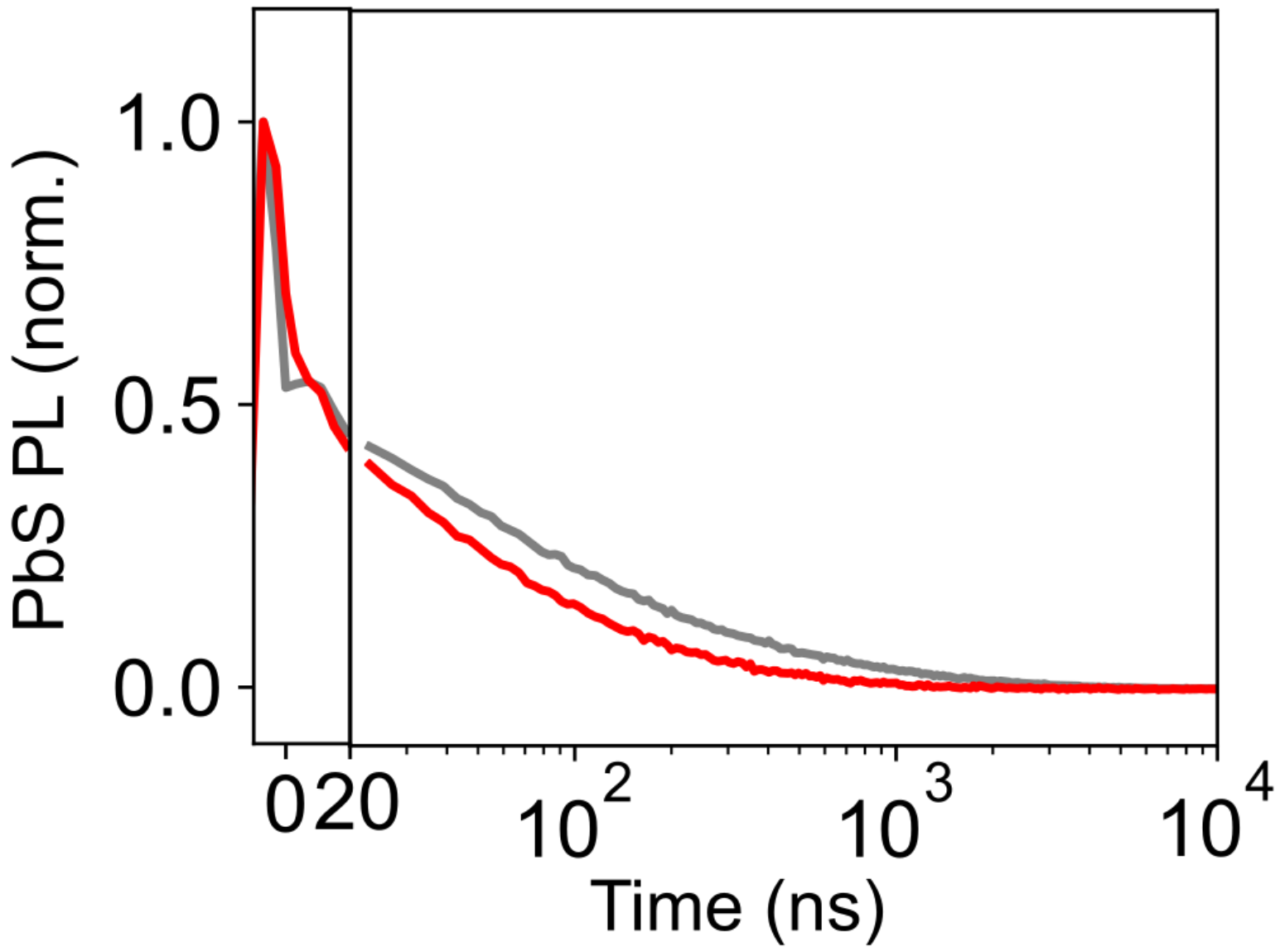

**Figure SI 19:** PbS QD PL kinetics for bare and TCA-treated QDs. The faster kinetics in PbS-TCA is due to energy transfer from QD to ligand.

Given the large bandgap of 800 nm (1.54 eV) QDs, the $\phi_{\mathrm{ET}}$ should approach unity. Furthermore, since in those QDs the reverse energy transfer from the ligand would be very endoergic, we anticipate that $\phi_{\mathrm{TET}}$ also approaches unity. Since these samples exhibited an IQE of 1.9%, we conclude that TTA proceeds with at least 15.2% yield, depending on the extent of BT of the final upconverted state (Table 6).

Evidence for less than 100% TTA yield is also shown in the observed ground-state bleach (GSB) lifetime of the TES-ADT (Figure SI 11a), which exceeds the observed UC lifetime by an order of magnitude. This implies that there is a subset of TES-ADT triplets which are unable to locate an annihilation partner.

The $\Phi_{\mathrm{UC}}$ differences between the different QD sizes must be due to a combination of $\phi_{\mathrm{ET}}$, $\phi_{\mathrm{TET}}$ and $\phi_{\mathrm{BT}}$. For the two films discussed in Table SI 6, we observe similar decrease in PLQY, suggesting similar $\phi_{\mathrm{BT}}$. Accordingly, since for the 1130 nm QDs, $\phi_{\mathrm{ET}}$ was determined to be 50.2%, and the overall yield $\Phi_{\mathrm{UC}}$ was measured to be 0.079%, we determine that the effective $\phi_{\mathrm{TET}}$ is 8.3% due to competition between reverse energy transfer from TCA ligands to the QD, and TET to the TES-ADT.

The proposed yields of each step in the photocycle are given in Table SI 6.

| QD abs. peak (nm) | $\phi_{ET}$ (%) | $\phi_{TET}$ (%) | $\phi_{TTA}(1-\phi_{BT})$ (%) | $\phi_{PL}$ (%) | $\Phi_{UC}$ (%) |
|---|---|---|---|---|---|
| 800 | 100 | 100 | 15.2 | 25 | 1.9 |
| 1130 | 50.2 | 8.3 | 15.2 | 25 | 0.079 |

**Table SI 6:** Breakdown of individual efficiency terms in the UC process for different QD sizes.

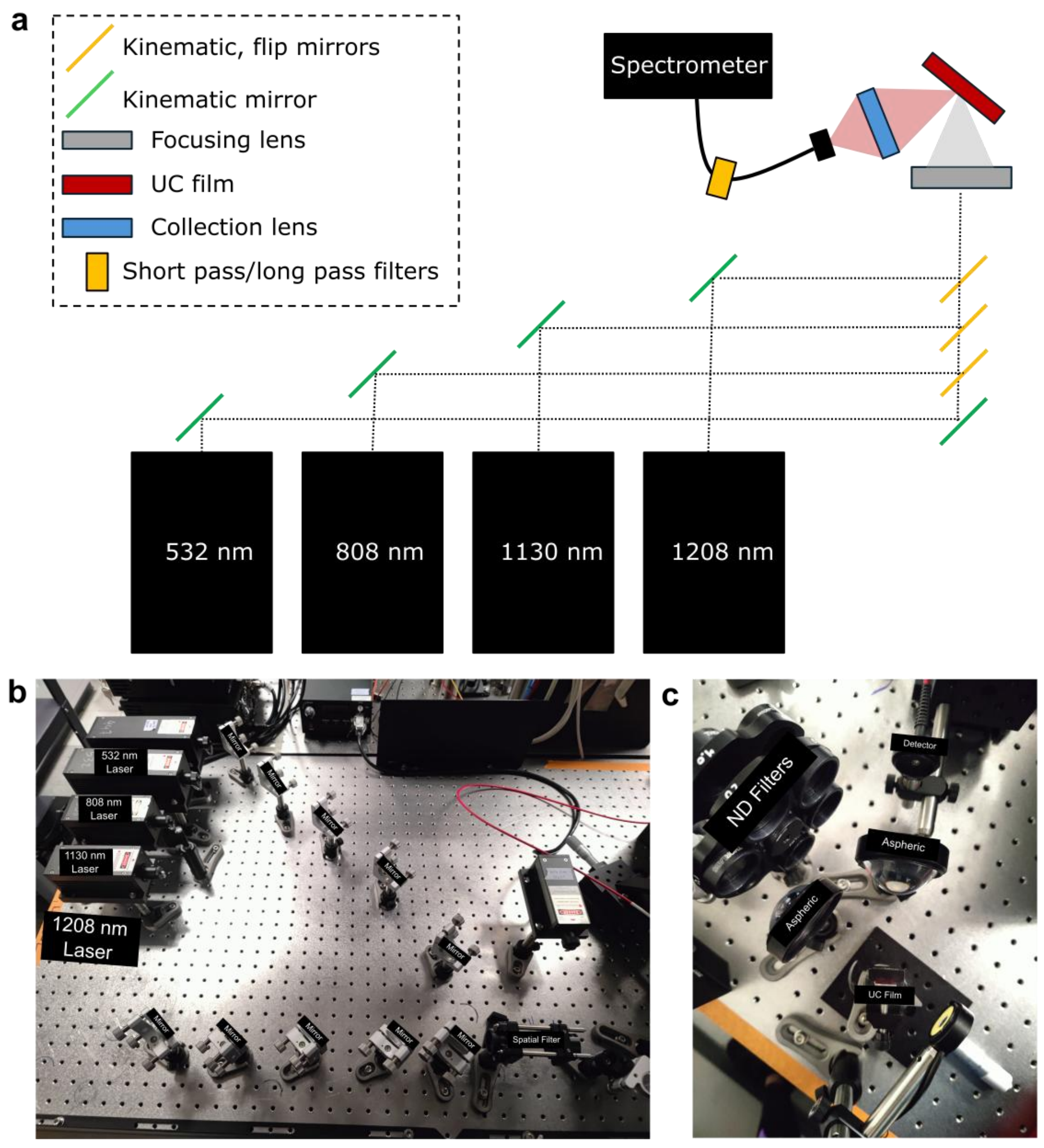

**Figure SI 20:** (a) Schematic and (b, c) pictures of the PL and UCPL setup.

| **PbS QD size used in UC film (nm)** | **$\lambda_{exc}$ (nm)** | **Threshold Intensity ($W/cm^2$)** |
|---|---|---|
| 800 | 808 | 0.38 |
| 900 | 808 | 0.41 |
| 1020 | 808 | 1.6 |
| 1120 | 808 | 14 |

**Table SI 7:** Table of measured threshold intensities for UC devices.

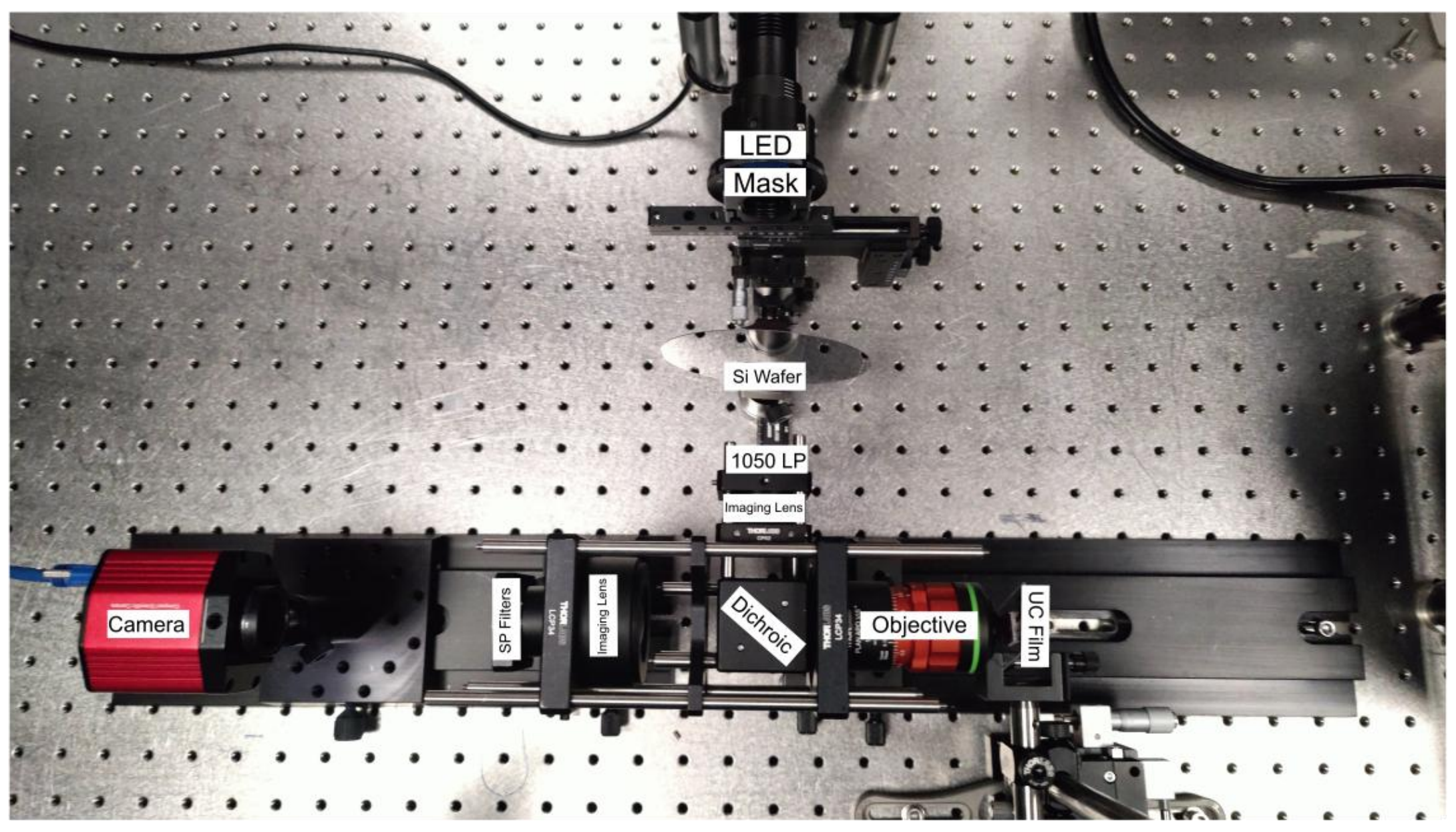

**Figure SI 21:** Picture of the setup used to image 1200 nm light with components labelled.

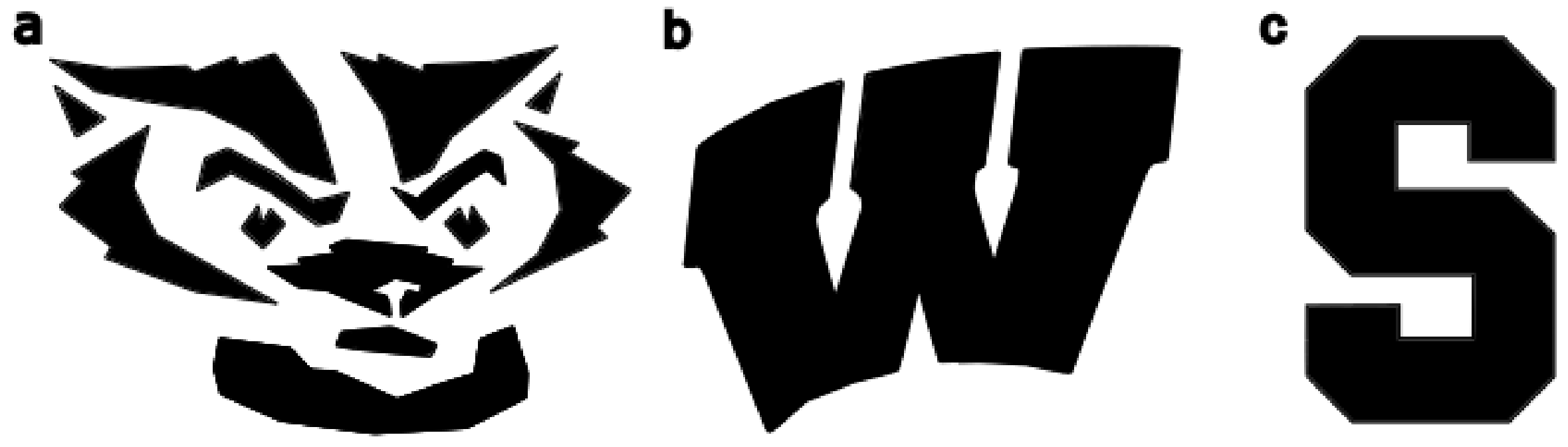

**Figure SI 22: (a)** Bucky Badger, **(b)** Wisconsin logo, and **(c)** Stanford logo designs used to fabricate the target masks which were imaged using 1200 nm light

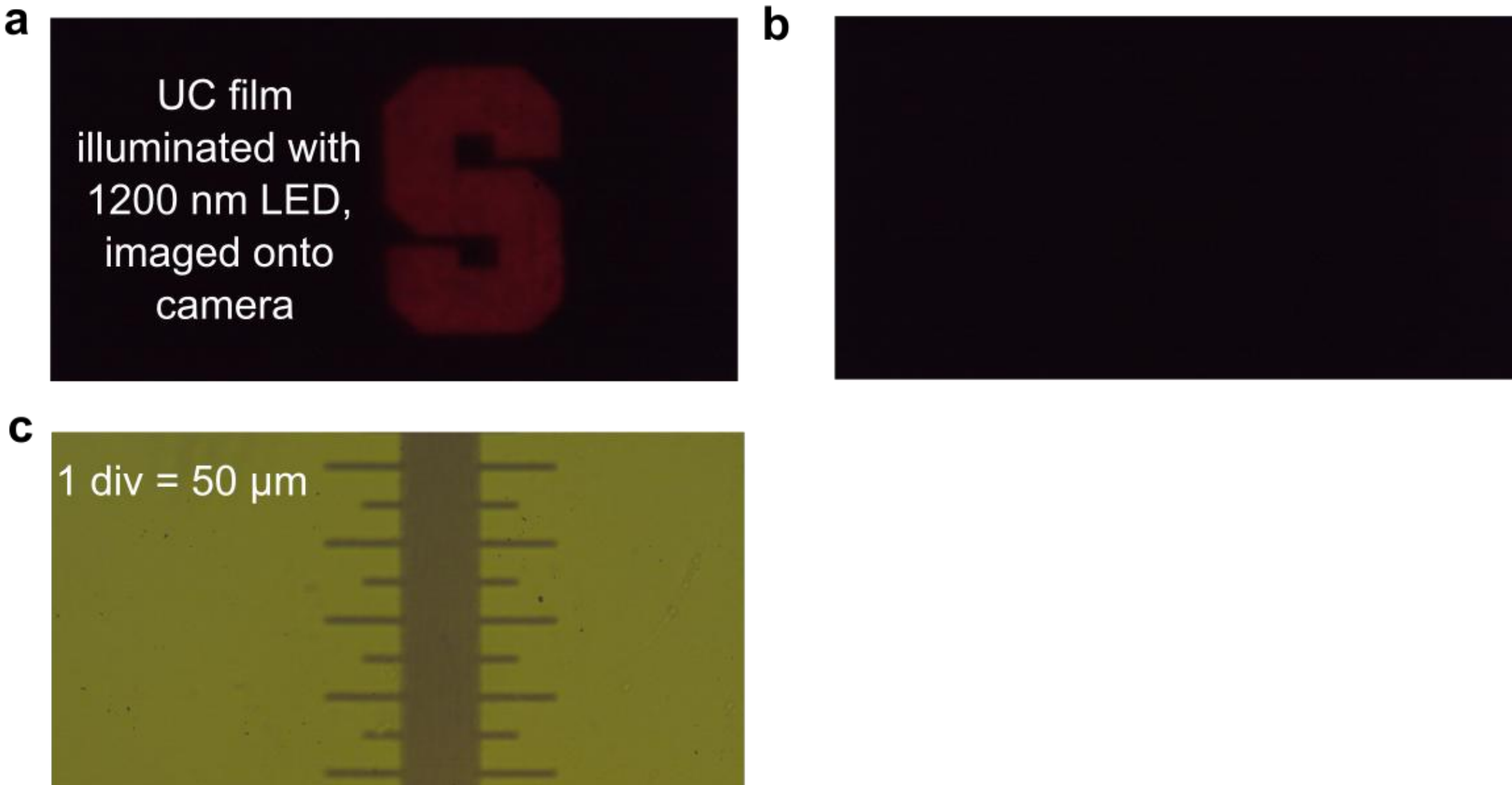

**Figure SI 23: (a)** Image of the upconverted Stanford logo taken using the imaging set up. **(b)** Image of the Stanford logo when the upconverter is replaced with a mirror. This demonstrates that NIR light is negligible compared to our upconverted light. Images in (a) and (b) were taken with gain of 15 dB and exposure of 2 seconds. **(c)** Image of the calibration slide used to assign scale bars to all images collected using the imaging setup. The upconverter was replaced with a backlit micrometer slide for this image.

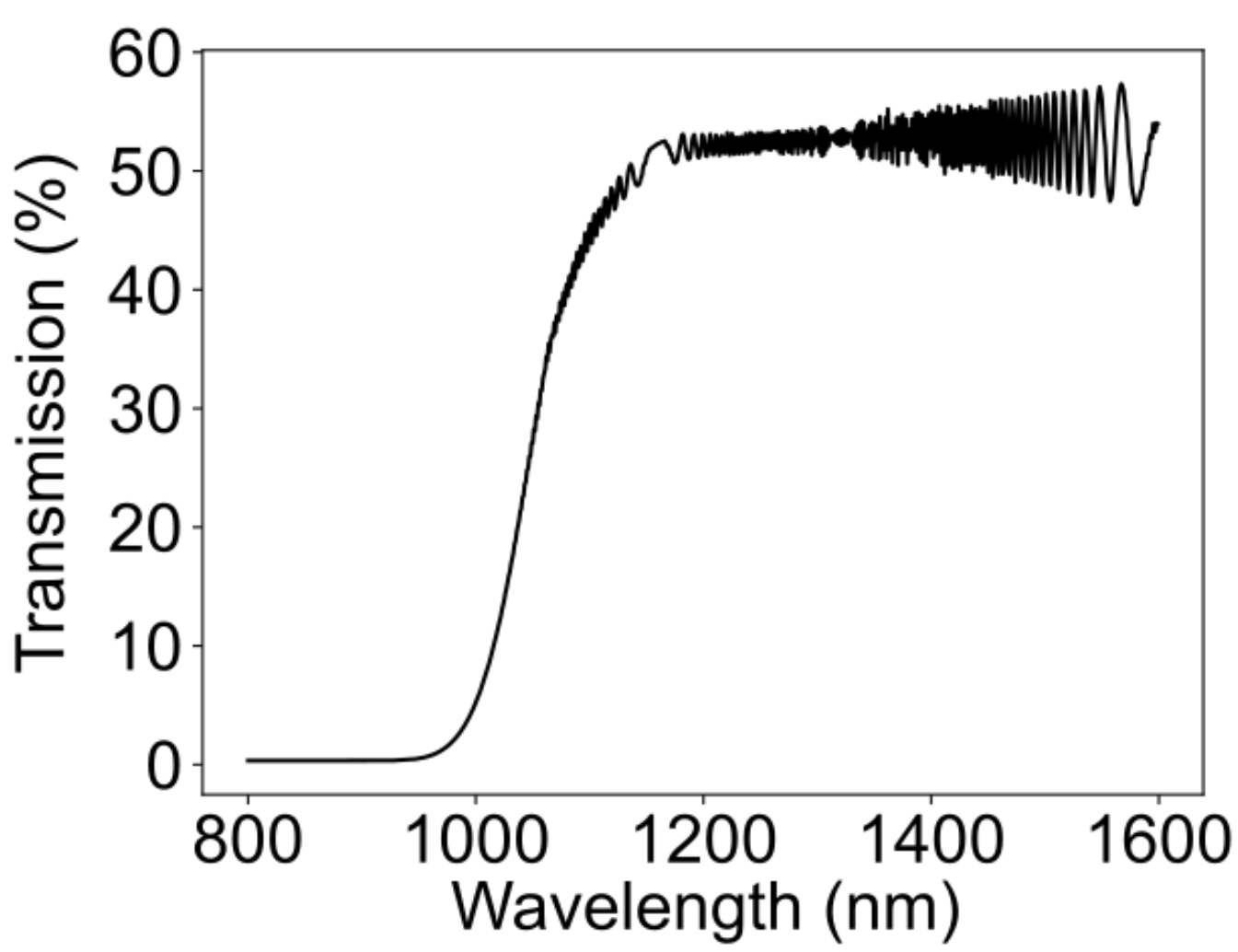

**Figure SI 24:** Transmission spectrum of the double-polished 350-micron silicon wafer used in Figure 4 to demonstrate imaging of sub-silicon bandgap photons.

**SI Note 6:** Transmission of 1200 nm light is limited to ~52% due to reflections at the two air-silicon interfaces. This results in a dimmer image (Stanford logo) in Figure 4f when the silicon wafer is introduced in front of the LED.

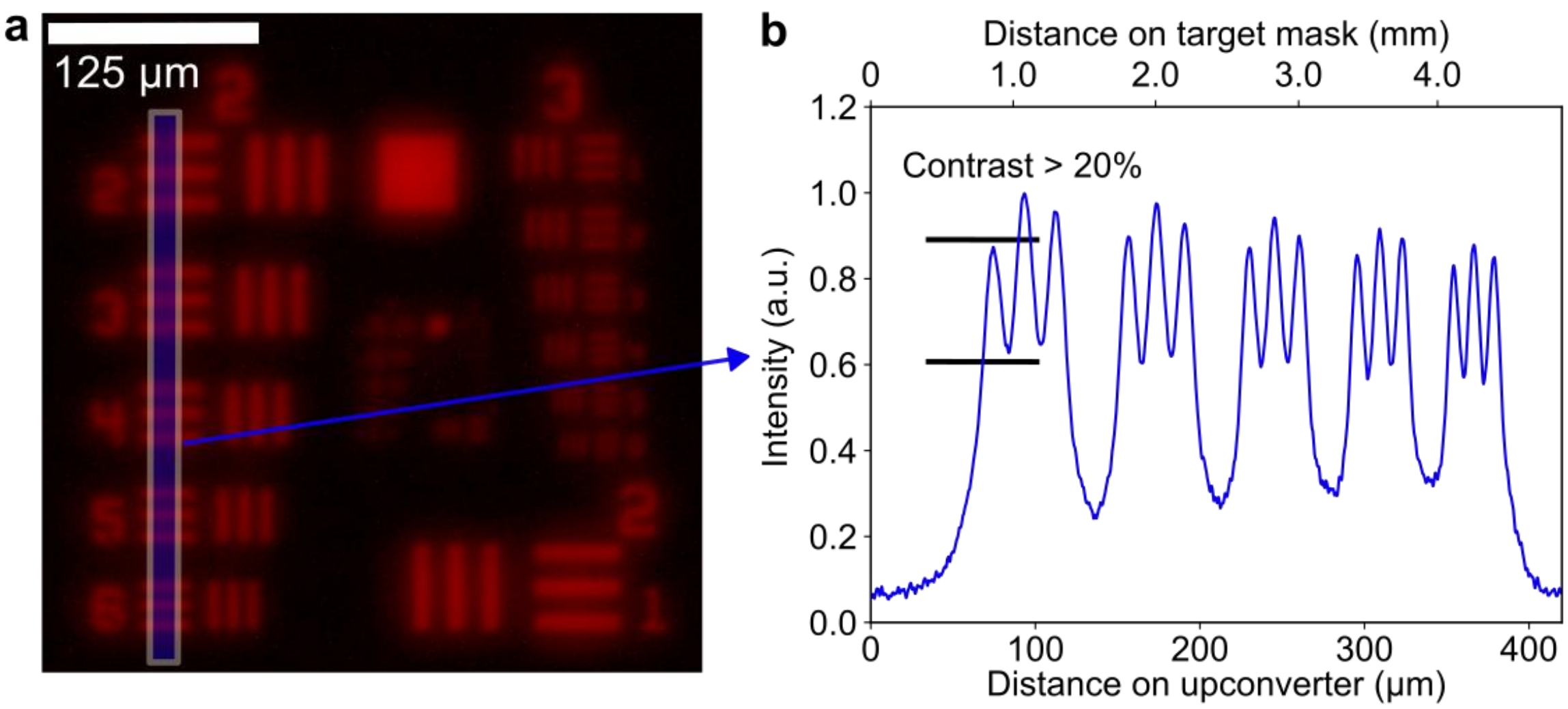

**Figure SI 25: (a)** Upconverted image from the projected NIR US Air Force (USAF) target taken using the imaging system in Figure 4a. **(b)** Plots across the horizontal line pairs in group 2, elements 2, 3, 4, 5, 6 were used to calculate the contrast of the image. The optical setup condenses macroscopic NIR features, and observes respective upconverted features on the intermediate image plane of at least 80 lp/mm. Further investigation is needed to determine the limiting spatial resolution of the upconversion process.

| Mask | Power at UC Film (mW) |
|---|---|
| Stanford S logo | 0.25 |
| Wisconsin W logo | 0.21 |
| Winsconsin Bucky Badger | 0.10 |
| US Air Force (USAF) target | 0.13 |

**Table SI 8:** Power of 1200 nm light measured at the UC film with different masks.

**SI Note 7:** The area of the Stanford logo at the UC film was measured to be 6.54E-04 $cm^2$ by replacing the UC film with a camera. Using this, the intensity of 1200 nm LED on the UC film was calculated to be 378 $mW/cm^2$.

**SI Note 8:** The intensity of the 1200 nm LED light at the mask was estimated by using an 800 µm (Thorlabs, P800K) pinhole in place of the imaging mask. We measure the power transmitted through the pinhole using a thermal power sensor (Thorlabs, S401C) and divide by the area of the pinhole aperture to estimate the intensity at the mask, which comes out to be ~20 $mW/cm^2$. We estimate that the intensity at the aperture of the first imaging lens (L1, Figure 4a) is likely to be lower than this intensity due to the divergence of the LED light. Additionally, we note that this measurement did not include the 1050 nm longpass filter shown in Figure SI 21, which would slightly lower the intensity.

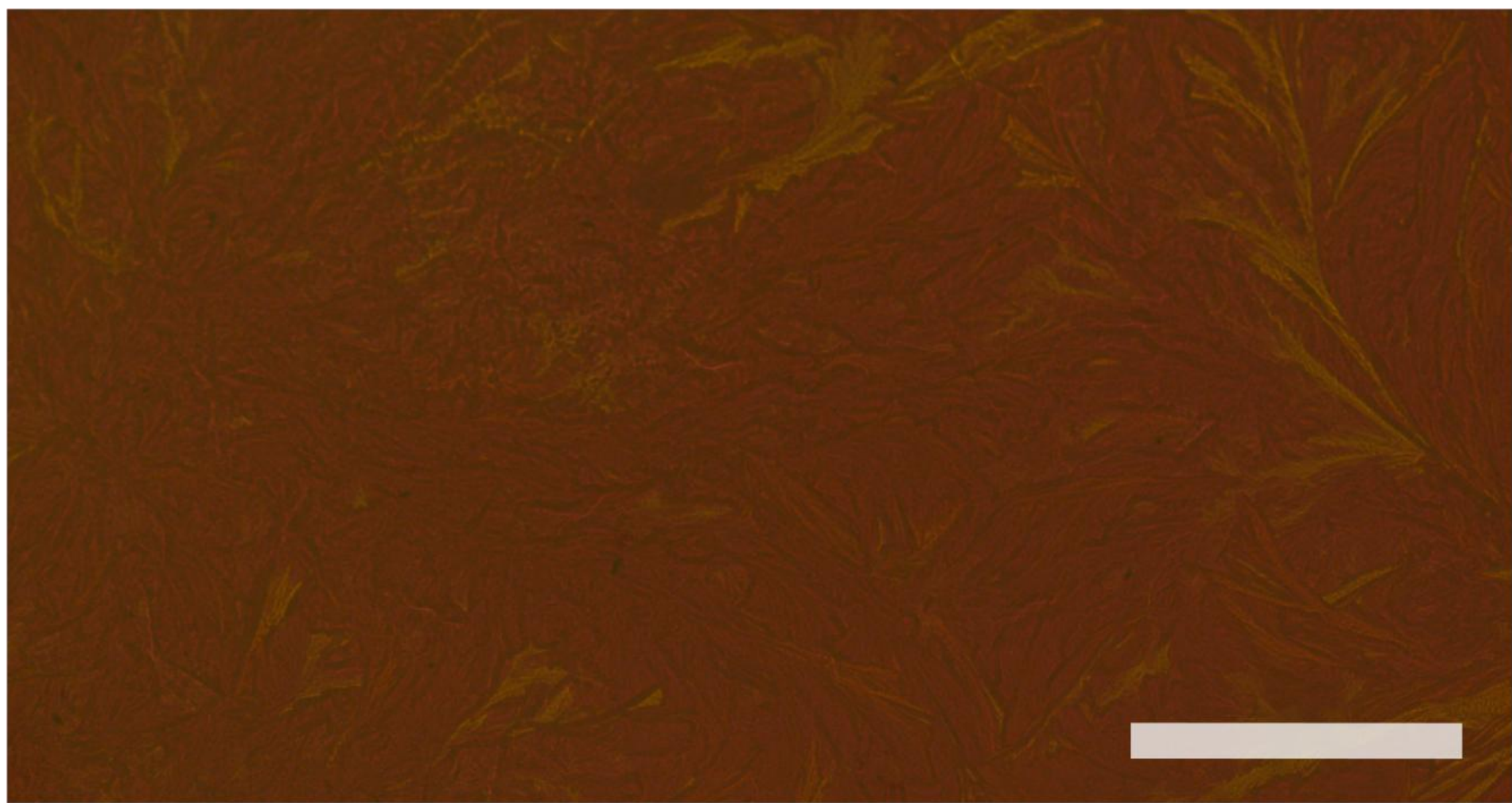

**Figure SI 26**: Backlit image of an UC film (no LiF/Al mirror on the film) a few days after fabrication shows crystallinity. The image was taken on the imaging setup from Figure 4a. The scale bar is 200 μm.

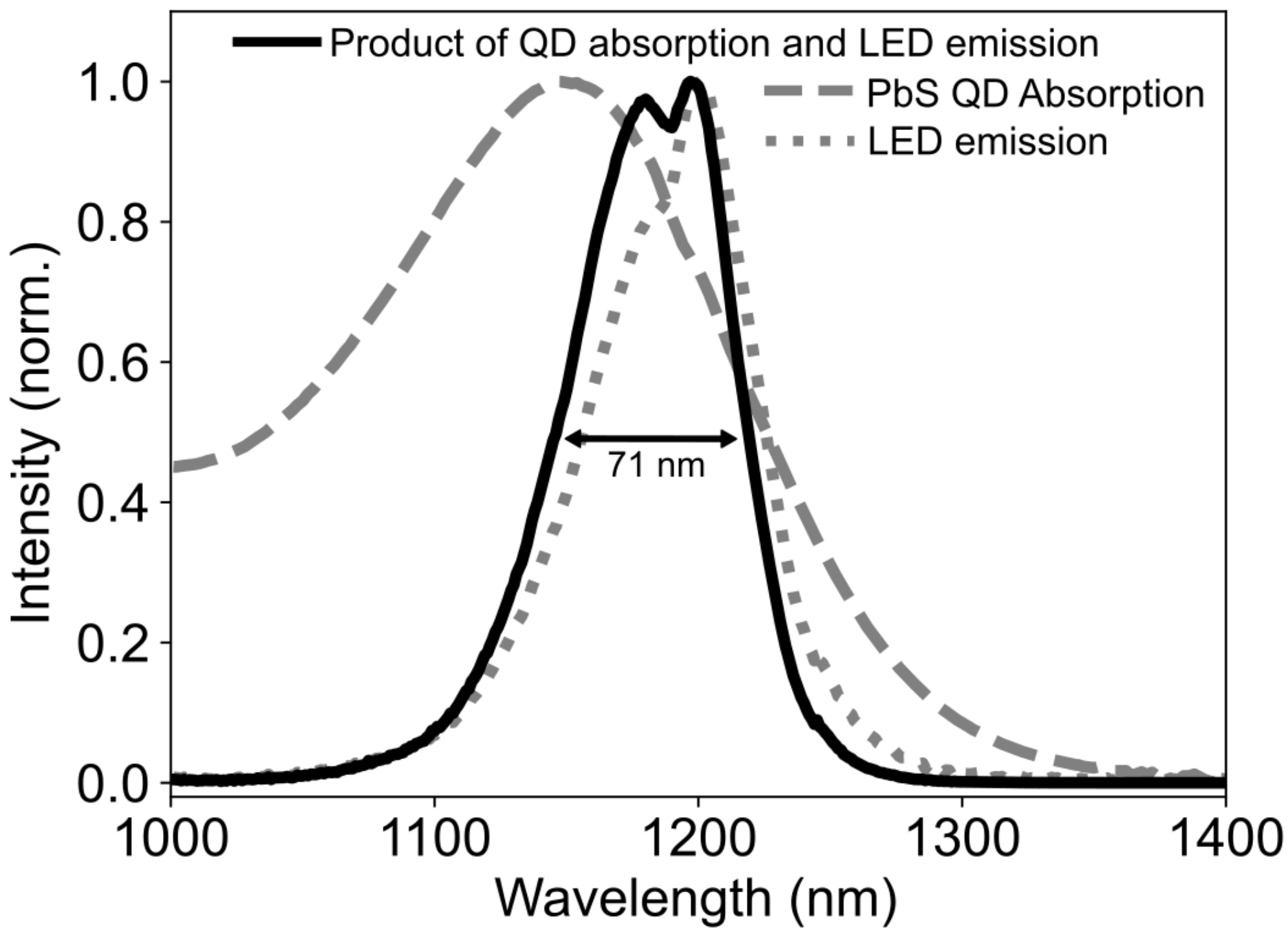

**Figure SI 27**: The wavelength of photons from the 1200 nm LED that are employed in UC-based imaging are highlighted by the black trace which is the product of the absorption spectrum of 1150 nm PbS QDs in hexanes (gray, dashed) and emission spectrum of 1200 nm LED (gray, dotted, raw data from Thorlabs datasheet).

**SI Note 9:** In this section we compare routes for UC imaging across the near-infrared (NIR, ~700–1550 nm) into the visible. Across these techniques, "efficiency" is reported under different definitions: quantum conversion efficiency for nonlinear upconversion processes, photon-to-photon (p–p) efficiency for biased upconversion devices, and quantum yield and internal quantum efficiency for photon upconversion. Hence, any direct numerical comparison should be interpreted with this in mind.

Nonlinear upconversion imaging was first demonstrated in 1968 by Midwinter[26]. Since then, the field has advanced rapidly, with several demonstrations of second- and third-order UC imaging across the NIR and MIR. Second order ($\chi^2$) processes such as second harmonic generation and sum-frequency generation in nonlinear crystals (e.g. periodically poled lithium niobate or potassium titanyl phosphate) have achieved high conversion efficiencies (~20-35%) under high intensity pump sources and long crystal path lengths[27–32]. However, the long path length that enables high efficiency also imposes a tighter phase-matching condition that restricts the angular acceptance cone of the upconversion process, fundamentally limiting the field of view for NIR imaging[31]. In recent years, researchers have proposed several strategies to enhance the field of view, including high-speed crystal rotation, engineered thermal gradients, and broadband pump sources, all with modest improvements in field of view[33–36]. A breakthrough came in 2022, when Huang et al. demonstrated for the first time, wide-field MIR upconversion imaging through chirped poling of lithium niobate crystal[37,38]. Although this advancement puts nonlinear upconversion imaging resolution and field of view at par with commercial imaging systems, it comes at the cost of reduced efficiency.

More recently, researchers have been trying to replace the bulk crystal with thin film of nonlinear material[39,40]. The reduced path length inside the nonlinear material relaxes the phase matching condition and hence removes the field of view limit associated with bulk crystals, but it also lowers the efficiency of the imaging process. Valenica et al. demonstrated that a nonlocal metasurface can be used to resonantly enhance the upconversion process and improve the conversion efficiency compared to a bare film, although the absolute efficiency remains low[41]. Silicon metasurfaces that support quasi-bound states in the continuum have been demonstrated for third-order ($\chi^3$) upconversion imaging such as third harmonic generation and four wave mixing[42–45]. These metasurfaces locally enhance the nonlinear field and hence increase the efficiency of the upconversion process, however, the same high-Q resonances can also introduce angle-dependence and limit the field of view, and the absolute efficiency remains challenging. In general, nonlinear processes exhibit great promise, however their dependence on a high-intensity pump/source, angular constraints, and narrow bandwidth limits their suitability for broadband and low-intensity NIR imaging applications such as night-vision.

Recently, organic upconverting devices have been demonstrated for NIR upconversion imaging[46–54]. These devices consist of an organic infrared photodetector coupled to an OLED in series, and when electrically biased, can sense low intensities of infrared light across the NIR-I window. There is no fundamental limit on field of view imposed by organic upconversion devices, and the limit is instead governed by active area of device (for proximity imaging) or set by the imaging optics and camera, which makes them a promising option for voltage powered NIR imaging and sensing.

Photon UC methods such as TTA and lanthanide-doped nanoparticles offer a passive alternative to the methods proposed above. In particular, lanthanide-doped nanoparticles have already been incorporated into contact lens to enable NIR vision in animals, and more recently humans, with full field of view[55–57]. However, one caveat is that these nanoparticles have a narrow bandwidth of operation in the NIR, as they rely on f-subshell transitions in rare-earth-doped ions to upconvert.

TTA, in contrast, can work across a broad wavelength range across NIR-I using broadband absorbing sensitizers. In this work, we push this working range into NIR-II with our demonstration of TTA-UC imaging of low-intensity incoherent 1200 nm light. We also note that there is no fundamental limit on the field of view for TTA-UC imaging and is instead set by the imaging optics and camera. Given this, we believe this work paves the way for the development of passive, broadband, and efficient NIR to visible imaging systems.

| Upconversion system | Medium used for upconversion | Bandwidth | Reported conversion efficiency | Field of view (Object size in diameter) |
| --- | --- | --- | --- | --- |
| **Second-harmonic generation**[27,58]<br>Reported efficiency is quantum conversion efficiency | 8 mm periodically poled potassium titanyl phosphate crystal (1342 nm to 671 nm)[27] (2018) | Narrowband, few nm | 0.04% | 6 mm |
| **Sum frequency generation** [29–35,38,40,41]<br>Utilizes pump source. Reported efficiency is quantum conversion efficiency | 20 mm periodically poled lithium niobate crystal<br>(1040 nm to 622 nm)[32] (2015) | Narrowband, few nm | 27% | ~3 mm |
| | 20 mm periodically poled lithium niobate crystal<br>(1563 nm to 660 nm)[33] (2018) | Narrowband, few nm | 35% | ~2.3 mm |
| | | | 7% [a] | ~6.6 mm[a] |
| | 2D material based nonlinear optical mirror using 45 nm of gallium selenide (1550 nm to 622 nm)[40] (2024) | Narrowband, few nm | 0.00001% | 5.6 mm |
| | 307 nm thick lithium niobate metasurface<br>(1530 nm to 550 nm)[41] (2024) | Narrowband, few nm | 0.0001%[b] | Not reported[c] |
| **Third-harmonic generation**[43–45]<br>Reported efficiency is quantum conversion efficiency | Silicon metasurface (1533 nm to 511 nm)[43] (2026) | Narrowband, few nm | 0.001%[b] | Not reported[c] |
| **Four wave mixing**[42,59]<br>Utilizes pump source.<br>Reported efficiency is quantum conversion efficiency | Hot rubidium vapor in vacuum cell (1529 nm to 780 nm)[59] (2012) | Narrowband, few nm | 0.045% | ~2 mm |
| | Silicon metasurface[42] (2024) | Broadband, 1000-4000 nm | $2 \times 10^{-6}$ % @ 2250 nm[b] | Not reported[c] |

| **Organic upconverting devices**[46–54]<br><br>Electrically biased. Reported efficiency is photon-to-photon conversion efficiency | Organic infrared detector + visible emitter (ClAlPc and BCzPh:CN-T2T: Ir(ppy)2(acac))[50] (2023) | Broadband, 700 – 1000 nm | Maximum of 31% @ 780 nm | Up to ~3 cm, limited by active device area and camera-lens system |
|---|---|---|---|---|
| | PbS QD based detector + organic visible emitter[49] (2026) | Broadband, 700-1100 nm | 23 % @ 850 nm | ~1 cm, limited by active device area and camera-lens system |
| **Lanthanide-doped nanoparticles**[55–57]<br>Reported efficiency is quantum yield | Upconversion microspheres using NaYF4:Yb embedded inside contact lens (980 nm to 550 nm)[56] (2025) | Narrowband, ~20 nm | Maximum of 9% @ 980 nm | Full field of view |
| **Triplet-triplet annihilation**[60]<br><br>Reported efficiency is internal quantum efficiency | 100 nm thin film of Y6/rubrene/DBP[60] (2025) | Broadband, 700- 1000 nm | 0.5% @ 808 nm | ~10 mm, limited by camera-lens system |
| | This work<br><br>3300 nm thin film of PbS/TES-ADT/DBP (2026) | Broadband, 800 – 1200 nm | 0.00086% @ 1200 nm<br><br>0.0029% @1130 nm<br><br>0.15% @ 808 nm | ~5 mm, limited by camera-lens system |

[a]Results reported after increasing the spectral width of the pump laser to increase field of view.
[b]Converted from power-to-power efficiency
[c]Possibly limited by metasurface area /camera-lens system/resonance angular response.

**Table SI 9**: Comparison of metrics across different representative NIR upconversion imaging and sensing systems.

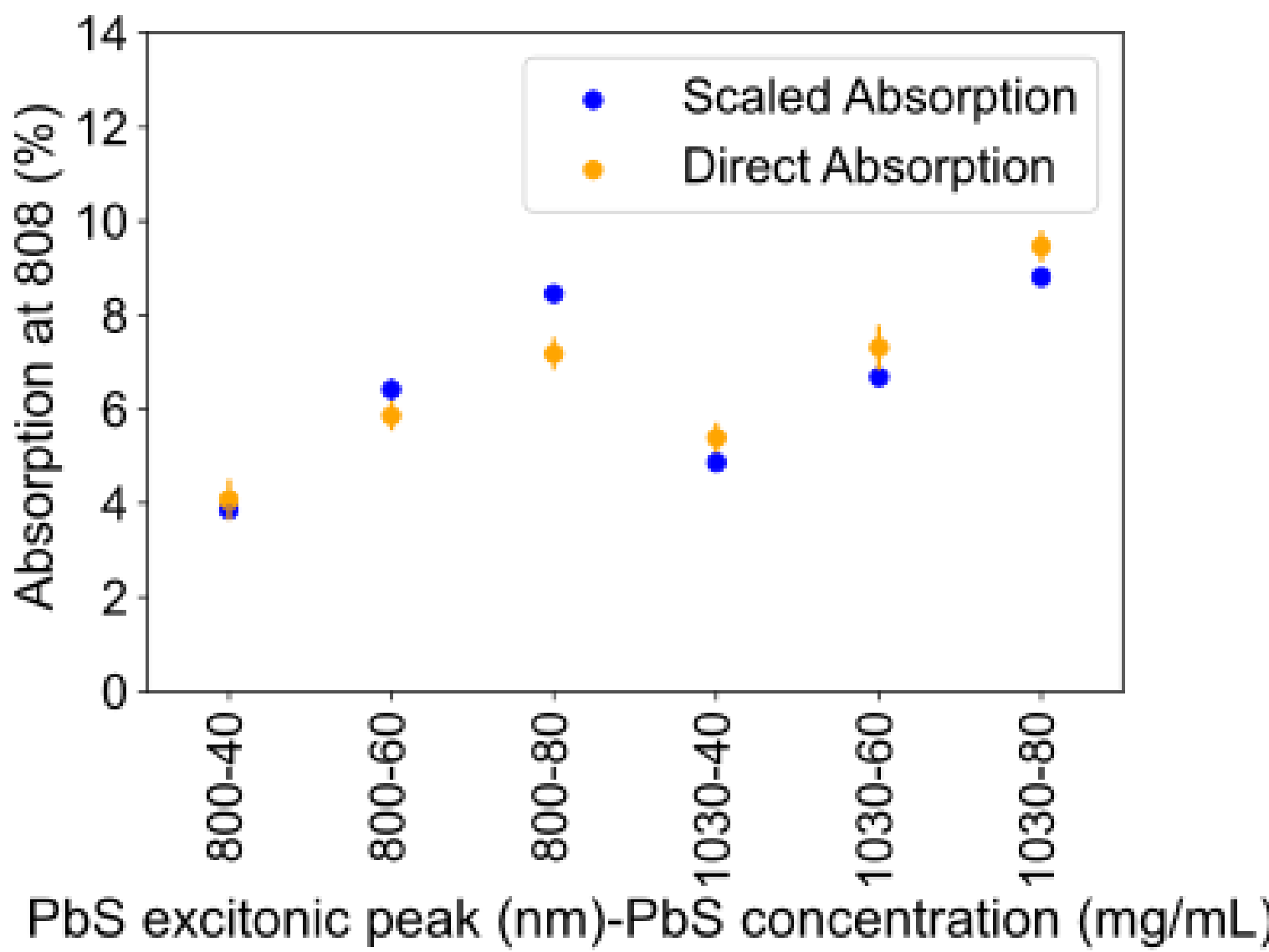

**Figure SI 28:** Percent absorption for several samples measured directly at 808 nm (yellow) compared to the scaled percent absorption at 808 nm based on absorption measurements at 735 nm (blue).

SI Note 10: Due to the wavelength restrictions of our silicon-based spectrometers, direct measurement of absorption at 1130 nm and 1208 nm using the integrating sphere was not possible. Additionally, full R/T data proved difficult to collect due to thin film interference and scattering artifacts. To corroborate that our absorption scaling methodology is valid, we have directly measured absorption at 808 nm of several films and compared it to the scaled absorption we calculate using the 735 nm absorption (Figure SI 28). We measured the absorption of three UC films each, of 800 nm and 1030 nm QDs with varying concentrations (40, 60, 80 mg/mL). Upon normalizing, the deviation between the two methodologies across these samples is only (2.7±0.6)%, suggesting no systematic error.

REFERENCES

1. Zhou, Y., Castellano, F. N., Schmidt, T. W. & Hanson, K. On the Quantum Yield of Photon Upconversion via Triplet–Triplet Annihilation. *ACS Energy Lett.* **5**, 2322–2326 (2020).
2. Radiunas, E. *et al.* CN-Tuning: A Pathway to Suppress Singlet Fission and Amplify Triplet-Triplet Annihilation Upconversion in Rubrene. *Adv. Opt. Mater.* **13**, 2403032 (2025).
3. Zhao, S. *et al.* Efficient Near-Infrared to Blue Photon Upconversion by Ultrafast Spin Flip and Triplet Energy Transfer at Organic/2D Semiconductor Interface. *Angew. Chem.* **137**, e202420070 (2025).
4. Izawa, S. & Hiramoto, M. Efficient solid-state photon upconversion enabled by triplet formation at an organic semiconductor interface. *Nat. Photonics* **15**, 895–900 (2021).
5. Duan, J. *et al.* Efficient solid-state infrared-to-visible photon upconversion on atomically thin monolayer semiconductors. *Sci. Adv.* **8**, eabq4935 (2022).
6. Ho, E. A., Soni, A., Zhai, F. & Wang, L. Pseudo-Solid-State Polymer Materials for QD-Sensitized NIR-I and NIR-II Upconversion Beyond the Silicon Bandgap. *Adv. Mater.* e12741 (2025) doi:10.1002/adma.202512741.
7. Nienhaus, L. *et al.* Triplet-Sensitization by Lead Halide Perovskite Thin Films for Near-Infrared-to-Visible Upconversion. *ACS Energy Lett.* **4**, 888–895 (2019).
8. Wang, L. *et al.* Interfacial Trap-Assisted Triplet Generation in Lead Halide Perovskite Sensitized Solid-State Upconversion. *Adv. Mater.* **33**, 2100854 (2021).
9. Geva, N. *et al.* A Heterogeneous Kinetics Model for Triplet Exciton Transfer in Solid-State Upconversion. *J. Phys. Chem. Lett.* **10**, 3147–3152 (2019).
10. Wu, M. *et al.* Solid-state infrared-to-visible upconversion sensitized by colloidal nanocrystals. *Nat. Photonics* **10**, 31–34 (2016).

11. Nienhaus, L. *et al.* Speed Limit for Triplet-Exciton Transfer in Solid-State PbS Nanocrystal-Sensitized Photon Upconversion. *ACS Nano* **11**, 7848–7857 (2017).

12. Wu, M., Jean, J., Bulović, V. & Baldo, M. A. Interference-enhanced infrared-to-visible upconversion in solid-state thin films sensitized by colloidal nanocrystals. *Appl. Phys. Lett.* **110**, 211101 (2017).

13. Hu, M. *et al.* Bulk Heterojunction Upconversion Thin Films Fabricated via One-Step Solution Deposition. *ACS Nano* **17**, 22642–22655 (2023).

14. Narayanan, P. *et al.* Alleviating Parasitic Back Energy Transfer Enhances Thin Film Upconversion. *Adv. Opt. Mater.* **13**, 2500252 (2025).

15. Bi, P. *et al.* Donor-acceptor bulk-heterojunction sensitizer for efficient solid-state infrared-to-visible photon up-conversion. *Nat. Commun.* **15**, 5719 (2024).

16. Amemori, S., Sasaki, Y., Yanai, N. & Kimizuka, N. Near-Infrared-to-Visible Photon Upconversion Sensitized by a Metal Complex with Spin-Forbidden yet Strong $S_0$ –$T_1$ Absorption. *J. Am. Chem. Soc.* **138**, 8702–8705 (2016).

17. Kinoshita, M. *et al.* Photon Upconverting Solid Films with Improved Efficiency for Endowing Perovskite Solar Cells with Near-Infrared Sensitivity. *ChemPhotoChem* **4**, 5271–5278 (2020).

18. Tripathi, N., Ando, M., Akai, T. & Kamada, K. Near-infrared-to-visible upconversion from 980 nm excitation band by binary solid of PbS quantum dot with directly attached emitter. *J. Mater. Chem. C* **10**, 4563–4567 (2022).

19. Tripathi, N. & Kamada, K. Enhanced Near-Infrared-to-Visible Upconversion by a Singlet Sink Approach in a Quantum-Dot-Sensitized Triplet–Triplet Annihilation System. *ACS Appl. Nano Mater.* **7**, 2950–2955 (2024).

20. Wu, M., Lin, T.-A., Tiepelt, J. O., Bulović, V. & Baldo, M. A. Nanocrystal-Sensitized Infrared-to-Visible Upconversion in a Microcavity under Subsolar Flux. *Nano Lett.* **21**, 1011–1016 (2021).

21. Moreels, I. *et al.* Size-Dependent Optical Properties of Colloidal PbS Quantum Dots. *ACS Nano* **3**, 3023–3030 (2009).

22. Kessler, M. L., Starr, H. E., Knauf, R. R., Rountree, K. J. & Dempsey, J. L. Exchange equilibria of carboxylate-terminated ligands at PbS nanocrystal surfaces. *Phys. Chem. Chem. Phys.* **20**, 23649–23655 (2018).

23. Giansante, C. *et al.* "Darker-than-Black" PbS Quantum Dots: Enhancing Optical Absorption of Colloidal Semiconductor Nanocrystals via Short Conjugated Ligands. *J. Am. Chem. Soc.* **137**, 1875–1886 (2015).

24. Giansante, C. Surface Chemistry Impact on the Light Absorption by Colloidal Quantum Dots. *Chem. – Eur. J.* **27**, 14358–14368 (2021).

25. Gholizadeh, E. M. *et al.* Photochemical upconversion of near-infrared light from below the silicon bandgap. *Nat. Photonics* **14**, 585–590 (2020).

26. Midwinter, J. E. IMAGE CONVERSION FROM 1.6 μ TO THE VISIBLE IN LITHIUM NIOBATE. *Appl. Phys. Lett.* **12**, 68–70 (1968).

27. Torregrosa, A. J., Maestre, H., Rico, M. L. & Capmany, J. Compact self-illuminated image upconversion system based on intracavity second-harmonic generation. *Opt. Lett.* **43**, 5050–5053 (2018).

28. Chalopin, B., Chiummo, A., Fabre, C., Maître, A. & Treps, N. Frequency doubling of low power images using a self-imaging cavity. *Opt. Express* **18**, 8033–8042 (2010).

29. Torregrosa, A. J., Maestre, H. & Capmany, J. Intra-cavity upconversion to 631 nm of images illuminated by an eye-safe ASE source at 1550 nm. *Opt. Lett.* **40**, 5315–5318 (2015).

30. Maestre, H., Torregrosa, A. J. & Capmany, J. IR Image Upconversion Under Dual-Wavelength Laser Illumination. *IEEE Photonics J.* **8**, 1–8 (2016).

31. Barh, A., Rodrigo, P. J., Meng, L., Pedersen, C. & Tidemand-Lichtenberg, P. Parametric upconversion imaging and its applications. *Adv. Opt. Photonics* **11**, 952–1019 (2019).

32. Tang, R. *et al.* Two-dimensional infrared and mid-infrared imaging by single-photon frequency upconversion. *J. Mod. Opt.* **62**, 1126–1131 (2015).

33. Demur, R. *et al.* Near-infrared to visible upconversion imaging using a broadband pump laser. *Opt. Express* **26**, 13252–13263 (2018).

34. Maestre, H., Torregrosa, A. J., Fernández-Pousa, C. R. & Capmany, J. IR-to-visible image upconverter under nonlinear crystal thermal gradient operation. *Opt. Express* **26**, 1133–1144 (2018).

35. Liu, S.-K. *et al.* Up-Conversion Imaging Processing With Field-of-View and Edge Enhancement. *Phys. Rev. Appl.* **11**, 044013 (2019).

36. Junaid, S. *et al.* Video-rate, mid-infrared hyperspectral upconversion imaging. *Optica* **6**, 702–708 (2019).

37. Fang, J. *et al.* Wide-field mid-infrared hyperspectral imaging beyond video rate. *Nat. Commun.* **15**, 1811 (2024).

38. Huang, K., Fang, J., Yan, M., Wu, E. & Zeng, H. Wide-field mid-infrared single-photon upconversion imaging. *Nat. Commun.* **13**, 1077 (2022).

39. Camacho-Morales, M. del R. *et al.* Infrared upconversion imaging in nonlinear metasurfaces. *Adv. Photonics* **3**, 036002 (2021).

40. Konkada Manattayil, J. *et al.* 2D Material Based Nonlinear Optical Mirror for Widefield Up-Conversion Imaging from Near Infrared to Visible Wavelengths. *Laser Photonics Rev.* **18**, 2400374 (2024).

41. Valencia Molina, L. *et al.* Enhanced Infrared Vision by Nonlinear Up-Conversion in Nonlocal Metasurfaces. *Adv. Mater.* **n/a**, 2402777.

42. Zheng, Z. *et al.* Broadband infrared imaging governed by guided-mode resonance in dielectric metasurfaces. *Light Sci. Appl.* **13**, 249 (2024).

43. Liu, T. *et al.* High-efficiency infrared upconversion imaging with nonlinear silicon metasurfaces empowered by quasi-bound states in the continuum. *Opto-Electron. Adv.* **9**, 250257–10 (2026).

44. Liu, T. *et al.* Polarization-Independent Enhancement of Third-Harmonic Generation Empowered by Doubly Degenerate Quasi-Bound States in the Continuum. *Nano Lett.* **25**, 3646–3652 (2025).

45. Zheng, Z. *et al.* Third-harmonic generation and imaging with resonant Si membrane metasurface. *Opto-Electron. Adv.* **6**, 220174–10 (2023).

46. Du, X. *et al.* Efficient Organic Upconversion Devices for Low Energy Consumption and High-Quality Noninvasive Imaging. *Adv. Mater.* **33**, 2102812 (2021).

47. Zhang, C.-F. *et al.* Infrared imaging visualization: organic material-based up-conversion devices. *J. Mater. Chem. C* **13**, 19149–19167 (2025).

48. Yang, D. *et al.* Near infrared to visible light organic up-conversion devices with photon-to-photon conversion efficiency approaching 30%. *Mater. Horiz.* **5**, 874–882 (2018).

49. Li, J.-L. *et al.* Photomultiplication Enables Efficient NIR Quantum Dot Organic Upconversion Devices. *ACS Photonics* **13**, 1456–1464 (2026).

50. Shih, C.-J. *et al.* Semi-Transparent, Pixel-Free Upconversion Goggles with Dual Audio-Visual Communication. *Adv. Sci.* **10**, 2302631 (2023).

51. Dong, S. *et al.* Ternary Compensation Enables High-sensitive Efficient Upconversion Device for NIR Visualization. *Adv. Mater.* **37**, 2412678 (2025).

52. Shih, C.-J. *et al.* Transparent organic upconversion device targeting high- grade infrared visual image. *Nano Energy* **86**, 106043 (2021).

53. Han, X. *et al.* Ultralow Turn-On Voltage Organic Upconversion Devices for High-Resolution Imaging Based on Near-Infrared Homotandem Photodetector. *Laser Photonics Rev.* **19**, 2401375 (2025).

54. Shih, C.-J. *et al.* Transparent organic upconversion devices displaying high-resolution, single-pixel, low-power infrared images perceived by human vision. *Sci. Adv.* **9**, eadd7526 (2023).

55. Ma, Y. *et al.* Near-infrared spatiotemporal color vision in humans enabled by upconversion contact lenses. *Cell* **188**, 3375-3388.e18 (2025).

56. Wang, D. *et al.* Ultralow-threshold upconversion infrared vision via a microsphere-mediated directional photofield. *Nat. Commun.* **16**, 5064 (2025).

57. Hu, Y. *et al.* Upconversion nanoparticles doped optical lens: let's see the near-infrared light. *J. Nanobiotechnology* **22**, 332 (2024).

58. Chalopin, B., Chiummo, A., Fabre, C., Maître, A. & Treps, N. Frequency doubling of low power images using a self-imaging cavity. *Opt. Express* **18**, 8033–8042 (2010).

59. Ding, D.-S. *et al.* Experimental up-conversion of images. *Phys. Rev. A* **86**, 033803 (2012).

60. Hamid, R. *et al.* Photonic Engineering Enables All-Passive Upconversion Imaging with Low-Intensity Near-Infrared Light. *Adv. Funct. Mater.* **36**, e15334 (2026).